# The AFLOW Library of Crystallographic Prototypes: Part 4

Hagen Eckert ,[1, 2] Simon Divilov ,[1, 2] Michael J. Mehl ,[1, 2, *] David Hicks ,[1, 2]
Adam C. Zettel ,[1, 2] Marco Esters ,[1, 2] Xiomara Campilongo ,[2] and Stefano Curtarolo [1, 2, †]

[1]*Department of Mechanical Engineering and Materials Science, Duke University, Durham, NC 27708, USA*
[2]*Center for Extreme Materials, Duke University, Durham, NC 27708, USA*
(Dated: February 15, 2024)

The AFLOW Library of Crystallographic Prototypes has been updated to include an additional 683 entries, which now reaches 1,783 prototypes. We have also made some changes to the presentation of the entries, including a more consistent definition of the AFLOW-prototype label and a better explanation of our choice of space group when the experimental data is ambiguous. A method is presented for users to submit new prototypes for the Encyclopedia. We also include a complete index linking to all the prototypes currently in the Library.

## I. INTRODUCTION

There are a variety of available resources to find crystallographic information about the structure of materials. These include the *Pauling File Project* [1], the *American Mineralogist Crystal Structure Database* (AMCSD) [2], the Inorganic Crystal Structure Database (ICSD) [3], the Cambridge Structural Database (CSD) [4], and our own AFLOW (Automatic FLOW for Materials Discovery) [5]. This list is by no means complete. Most of these databases group similar crystal structures under one prototype, a specific compound that defines the entire class. These prototypes are usually labeled by the chemical formula of the prototype structure, which, while helpful, does not guarantee a unique definition. In fact, there is no agreed-upon method for designating prototypes, though many have been proposed and are in use [6].

Our approach is the AFLOW-prototype label [7], an alphanumeric string describing stoichiometry, unit cell Pearson symbol and space group, and occupied Wyckoff positions for each atomic species of a given structure. The label is the same for all structures in the prototype class and is independent of the choice of prototype compound, although we always make one choice to use as an example. All structures in a prototype class have the same AFLOW-prototype label. An example is the rock salt structure: NaCl — halite — and hundreds of binary compounds [8, 9] all have the same prototype label, `AB_cF8_225_a_b`. The AFLOW-prototype label is not unique in its basic form. As a trivial example, the label `AB_cF8_225_b_a` also describes the rock salt structure. In other cases, two structures with different atomic arrangements might have the same prototype label. The construction of a unique label requires additional conditions to the naming convention that we explore in this article.

We use the AFLOW-prototype label to refer to structures in our Library of Crystallographic Prototypes. This database began with 298 structures [7] with later additions bringing the total to 1,100 [6, 10]. Here we report the addition of 683 new prototypes. Many of these were chosen by scanning the literature to find systems currently being explored by researchers. All have been included in the updated AFLOW-XtalFinder [11] and can be accessed directly from the `aflow++` software [12] and through the AFLOW.org web [5]. All the individual entries in the current library have been updated to include additional compounds that the same prototype can describe and to correct errors that unavoidably occur in a large project such as this. The library is now designed to be updated regularly, and we are accepting suggested prototypes from users.

This article describes the changes we made to the Library of Crystallographic Prototypes. The following section discusses the new structures added to the Library, changes in AFLOW-prototype labels and space groups, cross references with other databases, and extensions to the online version of the Library, the AFLOW Encyclopedia of Crystallographic Prototypes.

## II. CHANGES AND UPDATES TO THE LIBRARY OF CRYSTALLOGRAPHIC PROTOTYPES

The newest version of the Library includes several updates and changes in the way we index and present the prototypes. These are highlighted below.

### A. New Structures

We have added an additional 683 prototypes to the database, bringing the total number of entries to 1,783 at the time of submission. Newly added systems include:

- Aurivillius phases [15] at the request of users;
- $Ti_3Cu_4$ [16] and others, based on recent mentioned in literature [17];
- $ReB_2$ [18] and others, to extend previous class of entries, in this case, the $Re_xB_y$ system; and
- $U_2Co_3Si_5$ and other less-common structures [19], to provide more examples of structures in sparsely populated space groups.

* michael.mehl@duke.edu
† stefano@duke.edu

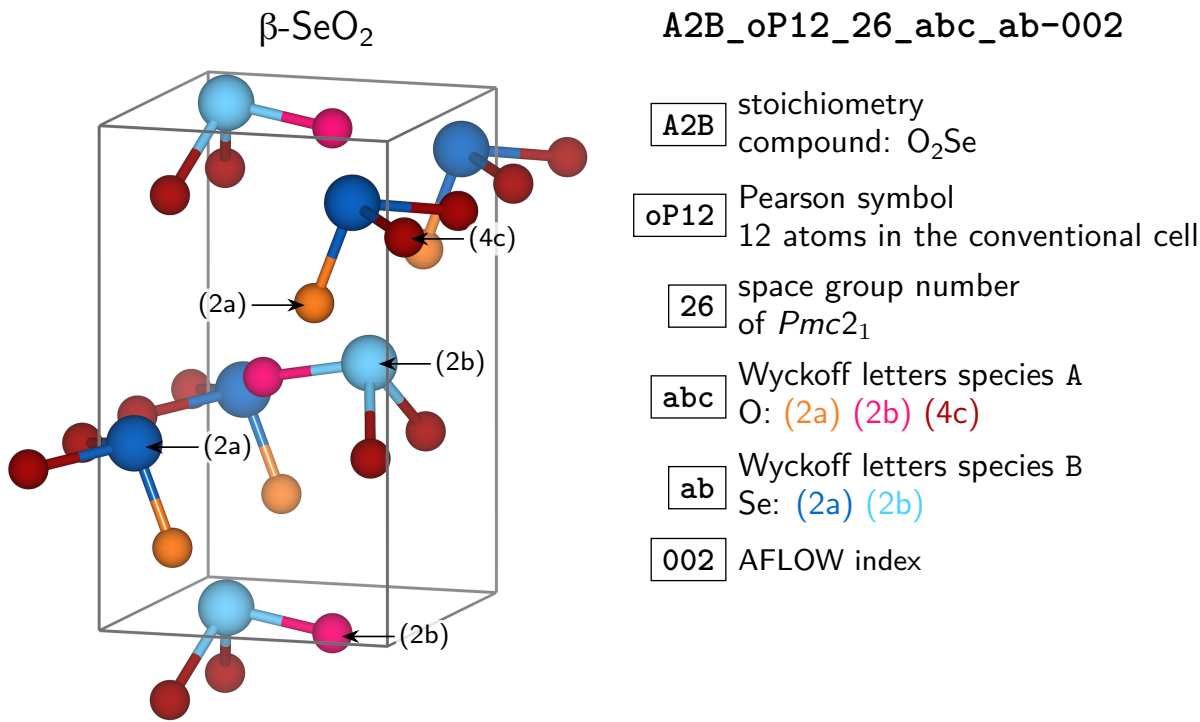

β-SeO₂

**A2B_oP12_26_abc_ab-002**

A2B — stoichiometry
compound: O₂Se

oP12 — Pearson symbol
12 atoms in the conventional cell

26 — space group number
of *Pmc*2₁

abc — Wyckoff letters species A
O: (2a) (2b) (4c)

ab — Wyckoff letters species B
Se: (2a) (2b)

002 — AFLOW index

FIG. 1. Deconstruction of the AFLOW-prototype label, describing the atoms in the conventional unit cell of the system. On the left oxygen is drawn using VESTA [13] in shades of red, while blue is used for selenium. The different Wyckoff positions share the same colors and are labeled in the figure. The stoichiometry term is determined by listing the elements in the crystal in alphabetical order, calling the first element A, the second B, etc. The multiplicity of each element is indicated by the number after its letter. This is followed by the Pearson Symbol and the space group number assigned in the International Tables of Crystallography [14]. The Wyckoff letters are then listed for each atom. If there are multiple instances of the same Wyckoff positions for an atom, it is indicated by a number placed *before* the Wyckoff letter. Finally, the AFLOW index concludes the label. The -002 here indicates that at least one other structure in the AFLOW database has the label A2B_oP12_26_abc_ab.

All structures currently in the Encyclopedia are listed in the Appendix, with new structures highlighted by a filled bullet point (•). Links to the individual entries are provided there. Going forward from this point, we plan frequent updates to the entries in the online Encyclopedia of Crystallographic Prototypes [20], as well as streamlined reports for archival literature.

**B. AFLOW Labels**

Any database of prototypes requires a unique method to catalog each entry. AFLOW and the Library of Crystallographic Prototypes are no exception. A unique label allows quick access to the Library's web pages and facilitates the generation of new structures within AFLOW. It would be helpful if there were a universally recognized method for labeling crystalline prototypes, but this goal has yet to be achieved.

The Library originated as a web page at the U.S. Naval Research Laboratory (NRL) [21]. There a prototype was usually referenced by its *Strukturbericht* designation [22–28], if one existed, or the chemical formula. These methods are not sustainable for large-scale databases: the practice of defining new *Strukturbericht* labels ceased early in the post World War II period. The seemingly obvious idea of using chemical formulas fails as many of these entities, e.g. WO₃, can be found in multiple phases even under ambient conditions, with many of the phases designated as the prototypes for their class. AFLOW circumvents this problem by introducing the prototype label [7] . The method is outlined in Fig. 1. Consider a chemical compound, here β-SeO₂ [29]. We have a complete description of this structure from the literature [30]. This includes the space group, the lattice constants, the Wyckoff positions occupied by each species, and the numerical values used to locate each atom associated with a given Wyckoff position. The prototype label is generated from this information, to wit:

- The chemical formula is written in alphabetical order: SeO₂ → O₂Se. Assign the letter A to the first element and the letter B to the second. The stoichiometry of the compound is given by a number (if not 1) after the letter. Thus SeO₂ becomes A2B. This construction allows other compounds with the same structure to use this label.

- This compound has a primitive orthorhombic lat-

tice with twelve atoms in the unit cell. The Pearson symbol [31] for this structure is then `oP12`.

- The compound is in space group $Pmc2_1$, which is #26 in the International Tables of Crystallography [14].

- The selenium atoms occupy (2a), (2b), and (4c) Wyckoff positions [32], while the oxygen atoms are on (2a) and (2b) sites. The numbers represent the number of atoms associated with each Wyckoff position, but they are redundant: the Wyckoff label (4c) and the Wyckoff letter `c` convey the same information.

The prototype label is constructed from these pieces: the abstract chemical formula, the Pearson symbol, the space group number, and the Wyckoff letters of each element, all separated by underscores. This gives us the prototype label `A2B_oP12_26_abc_ab`.

Now consider a slightly more complicated case, the $Na_2PrO_3$ structure [33]. This is a base-centered monoclinic structure, space group $C2/c$ #15. The sodium atoms are on (2a), (4e), and (8f) Wyckoff sites, praseodymium is found on two (4e) sites, and the oxygen atoms occupy three (8e) sites. When one atomic species occupies multiple Wyckoff positions with the same letter, we indicate this by placing the number of sites *before* the Wyckoff letter, so the prototype label is `A2B3C_mC48_15_aef_3f_2e`.

Once the label is constructed, and with knowledge of the atomic species and the numerical values of the positional parameters, AFLOW can generate the entire crystal structure. Using experimental data [30] for β-$SeO_2$, the AFLOW command:

```
aflow --proto=A2B_oP12_26_abc_ab:O:Se
--params=5.0722,0.88135,1.48474,0.746,0.672,
0.1219,0.3748,0.620,-0.039,0.2516,0.000,
0.247,0.152,0.841
```

generates the VASP POSCAR file for this structure. Adding the flag `--cif` to the command will instead produce a Crystallographic Information File (CIF) [34]. This process works even if the AFLOW-prototype label is not in the database.

The procedure outlined above is sufficient to generate a crystal structure for any periodic crystal. Once a prototype is incorporated into the database, however, problems can arise because the labeling still needs to produce unique labels. The first difficulty arises when two or more distinct crystal structures have the same label. For example, α-gallium (*Strukturbericht* designation $A11$), crystalline molecular iodine ($A14$), and black phosphorous ($A17$) all occupy a single (8f) Wyckoff position in space group $Cmca$ #64, giving all of these structures the prototype label `A_oC8_64_f`. The crystals are very different, however, as seen in Fig. 2, which compares α-Ga and molecular $I_2$. While the structures have the same label, they cannot be said to have the same prototype. In the past, we distinguished these structures using arbitrary labels, in this case α-Ga [35], I [36], and P [37], but this is generally unwieldy. Our new method distinguishes these structures by adding a three-digit suffix to the label. This index number is determined by the order in which the structure was added to the AFLOW database [38]. In this case,

- α-Ga is `A_oC8_64_f-001`,
- phosphorous is `A_oC8_64_f-002`, and
- molecular iodine is `A_oC8_64_f-003`.

As a further example, again consider β-$SeO_2$, AFLOW label `A2B_oP12_26_abc_ab`. This prototype label is the same for both β-$SeO_2$ and the high-pressure (70 GPa) $H_2S$ structure [39]. In previous versions of the Library we distinguished between the two structures by placing the chemical formula after the label, e.g. `A2B_oP12_26_abc_ab.H2S` and `A2B_oP12_26_abc_ab.SeO2`, respectively. $H_2S$ was added to our database before β-$SeO_2$, so the new labels become `A2B_oP12_26_abc_ab-001` and `A2B_oP12_26_abc_ab-002`, respectively.

While we will always use the new labels in future work, the Library will maintain backward compatibility. In particular, this will allow links in our previously published reports to continue to function. Thus a reference to the label `A2B_oP12_26_abc_ab.H2S` will link to $H_2S$ (`A2B_oP12_26_abc_ab-001`), while `A2B_oP12_26_abc_ab.SeO2` will link to β-$SeO_2$ (`A2B_oP12_26_abc_ab-002`). Similarly both `A_oC8_64_f.alpha-Ga` and `A_oC8_64_f-001` will resolve as $A11$ α-gallium.

One could hope that the above criteria would produce a unique prototype label for every structure. This turns out to not be the case, as many space groups permit several different choices of origin [40] and orientation [41] to describe the same structure. Each choice of origin and orientation will produce a different set of occupied Wyckoff positions.

For example, consider the compound $UTe_2$ [42], which has an orthorhombic structure in space group $Immm$ #71. Hutanu et al. [43] describe this structure in an orientation with lattice constants $c > b > a$. Using this orientation, the crystal has the AFLOW-prototype label `A2B_oI32_71_hj_i`. Alternatively, since space group $Immm$ allows us to make any permutation of the axes we wish to describe a structure, we could instead rotate the system so that $a > b > c$. In that case, the AFLOW-prototype label becomes `A2B_oI32_71_eh_f`. Other permutations give other possible labels. Still, more choices occur because the space group allows an origin shift of one-half of the lattice constant along any of the principle axes.

To avoid confusion, AFLOW versions beginning with 3.2.14 fix the origin and orientation of the unit cell by

- assigning a numerical score to each Wyckoff letter: $a = 1$, $b = 2$, etc., and take the sum over all Wyckoff positions. Only labels with the smallest sum will be considered.

- If there are still multiple permutations of Wyckoff

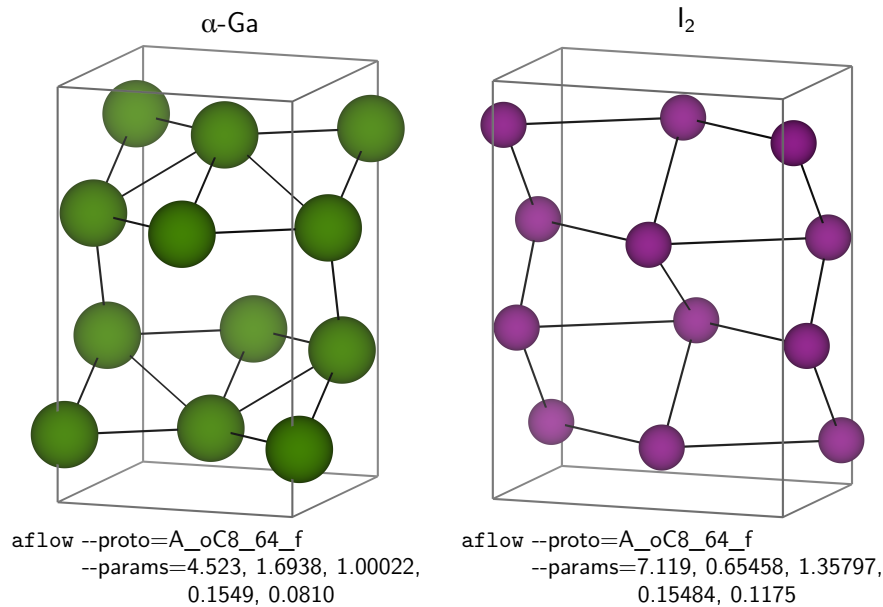

α-Ga

I₂

```
aflow --proto=A_oC8_64_f
    --params=4.523, 1.6938, 1.00022,
        0.1549, 0.0810
```

```
aflow --proto=A_oC8_64_f
    --params=7.119, 0.65458, 1.35797,
        0.15484, 0.1175
```

FIG. 2. Crystal structures of α-Ga (left) and molecular iodine (right), drawn using VESTA [13]. Both structures use are in the same space group with a single (8f) occupied Wyckoff position, so they have the same AFLOW-prototype label, `A_oC8_64_f`. The boxes delineated the conventional unit cells for each structure, and black lines show the connection to the nearest neighbors, to guide the eye. The AFLOW command below each structure generates that structure.

positions with this sum, chose the one where the first atomic species has the smallest Wyckoff letter.

In our UTe₂ example, this gives us the label `A2B_oI32_71_eh_f-001`, the suffix signaling that this is the first structure we have found with this label. AFLOW will always report this as the label for UTe₂, even if it is generated using data from another orientation.

Combining AFLOW-prototype label plus suffix can generate unwieldy labels. We introduce a unique identifier (UID) for all published prototypes to alleviate this. This four-character UID is created using a similar approach to the AUID used for AFLOW database entries. The prototype label is hashed by utilizing a 64-bit variant of the cyclic redundancy check (CRC64) [44] with "Jones" coefficients. The resulting 64-bit integer is then represented as a string using a set of 34 characters (uppercase alphanumeric without the letters I and O to avoid confusion). As the number of anticipated prototypes is considerably smaller than the number of possible entries in the AFLOW database, just a fraction of the 64-bit long hash is utilized. With just the first four characters of the produced hash, over 1.3 million prototypes could be differentiated. While this is ample space to describe all foreseeable prototypes, the convenience of a smaller UID leads to a high risk of hash collisions. As creating a new label for a prototype depends on a database lookup to determine the suffix number, a collision can be avoided at this point by using the result of the first attempt as new input for the UID creation. This process can be repeated until a unique UID is found. The created UIDs are shown in the index in the Appendix and on the prototype index page. If the UID is known, the structure may be retrieved from the Library using the link `http://aflow.org/p/UID`, where UID is replaced by the four-character UID for the structure.

## C. Assignment of the Space Group

By convention, a structure that could be described in multiple space groups is placed in the space group with the highest symmetry as determined by its index number in the International Tables. In practice, structures with relatively high symmetry are frequently reported as having lower symmetry [45–47]. This often happens when a reference places the structure in a low-symmetry space group, but assigns lattice and Wyckoff parameters consistent with a higher symmetry. An example is PtSn [48]. The original publication [49] placed this system in space group $Aea2$ (#41) with a primitive orthorhombic cell. The platinum atoms occupy the (2a) site. The tin atoms occupy two (2b) Wyckoff sites, with $x_1 = y_2$, $x_2 = y_1$, and $z_2 = -z_1$. These relations make the atomic positions consistent with the higher symmetry space group $Ccce$ (#68), with a base-centered orthorhombic primitive cell. In the former case, the two tin sites are nominally independent, but in the latter case, all tin atoms are on a single (16i) Wyckoff position.

Our approach to cases like this has varied. We usually place the problematic structure in the higher symmetry space group, especially if it was a structure highlighted by Cenzual et al. [45, 46] or similar works. In the case of PtSn$_4$ (and a few other structures), we decided to emphasize the nature of the lower symmetry space group, $Aea2$, as there are relatively few structures in that group. We did this by slightly changing the lattice constants and/or atomic coordinates. In either case, the comments accompanying each entry discuss the situation: the published space group, the higher symmetry space group, and our choice of the space group to display.

Another problem arises when the experimentally reported structure is correctly placed in a low-symmetry space group. However, a small amount of uncertainty in the atomic positions or lattice parameters would make the structure consistent with a structure with a higher symmetry. In this case, electronic structure programs such as VASP assume that the structure will relax to the higher symmetry structure and so start calculations by making this assumption. By default AFLOW follows the same convention, but we felt it best to use the originally reported structure for our purposes. This requires us to change the default tolerance when calling AFLOW. One such case is α-FeSe [50]. This structure was reported in space group $Cmme$ (#67) with a base-centered orthorhombic lattice [51], with the lattice constants $a$ and $b$ differing by less than 0.3 %. As a result, the default tolerance set in AFLOW, as well as electronic structure programs such as VASP, place this in the higher symmetry tetragonal space grep $P4/nmm$ (#129). Tightening the AFLOW tolerance eventually reveals the reported space group. In this type of case, we uniformly use the reported space group. However, we note the problem in the comment section for the prototype.

### D. Cross Reference With Other Databases

The Inorganic Crystal Structure Database (ICSD) [3] is perhaps the largest collection of published inorganic crystal structures. We currently provide the corresponding ICSD identification number for each structure in the Library, when available. If the structure is not in the ICSD, we then search through the Cambridge Crystallographic Data Centre (CCDC) [52] database, which also searches the Cambridge Structural Database (CSD) [4]. We include the CCDC identification number if the structure is found there. At the present time we have 1,632 entries with ICSD or CCDC indentifications. The remaining 151 structures are mostly hypothetical, computationally predicted, or obsolete structures not included in the ICSD or CCDC databases.

### E. Extensions

We have made several improvements to the library and AFLOW. For example, AFLOW-XtalFinder [11] has been updated to include all of the structures currently in the library. When a user enters a structure, it is automat- ically compared to all existing prototype entries. If a matching entry is found, the relevant prototype information is returned (e.g., prototype label and parameters, *Strukturbericht* designation, symmetry descriptions, Inorganic Crystal Structure Database (ICSD) [3] or Cambridge Structural Database (CSD) [53] identifier, if available, and a weblink to the entry. If no matching entry is found – signaling a new prototype – users can report the structure by providing the following information in an online form:

  **i.** Geometry file (in any standardized format),
  **ii.** Reference(s) that reported the structure,
  **iii.** Supplemental reference(s) used to find the structure (e.g., catalogs or online databases),
  **iv.** The ICSD or CCDC entry number, if available.
  **v.** Any useful comments about the prototype, such as its relationship to other entries in the database and
  **vi.** The uploader's contact information (to credit the contributor on that structure's entry page, if desired).

Unique prototype suggestions will be collected for future enrichment of the encyclopedia. While previous library versions were only updated upon publication of a new article, we plan to switch to incremental updates. As new structures are added to the system, the online library will be directly expanded, with periodic reports such as this one focusing on additions and improvements to the library.

### III. CONCLUSION

The AFLOW Library of Crystallographic Properties has been updated to provide users with complete crystallographic information for 1,783 structures, with future expansion planned. This information can be used to construct input files for many electronic structure programs.

The updated Library changes the default definition of the AFLOW-prototype label to make it easier for users and programmers to construct a unique label for each prototype. Structures that can be placed in multiple space groups have been highlighted, and the comments for each structure outline why we have made that particular choice. Each entry now includes the ICSD or CCDC identification number, when available. Finally, we now provide an interface for users to suggest new structures to be added to the Library.

### Appendix A: Index of Prototypes Ordered by Space Groups

This index lists all prototype structures in the current Encyclopedia of Crystallographic Prototypes. En-tries headed by a filled bullet point (●) are new to this edition of the Encyclopedia, while previous entries are designated by an empty one (○). Clicking on the UID or AFLOW-prototype label will open a web browser window showing the details of that structure. A PDF version of the page can be obtained by clicking the "PDF version" link on the web page.

**P1 (1):**
- ○ $FeS_2$ ($P1$):
  A2B_aP12_1_4a_8a-001 — RHYA
- ○ $AsKSe_2$ ($P1$):
  ABC2_aP16_1_4a_4a_8a-001 — GZNE
- ○ $NaC_5H_{11}O_8S$:
  A5B11C D8E_aP26_1_5a_11a_a_8a_a-001 — WTTR

**P$\overline{1}$ (2):**
- ○ Cf:
  A_aP4_2_aci-001 — G18C
- ○ $P_2I_4$:
  A2B_aP6_2_2i_i-001 — V85G
- ○ $H_2S$ (90 GPa):
  A2B_aP6_2_aei_i-001 — Q769
- ● Hexamethylbenzene II ($C_{12}H_{18}$):
  A_aP12_2_6i-001 — 7LZP
- ● $\gamma$-$CuZrF_6$:
  AB12C_aP14_2_b_6i_c-001 — 5H3G
- ○ TaTi (BCC SQS-16):
  AB_aP16_2_4i_4i-001 — BQ6Y
- ○ $Co_2B_2O_5$:
  A2B2C5_aP18_2_2i_2i_5i-001 — RJNZ
- ● $Na_4SiO_4$:
  A4B4C_aP18_2_4i_4i_i-002 — A8AW
- ● $Cu_3(P_2O_6OH)_2$:
  A3B12C2D4_aP21_2_ai_6i_i_2i-001 — L1KQ
- ● Low Temperature "White" Phosphorous:
  A_aP24_2_12i-001 — YLP3
- ○ Albite ($NaAlSi_3O_8$, $S6_8$):
  ABC8D3_aP26_2_i_i_8i_3i-001 — 8BAG
- ○ Boric Acid ($H_3BO_3$, $G5_1$):
  ABC3_aP28_2_2i_6i_6i-001 — H1GD
- ○ Wollastonite ($CaSiO_3$):
  A3C_aP30_2_3i_9i_3i-001 — 89U5
- ○ Kyanite ($Al_2SiO_5$, $S0_1$):
  A2B5C_aP32_2_4i_10i_2i-001 — 4TPV
- ○ $\delta$-$WO_3$:
  A3B_aP32_2_12i_4i-001 — 9JTN
- ○ $Co_3(SeO_3)_3$·$H_2O$:
  A3B2C10D3_aP36_2_be2i_2i_10i_3i-001 — 0ZM1
- ○ Chalcanthite ($CuSO_4$·$5H_2O$, $H4_{10}$):
  A B10C9D_aP42_2_bc_10i_9i_i-001 — NQQF
- ○ $\alpha$-$Ho_2Si_2O_7$:
  A2B7C2_aP44_2_4i_14i_4i-001 — VN5Q
- ○ $Ni(NO_3)(H_2O)_3$:
  A12B2CD12_aP52_2_12i_2i_i_12i-001 — 6GLF
- ● Pentacene ($C_{11}H_7$):
  A11B7_aP72_2_22i_14i-001 — C9KZ

**P2 (3):**
- ○ Predicted $SiO_2$ ($P2$):
  A2B_mP12_3_ab3e_2e-001 — QAF3
- ● $Pb_3TeCo_3P_2O_{14}$:
  A3B14C2D3E_mP138_3_3c3d6e_42e_6e_3a3b6e_3a3b-001 — TBT2

**P$2_1$ (4):**

High-pressure Te:
A_mP4_4_2a-001 — H0F1
- ● $\alpha$-BiPd:
  AB_mP16_4_4a_4a-001 — QNKQ
- ○ $Li_2SO_4$·$H_2O$ ($H4_8$):
  A2B2C5D_mP20_4_2a_2a_5a_a-001 — 674K
- ○ $Ca_3UO_6$:
  A3B6C_mP20_4_3a_6a_a-001 — T9UP
- ● Monoclinic (I) $Li_2FeSiO_4$:
  A2B C4D_mP32_4_2a_4a_8a_2a-001 — TK4X
- ○ $W_2O_3(PO_4)_2$:
  A11B2C2_mP60_4_22a_4a_4a-001 — Z0HK
- ● Monoclinic $KH_2PO_4$:
  A5B2C8D2_mP68_4_10a_4a_16a_4a-001 — MD4G

**C2 (5):**
- ○ $(Ba,Ca)CO_3$ ("$C2$"):
  ABC3_mC10_5_b_a_ac-001 — 6HGL
- ○ A19 Po (*Obsolete*):
  A_mC12_5_3c-001 — FECP
- ○ $NbAs_2$:
  A2B_mC12_5_2c_c-001 — 0898
- ○ $D0_{15}$ ($AlCl_3$) (*Obsolete*):
  AB3_mC16_5_c_3c-001 — RSL8
- ● $CrPS_4$:
  ABC4_mC24_5_2a_c_4c-001 — VB2Y
- ○ $Rb_2CaCu_6(PO_4)_4O_2$:
  A B6C18D4E2_mC62_5_b_2a2c_9c_2c_c-001 — 74YK
- ● $Al_{17}Mo_4$:
  A17B4_mC84_5_ab16c_4c-001 — QL9A
- ○ Bassanite [$CaSO_4(H_2O)_{0.5}$, $H4_7$]:
  A2B2C9D2_mC90_5_ab2c_3c_a13c_3c-001 — Q14S
- ● $\alpha$-$Bi_4V_2O_{11}$:
  A3B9C2_mC112_5_6c_2a2b16c_4c-001 — JBQX
- ● High Temperature Monoclinic TlS:
  AB_mC256_5_2a2b30c_32c-001 — J6WY

**Pm (6):**
- ○ Tetrataenite (FeNi):
  AB_mP4_6_2a_2b-001 — X3MC
- ○ $Mo_8P_5$:
  A8B5_mP13_6_5a3b_2a3b-001 — LEL6
- ○ $Ta_5Ti_{11}$ (BCC SQS-16):
  A5B11_mP16_6_2abc_2a3b3c-001 — EUM5

**Pc (7):**
- ○ $H_2S$ IV:
  A2B_mP12_7_4a_2a-001 — UVJS
- ○ Calaverite ($AuTe_2$):
  AB2_mP12_7_2a_4a-001 — G433
- ○ $\epsilon$-$WO_3$ (low-temperature):
  A3B_mP16_7_6a_2a-001 — SUKZ
- ○ $BaAs_2$:
  A2B_mP18_7_6a_3a-001 — 34FL
- ○ $Rh_2Ga_9$:
  A9B2_mP22_7_9a_2a-001 — 2DRY
- ● Monoclinic $Bi_4Ti_3O_{12}$ $m = 3$ Aurivillius:
  A4B12C3_mP38_7_4a_12a_3a-001 — S41T
- ○ $Na_2Ca_6Si_4O_{15}$:
  A6B2C15D4_mP54_7_6a_2a_15a_4a-001 — 5V2T
- ○ Low Temperature $Mo_8O_{23}$:
  A8B23_mP124_7_16a_46a-001 — 5T76

**Cm (8):**
- ○ $F5_{11}$ ($KNO_2$) (*Obsolete*):
  ABC2_mC8_8_a_a_b-001 — X9AR
- ○ Monoclinic PZT [$Pb(Zr_xTi_{1-x})O_3$]:

A3BC_mC10_8_ab_a_a-001    XHKF

- Base-Centered Monoclinic La$_2$CuO$_4$:
  A2C4_mC28_8_2a_4a_4a2b-001    T7TZ
- Al$_4$W:
  A4B_mC30_8_2a5b_ab-001    10PS
- TaTi$_3$-II (BCC SQS-16):
  AB3_mC32_8_4a_4a4b-001    P5PQ
- TaTi$_3$-I (BCC SQS-16):
  AB3_mC32_8_4a_12a-001    QVN8
- Monoclinic Co$_4$Al$_{13}$:
  A13B4_mC102_8_17a11b_8a2b-001    DU5A

**Cc (9):**
- H$_3$Cl (20 GPa):
  AB3_mC16_9_a_3a-001    QYSB
- Room Temperature Ga$_2$Se$_3$:
  A2B3_mC20_9_2a_3a-001    4F67
- In$_2$Te$_5$ (I):
  A2B5_mC28_9_2a_5a-001    FDOC
- α-P$_3$N$_5$:
  A5B3_mC32_9_5a_3a-001    9173
- CuInP$_2$S$_6$:
  A2B2C2D6_mC44_9_2a_a_2a_6a-001    G59D
- GeCd$_4$S$_6$:
  A4BC6_mC44_9_4a_a_6a-001    ZP3W
- Chrysotile [Mg$_3$Si$_2$O$_5$(OH)$_4$]:
  A3B5C4D2_mC56_9_3a_5a_4a_2a-001    BT10
- Nacrite [Al$_2$Si$_2$O$_5$(OH)$_4$, $S5_4$]:
  A2B4C9D2_mC68_9_2a_4a_9a_2a-001    SDY7
- Monoclinic (Cc) Low Tridymite (SiO$_2$):
  A2B_mC144_9_24a_12a-001    MKOW
- Cs$_6$W$_{11}$O$_{36}$:
  A6B36C11_mC212_9_6a_36a_11a-001    A3V8
- α-LiZnPO$_4$:
  AB4CD_mC224_9_8a_32a_8a_8a-001    L704

**P2/m (10):**
- δ-Pd$_2$Cl:
  A2B_mP6_10_mn_bc-001    ZWXH
- α-LiSn:
  AB_mP6_10_bn_cm-001    QVV1
- Muthmannite (AuAgTe$_2$):
  ABC2_mP8_10_ac_eh_mn-001    ALSR
- S-carbon:
  A_mP8_10_2m2n-001    XRQ1
- H$_3$Cl (400 GPa):
  AB3_mP16_10_mn_3m3n-001    1EYW
- Hulsite [(Fe$_{1.315}$Mg$_{0.56}$Sn$_{0.1}$)BO$_5$]:
  A2B2C3D10E_mP18_10_m_ac_en_3m2n_g-001    8URQ

**P2$_1$/m (11):**
- NiTi:
  AB_mP4_11_e_e-001    WBFX
- YO(OH):
  ABC_mP6_11_e_e_e-001    LNCX
- CaSb$_2$:
  AB2_mP6_11_e_2e-001    5Q2X
- ZrSe$_3$:
  A3B_mP8_11_3e_e-001    54KB
- EuFeAs$_2$:
  A2BC_mP8_11_2e_e_e-001    C60C
- KClO$_3$ ($G0_6$):
  ABC3_mP10_11_e_e_ef-001    RT35
- Mo$_2$S$_3$:
  A2B3_mP10_11_2e_3e-001    MUP5
- SbAsO$_2$:

AB4C_mP12_11_e_2ef_e-001    C55J
- α-Pu:
  A_mP16_11_8e-001    2MX7
- β-NbPt$_3$:
  AB3_mP16_11_2e_2e2f-001    A6R9
- K$_2$S$_2$O$_5$ ($K0_1$):
  A2B5C2_mP18_11_2e_e2f_2e-001    YYGA
- Barytocalcite (BaCa(CO$_3$)$_2$):
  AB2CD6_mP20_11_2e_2e_e_2e2f-001    YSME
- Squaric Acid (H$_2$C$_4$O$_4$):
  A2BC2_mP20_11_4e_2e_4e-001    0Y6Q
- Li$_7$Sn$_3$:
  A7B3_mP20_11_7e_3e-001    YDAL
- y-Y$_2$Si$_2$O$_7$:
  A7B2C2_mP22_11_3e2f_2e_ab-001    3RSG
- Pt$_6$Si$_5$:
  A6B5_mP22_11_6e_5e-001    H63U
- NbSe$_3$:
  AB3_mP24_11_3e_9e-001    4LKF
- Ho$_2$S$_3$:
  A2B3_mP30_11_6e_9e-001    TWS9

**C2/m (12):**
- α-O$_2$:
  A_mC4_12_i-001    M3W6
- Calaverite (AuTe$_2$, C34):
  AB2_mC6_12_a_i-001    1QX2
- Tolbachite (CuCl$_2$):
  A2B_mC6_12_i_a-001    PNDR
- NiO$_2$:
  AB2_mC6_12_a_i-003    DZY0
- α-HgO$_2$:
  AB2_mC6_12_a_i-004    RMCX
- CaC$_2$-III:
  A2B_mC12_12_2i_i-004    2JVM
- α-Bi$_2$Pd:
  A2B_mC12_12_2i_i-006    0NDQ
- OsGe$_2$:
  A2B_mC12_12_2i_i-007    RQLL
- RuU$_2$:
  AB2_mC12_12_i_aci-001    6TPG
- GdCBr:
  ABC_mC12_12_i_i_i-003    69PJ
- Au$_5$Mn$_2$:
  A5B2_mC14_12_a2i_i-001    9H7D
- δ-Ni$_3$Sn$_4$ ($D7_d$):
  A3B4_mC14_12_ai_2i-001    4WTF
- Brezinaite (Cr$_3$S$_4$):
  A3B4_mC14_12_ai_2i-002    17US
- AlCl$_3$:
  AB3_mC16_12_g_ij-001    2SQ0
- M-carbon:
  A_mC16_12_4i-001    MTF0
- Monoclinic FeTlSe$_2$:
  AB2C_mC16_12_g_2i_i-001    JGDH
- SrN:
  AB_mC16_12_2i_2i-001    ASFW
- β-BiI:
  AB_mC16_12_2i_2i-002    DA8J
- γ-BiI:
  AB_mC16_12_2i_2i-003    6RE9
- CoGe:
  AB_mC16_12_aci_2i-001    7BKU
- K$_2$Ti$_2$O$_5$:

A2B5C2_mC18_12_i_a2i_i-001 — JYHJ

○ NbTe$_2$:
AB2_mC18_12_ai_3i-002 — FY6X

○ Ta$_2$PdSe$_6$:
AB6C2_mC18_12_a_3i_i-004 — H8PR

● $\beta$-Ga$_2$O$_3$:
A2B3_mC20_12_2i_3i-001 — LKDZ

● MnPS$_3$:
ABC3_mC20_12_g_i_ij-001 — UPGR

● $\alpha$-As$_2$Te$_3$:
A2B3_mC20_12_2i_3i-003 — 8W6Z

● Au$_2$P$_3$:
A2B3_mC20_12_eh_ij-001 — PQQZ

● Mo$_2$As$_3$:
A3B2_mC20_12_3i_2i-001 — 8QA4

● Gd$_2$Cl$_3$:
A3B2_mC20_12_3i_2i-002 — TUQ7

○ Thortveitite ([Sc,Y]$_2$Si$_2$O$_7$, $S2_1$):
A7B2C2_mC22_12_aij_h_i-001 — 0J1F

○ Al$_8$Mo$_3$:
A8B3_mC22_12_4i_ai-001 — 1Y1T

○ LiOH·H$_2$O ($B36$):
A3BC2_mC24_12_ij_g_hi-001 — T76D

○ AlNbO$_4$:
ABC4_mC24_12_i_i_4i-001 — HK1Y

○ SiAs:
AB_mC24_12_3i_3i-001 — S6UY

● Li$_2$MnO$_3$:
A2BC3_mC24_12_acg_h_ij-002 — R3L4

○ Nb$_2$Se:
A2B_mC24_12_4i_2i-004 — 2691

● Ag$_3$LiIr$_2$O$_6$:
A3B2CD6_mC24_12_ag_h_c_ij-001 — ZSJ2

○ Li$_3$Zn$_2$SbO$_6$:
A3B6CD2_mC24_12_ag_ij_c_h-001 — 72ZM

● Y$_5$S$_7$:
A7B5_mC24_12_a3i_c2i-001 — G5Z0

● Pd$_5$As:
AB5_mC24_12_i_g2ij-001 — XSQ7

● Predicted $\zeta$-LiAlO$_2$:
ABC2_mC24_12_ah_cg_ij-001 — JWKX

● V$_5$S$_8$:
A8B5_mC26_12_2ij_ahi-001 — ZQW6

● Hollandite (BaMn$_8$O$_{16}$):
AB4C8_mC26_12_a_2i_4i-001 — 50AM

● TlCr$_5$Se$_8$:
A5B8C_mC28_12_a2i_4i_c-001 — AFXE

● Au$_{11}$Mn$_3$:
A11B3_mC28_12_a5i_ci-001 — J58A

● CoZn$_{13}$:
AB13_mC28_12_a_c2i2j-001 — 6H3E

● PuNi$_4$:
A4B_mC30_12_gi2j_ai-001 — WBF9

● Cr$_7$Se$_8$:
A7B8_mC30_12_aehi_2ij-001 — H4M1

○ Dolerophanite [Cu$_2$O(SO$_4$)]:
A2B5C_mC32_12_ei_3ij_i-001 — H100

○ Sc$_3$RhC$_4$:
A4BC3_mC32_12_2j_i_ghi-002 — 2TGL

● $\alpha$-BiI:
A3B2_mC32_12_4i_4i-001 — DK3T

○ $\beta$-Pu:
A_mC34_12_ah3i2j-001 — VV4N

○ Os$_4$Al$_{13}$:

A13B4_mC34_12_a6i_2i-001 — XT4N

● Metastable PbV$_2$O$_6$:
A6BC2_mC36_12_6i_i_i2i-001 — X6VQ

○ Sr$_2$NiTeO$_6$:
AB6C2D_mC40_12_ac_gh4i_j_bd-001 — 79Y1

● Pr$_5$Co$_2$Ge$_4$:
A2B3C5_mC40_12_2i_3i_5i-001 — 422U

○ Bischofite (MgCl$_2$·6H$_2$O, $J1_7$):
A2B12CD6_mC42_12_i_2i2j_a_ij-001 — 3SX5

● Nb$_7$P$_4$:
A7B4_mC44_12_ac6i_4i-001 — EWT8

○ $D2_2$ (MgZn$_5$?) (*Problematic*):
AB5_mC48_12_2i_ac5i2j-001 — 5R9K

○ Sanidine (KAlSi$_3$O$_8$, $S6_7$):
A8C4_mC52_12_i_gi3j_2j-001 — JF0M

● Radtkeite (Hg$_3$S$_2$ClI):
AB3CD2_mC56_12_2i_eg2ij_2i_2j-001 — 4UG2

● NiBi:
A16B17_mC66_12_4i2j_aeh4ij-001 — V4P2

● Tremolite (Ca$_2$Mg$_5$Si$_8$O$_{22}$(OH)$_2$ $S4_2$):
A2B2C5D24E8_mC82_12_h_i_agh_2i5j_2j-001 — BQ7Y

● Monoclinic Nb$_{12}$O$_{29}$:
A12B29_mC82_12_6i_a14i-001 — BPW1

● Staurolite (H$_2$Al$_5$Fe$_2$Si$_4$O$_{12}$):
A5B2C2D10E2_mC84_12_acghj_bdi_2i_5j_j-001 — QHVX

○ Al$_{13}$Fe$_4$:
A13B4_mC102_12_ah8i5j_4ij-001 — U8YR

○ Al$_{45}$V$_7$:
A45B7_mC104_12_a8i7j_cij-001 — 4Z2U

● Manganese-leonite [K$_2$Mn(SO$_4$)$_2$·4H$_2$O, $H4_{23}$]:
A8B2CD15E2_mC112_12_2i3j_j_ac_g4i5j_2i-001 — KYS8

○ Chrysotile [Mg$_6$O$_{11}$Si$_4$(H$_2$O)(OH)$_6$, $S4_5$]:
A6BC11D6E4_mC112_12_e_gi2j_i5j_2i2j_2j-001 — CNLR

## P2/c (13):

○ Sylvanite (AgAuTe$_4$, $E1_b$):
ABC4_mP12_13_e_a_2g-001 — 5BWB

● H$_2$S (15 GPa):
A2B_mP12_13_2g_ef-001 — VYK4

● Huanzalaite (MgWO$_4$, $H0_6$):
AB4C_mP12_13_e_2g_f-003 — 2SHQ

● AgCrP$_2$S$_6$:
ABC2D6_mP20_13_e_e_g_3g-001 — NDG5

● NaIn(WO$_4$)$_2$:
ABC8D2_mP24_13_e_f_4g_g-001 — 7246

● Cu(PtS$_2$)$_2$:
AB2C4_mP28_13_e_efg_4g-001 — HFY3

● Room Temperature V$_3$O$_5$:
A5B3_mP32_13_ef4g_ab2g-001 — 3J3M

● Rosickýite ($\gamma$-monoclinic Sulfur):
A_mP32_13_8g-001 — Y7J6

● Monoclinic (II) Ba$_2$CoIr$_2$O$_9$:
A3BC2D9_mP60_13_ef2g_ab_2g_ef8g-001 — 2BLQ

○ Approximate High Temperature Mo$_8$O$_{23}$:
A8B23_mP62_13_4g_a11g-001 — UWFW

● Monoclinic (Hittorf's) Phosphorus:
A_mP84_13_21g-001 — SEQF

## P2$_1$/c (14):

○ $\gamma$-PdCl$_2$:
A2B_mP6_14_e_a-001 — K9AG

● CdP$_4$:
AB4_mP10_14_a_2e-001 — L3ZV

○ Baddeleyite (ZrO$_2$, $C43$):
A2B_mP12_14_2e_e-001 — 5CSN

○ Arsenopyrite (FeAsS, $E0_7$):
ABC_mP12_14_e_e_e-002     RH3X

● Acanthite ($Ag_2S$):
A2B_mP12_14_2e_e-011     0GFL

● $NdAs_2$:
A2B_mP12_14_2e_e-012     5BM7

○ $HgCl_2 \cdot 2HgO$:
A2B3C2_mP14_14_ae_ae_e-001     AR8A

○ "$P2_1/c$"-$B_2H_6$:
AB3_mP16_14_e_3e-002     3XCA

○ $\beta$-$B_2H_6$:
AB3_mP16_14_e_3e-003     WZYX

○ $KNO_2$ III:
ABC2_mP16_14_e_e_2e-003     JSKP

○ Manganite ($\gamma$-MnO(OH), $E0_6$):
ABC2_mP16_14_e_e_2e-004     E23Q

○ Cu(OH)Cl:
ABCD_mP16_14_e_e_e_e-001     CH4P

○ $\alpha$-ICl:
AB_mP16_14_2e_2e-001     HU5N

○ LiAs:
AB_mP16_14_2e_2e-002     LRN3

● Luberoite ($Pt_5Se_4$):
A5B4_mP18_14_a2e_2e-001     1FWE

● Phase I $K_2SnCl_6$:
A6B2C_mP18_14_3e_e_a-001     C2DK

○ Orpiment ($As_2S_3$, $D5_f$):
A2B3_mP20_14_2e_3e-002     4PSJ

○ Sanguite ($KCuCl_3$):
A3BC_mP20_14_3e_e_e-001     M9HY

○ $Sr_2MnTeO_6$:
A2BC2D_mP20_14_a_3e_e_b-001     QXEE

○ Cryolite ($Na_3AlF_6$, $J2_6$):
AB6C3_mP20_14_a_3e_be-001     G2BV

○ $\alpha$-$SrRh_2As_2$:
A2B2C_mP20_14_2e_2e_e-001     1XXN

○ $\alpha$-$Bi_2O_3$:
A2B3_mP20_14_2e_3e-003     7G59

● $Y_2OS_2$:
A2BC2_mP20_14_e_2e_2e-001     A73U

○ $Sb_4O_5Cl_2$:
A2B5C4_mP22_14_a2e_2e-001     YL8Z

○ $K_2Ni(CN)_4$:
A4B2C4D_mP22_14_2e_e_2e_a-001     RARL

○ $Co_2Al_9$ ($D8_d$):
A9B2_mP22_14_a4e_e-001     T95J

○ $Er_2Si_2O_7$:
A2B7C2_mP22_14_a3e_e-001     WKQU

● Juangodoyite [$Na_2Cu(CO_3)_2$]:
A2BC2D6_mP22_14_e_a_e_3e-001     DUCH

○ $\gamma$-$Y_2Si_2O_7$:
A4BC_mP24_14_4e_e_e-001     3TA1

○ Anhydrous $KAuBr_4$:
A4BC_mP24_14_ab_4e_e-001     2L05

○ Ammonium Persulfate [$(NH_4)SO_4$, $K4_1$]:
A4BC_mP24_14_4e_e_e-001     1E0C

○ Monasite ($LaPO_4$):
A4BC_mP24_14_4e_e_e-002     G325

○ $AgMnO_4$ ($H0_9$):
ABC4_mP24_14_e_e_4e-001     8H0D

○ Nahcolite ($NaHCO_3$, $G0_{12}$):
ABCD3_mP24_14_e_e_e_3e-001     KV55

○ $\epsilon$-1,2,3,4,5,6-Hexachloro$cyclo$hexane ($C_6Cl_6$):
AB_mP24_14_3e_3e-001     7TV0

● Monoclinic (black) $ZnP_2$:
A2B_mP24_14_4e_2e-001     BW3Z

● $\beta$-$Tl_2TeO_3$:
A3BC2_mP24_14_3e_e_2e-001     8M4S

● $AgTe_2Tl_3$:
AB2C3_mP24_14_2e_3e-001     XDSH

● $CuTeO_4$:
AB4C_mP24_14_ac_4e_e-001     LRGD

● $GaPS_4$:
ABC4_mP24_14_e_e_4e-002     VQV1

○ Monoclinic $Cu_2SeO_3$:
A2B4C_mP28_14_abe_4e_e-001     D0DK

● $KICl_4 \cdot H_2O$ ($H0_{10}$):
A4BCD_mP28_14_4e_e_e-001     YLSG

● Larnite ($\beta$-$Ca_2SiO_4$):
A2B4C_mP28_14_2e_4e_e-001     QE0P

● Pyrostilpnite ($Ag_3SbS_3$):
A3B3C_mP28_14_3e_3e_e-001     8N0K

● $Tl_4S_3$:
A3B4_mP28_14_3e_4e-001     6ZEC

● $Pd_6P$:
AB6_mP28_14_e_6e-001     VTUH

● Azurite [$Cu_3(CO_3)_2(OH)_2$, $G74$]:
A2B3C2D8_mP30_14_e_ae_e_4e-001     QANU

● $\beta$-Se ($A_l$):
A_mP32_14_8e-001     QMXV

● $Ca_2UO_5$:
A2B5C_mP32_14_2e_5e_ab-001     E7S8

● $Gd_2SiO_5$ ($RE_2SiO_5$ X1):
A2B5C_mP32_14_2e_5e_e-001     HBLG

○ $\gamma$-$WO_3$:
A3B_mP32_14_6e_2e-001     S4UY

○ $KAuBr_4 \cdot 2H_2O$ ($H4_{19}$):
AB4C2D_mP32_14_e_4e_2e-001     SAP5

● Pararealgar (AsS):
AB_mP32_14_4e_4e-001     WH8O

● Realgar (AsS, $B_l$):
AB_mP32_14_4e_4e-002     R9YN

● Monoclinic (II) $Li_2FeSiO_4$:
A2B2C4D_mP32_14_2e_e_4e_e-001     U7VZ

● Lorándite ($TlAsS_2$):
A2BC_mP32_14_2e_2e-001     QZUF

● Monoclinic $ClF_3$:
AB3_mP32_14_2e_6e-001     4D8B

● NaGe:
AB_mP32_14_4e_4e-003     05NC

● NS:
AB_mP32_14_4e_4e-004     YXLR

○ $\alpha$-monoclinic Selenium:
A_mP32_14_8e-002     VEFR

● N-Hydroxyurea ($CH_4N_2O_2$):
A4B4C2D2_mP36_14_4e_4e_2e_2e-001     XWQW

○ $K_2NbF_7$ ($K6_2$):
A7B2C_mP40_14_7e_2e_e-001     TF3J

● $\alpha$-$Ca_2P_2O_7$:
A2B7C2_mP44_14_2e_7e_2e-001     9VRC

● $\alpha$-$Na_2CuP_2O_7$:
A2B7C7D2_mP48_14_e_7e_7e_2e-001     8Z5C

● Clinometaborite ($\beta$-$HBO_2$, monoclinic):
ABC2_mP48_14_3e_3e_6e-001     XNVP

● $Ce_7Pd_4Ge_2$:
A7B2C4_mP52_14_7e_2e_4e-001     N56L

● $K_2Pt(SCN)_6 \cdot 2H_2O$:
A6B4C2D6E2FG6_mP54_14_3e_2e_e_e_3e_e_a_3e-001     0UZV

- Cs₁₁O₃:

$Cs_{11}O_3$:
`A11B3_mP56_14_11e_3e-001`                `2343`
○ Parawollastonite (CaSiO₃, $S3_3$(II)):
`A3BC_mP60_14_3e_9e_3e-001`              `XADC`
○ α-Se ($A_k$):
`A_mP64_14_16e-001`                      `TBFR`
● β-monoclinic Sulfur:
`A_mP64_14_16e-002`                      `AYZW`
● Monoclinic Fe₂(SO₄)₃:
`A2B12C3_mP68_14_2e_12e_3e-001`          `LK3Z`
○ Tutton salt [Cu(NH₄)₂(SO₄)₂·H₂O, $H4_4$]:
`A20C2D14E2_mP78_14_a_10e_e_7e_e-001`    `KT73`
○ Manganese-leonite 110K [K₂Mn(SO₄)₂·4H₂O]:
`A8B2CD12E2_mP100_14_8e_2e_ab_12e_2e-001` `V1SG`
○ α-Toluene (C₇H₈):
`A7B8_mP120_14_14e_16e-001`              `NQ3P`
● High temperature Bi₂MoO₆:
`A2BC6_mP144_14_8e_4e_24e-001`           `UYPX`
● Ag(tcm)(phz)₁/₂ (AgC₁₀NH₄H₄):
`A10C4D4_mP152_14_2e_20e_8e_8e-001`      `JN9W`
● Monoclinic 2-4-6 Trinitrotoluene (C₇H₅N₃O₆):
`A7B5C3D6_mP168_14_14e_10e_6e_12e-001`   `MUZR`
● [Zn₂(Benzoato)₄(Caffeine)₂]·2 Caffeine (C₃₀H₃₀N₈O₈Zn):
`A30B30C8D8E_mP308_14_30e_30e_8e_8e_e-001` `6QN7`

**C2/c (15):**

○ β-Ga (*Obsolete*):
`A_mC4_15_e-001`                         `BRYB`
○ Tenorite (CuO, *B26*):
`AB_mC8_15_a_e-001`                      `HRJX`
● CrS:
`AB_mC8_15_a_e-003`                      `H6Z8`
● ThC₂ ($C_g$):
`A2B_mC12_15_f_e-001`                    `SKBR`
● PdP₂:
`A2B_mC12_15_f_a-001`                    `SPK1`
○ H₃Cl (50 GPa):
`A3B_mC16_15_e_af-001`                   `T96Y`
○ KFeS₂ ($F5_a$):
`ABC2_mC16_15_e_e_f-004`                 `G8C6`
● Ag₂PbO₂:
`A2B2C_mC20_15_ac_f_e-001`               `8GWY`
● CrP₄:
`AB4_mC20_15_e_e2f-001`                  `2S3X`
● H-III (300 GPa):
`A_mC24_15_2e2f-001`                     `MULK`
○ Clinocervantite (β-Sb₂O₄):
`A2B_mC24_15_2f_ae-001`                  `YM7H`
● Zabuyelite (Li₂CO₃):
`A2C3_mC24_15_e_f_ef-001`                `JF5W`
● B₂Pd₅:
`A2B5_mC28_15_f_e2f-001`                 `CZ5K`
○ ζ-Nb₂O₅ (B-Nb₂O₅):
`A2B5_mC28_15_f_e2f-003`                 `S3CP`
● Monoclinic Ni₄B₃:
`A3B4_mC28_15_ef_2f-001`                 `XJ9R`
○ Ta₂NiSe₅:
`AB5C2_mC32_15_e_e2f_f-001`              `YFLU`
○ Titanite (CaTiSiO₅, $S0_6$):
`AB5CD_mC32_15_e_e2f_e_a-001`            `EDTK`
● Oxyvanite (V₃O₅):
`A5B3_mC32_15_e2f_af-003`                `C32G`
● Miargyrite (AgSbS₂):
`AB2C_mC32_15_ae_2f_f-001`               `YUVG`

● β′-LiFeO₂:
`ABC2_mC32_15_2e_2e_2f-002`              `VG40`
● NSe:
`AB_mC32_15_2ef_2f-001`                  `RDBP`
● NaSi:
`AB_mC32_15_2f_2f-001`                   `58NH`
○ Rb₂C₂O₄·H₂O:
`A2BC4D2_mC36_15_f_e_2f_f-001`           `RBDM`
● Foordite (SnNb₂O₆):
`A2B6C_mC36_15_f_3f_e-001`               `6QBH`
○ Esseneite (CaFeSi₂O₆):
`ABC6D2_mC40_15_e_e_3f_f-001`            `5WH8`
○ Diopside [CaMg(SiO₃)₂, $S4_1$]:
`ABC6D2_mC40_15_e_e_3f_f-002`            `J45L`
● Lu₂Co₃Si₅:
`A3B2C5_mC40_15_ef_f_3ef-001`            `1Q2N`
● VS₄:
`A4B_mC40_15_4f_f-001`                   `6M9R`
● CoGeO₃:
`ABC3_mC40_15_2e_f_3f-001`               `6YLQ`
● α-Zn₂V₂O₇:
`A7B2C2_mC44_15_e3f_f_f-001`             `JMLC`
● η′-Cu₆Sn₅:
`A6B5_mC44_15_ae2f_e2f-001`              `LPH7`
● Coesite (SiO₂):
`A2B_mC48_15_ae3f_2f-001`                `QHEE`
● Na₂PrO₃:
`A2B3C_mC48_15_aef_3f_2e-001`            `NFQU`
● Gypsum (CaSO₄·2H₂O, $H4_6$):
`AB4C6D_mC48_15_e_2f_3f_e-001`           `8924`
● α-SnF₂:
`A2B_mC48_15_4f_2f-001`                  `YKEJ`
● β-Na₂CuP₂O₇:
`A2C7D2_mC48_15_a_f_e3f_f-001`           `9DDX`
● Wodginite (LiFe(WO₄)₂):
`ABC8D2_mC48_15_e_e_4f_f-001`            `013B`
○ BaNi(CN)₄·4H₂O ($H4_{22}$):
`AB4C4D4E_mC56_15_e_2f_2f_2f_a-001`      `QDTB`
○ Xanthoconite (Ag₃AsS₃):
`A3BC3_mC56_15_3f_f_3f-001`              `HRK9`
● Ta₆S:
`AB6_mC56_15_f_6f-001`                   `HXME`
● Monoclinic (I) Ba₃CoIr₂O₉:
`A3BC2D9_mC60_15_ef_a_f_e4f-001`         `G3XE`
● Monoclinic Pyrrhotite (Fe₇S₈):
`A7B8_mC60_15_e3f_4f-001`                `L5GZ`
● Y₂SiO₅ (RE₂SiO₅ X2):
`A5BC2_mC64_15_5f_f_2f-001`              `G8V5`
○ Pyrophyllite [AlSi₂O₅(OH), $S5_6$]:
`AB5CD2_mC72_15_f_5f_f_2f-001`           `J4ZX`
○ Rhodan Hydrate (H₂C₂N₂S₂O):
`A2B2C2DE2_mC72_15_2f_2f_2f_f_2f-001`    `UX1J`
○ Muscovite (KH₂Al₃Si₃O₁₂, $S5_1$):
`A2BC10D2E4_mC76_15_f_e_5f_f_2f-001`     `8J4A`
○ Alluaudite [NaMnFe₂(PO₄)₃]:
`A2BCD12E3_mC76_15_f_e_a_6f_ef-001`      `8ZWK`
● In₂Te₅ (II):
`A2B5_mC84_15_3f_e7f-001`                `A3RY`
● Smithite (AgAsS₂):
`ABC2_mC96_15_2e2f_3f_6f-001`            `69FP`
● Clinochlore [Mg₅(Mg₂Al)(Si₃Al)O₁₀(OH)₈, $S5_5$]:
`A3B5C4D2_mC112_15_a3ef_5f_4f_2f-001`    `RDRH`
● Eudidymite (BeHNaO₈Si₃):
`A2B4C2D17E6_mC124_15_f_2f_f_e8f_3f-001` `AU9C`

- MoP$_3$SiO$_{11}$ (Monoclinic Model):
  AB11C3D_mC128_15_f_ae10f_3f_f-001    **2HE4**
- Catapleiite (Na$_2$ZrSi$_3$O$_9$·2H$_2$O):
  A2B3C9D3E_mC144_15_2f_abcef_9f_3f_de-001    **L4VZ**
- Low Temperature Cu$_2$Se:
  A2B_mC144_15_12f_6f-001    **F9TK**
- Rb$_2$Cu2(MoO$_4$)$_3$:
  A2B3C12D2_mC152_15_2f_3f_12f_aef-001    **A42Z**
- (CdSO$_4$)$_3$·8H$_2$O (*H4$_{20}$*):
  A3B16C20D3_mC168_15_ef_8f_10f_ef-001    **VXAE**
- Manganese-leonite 185K [K$_2$Mn(SO$_4$)$_2$·4H$_2$O]:
  A8B2CD12E2_mC200_15_8f_2f_ae_2e11f_2f-001    **5SVW**

**P222 (16):**
- AlPS$_4$:
  ABC4_oP12_16_ae_bd_2u-001    **8PL6**

**P222$_1$ (17):**
- $\alpha$-Naumannite (Ag$_2$Se):
  A2B_oP12_17_abe_e-001    **G353**
- NaNbO$_3$:
  ABC3_oP40_17_abcd_2e_abcd4e-001    **D6BJ**

**P2$_1$2$_1$2 (18):**
- $\gamma$-TeO$_2$ (*Erroneous*):
  A2B_oP12_18_2c_c-001    **4QQ6**
- BaS$_3$ (original *D0$_{17}$*):
  AB3_oP16_18_ab_3c-001    **69TS**
- Diamminetriamidodizinc Chloride
  ([Zn$_2$(NH$_3$)$_2$(NH$_2$)$_3$]Cl):
  A12C5D2_oP40_18_a_6c_b2c_c-001    **5HSN**

**P2$_1$2$_1$2$_1$ (19):**
- Naumannite (Ag$_2$Se II):
  A2B_oP12_19_2a_a-001    **LU1L**
- $\beta$-SnF$_2$:
  A2B_oP12_19_2a_a-002    **RZ21**
- H$_3$Cl (100 GPa):
  AB3_oP16_19_a_3a-001    **WTE4**
- NaP:
  AB_oP16_19_2a_2a-001    **MEBJ**
- Wülfingite ($\epsilon$-Zn(OH)$_2$, C31):
  A2B2C_oP20_19_2a_2a_a-001    **HY2J**
- NaAlCl$_4$:
  AB4C_oP24_19_a_4a_a-001    **43KD**
- Nd$_2$Si$_2$O$_7$:
  A2B7C2_oP44_19_2a_7a_2a-001    **TYGX**
- Ferroelectric (NH$_4$)H$_2$PO$_4$:
  A6BC4D_oP48_19_6a_a_4a_a-001    **P6ME**
- $\beta$-Arabinose (CO$_2$H)$_{20}$:
  A2B2C_oP80_19_5a_10a_5a-001    **K9SN**
- Ca-Malate-dihydrate (CaC$_4$H$_4$O$_5$·2H$_2$O):
  A4BC8D7_oP80_19_4a_a_8a_7a-001    **RRH1**
- Morenosite (NiSO$_4$·7H$_2$O, *H4$_{12}$*):
  A14BC11D_oP108_19_14a_a_11a_a-001    **S9KY**

**C222$_1$ (20):**
- *D0$_7$* (CrO$_3$) (*Obsolete*):
  AB3_oC16_20_a_bc-001    **KCMH**
- Orthorhombic (High) Tridymite (SiO$_2$):
  A2B_oC24_20_abc_c-001    **NDBQ**
- AlPO$_4$ "low cristobalite type":
  AB4C_oC24_20_a_2c_b-001    **HRBV**
- K$_3$ (Tl$_2$AlF$_5$):
  A5B2C_oC32_20_a_2abc_c-001    **XBH3**
- Na$_2$Tl:
  A2B_oC48_20_ab3c_2c-001    **PRYE**

**C222 (21):**

---

- Ta$_2$H:
  AB2_oC6_21_a_k-001    **6XPU**
- HoSb$_2$:
  AB2_oC6_21_a_k-002    **DD5U**
- Godlevskite (Ni$_9$S$_8$):
  A9B8_oC68_21_acehik2l_4l-001    **L6M4**

**F222 (22):**
- Low temperature FeS:
  AB_oF8_22_a_c-001    **H691**
- CeRu$_2$B$_2$:
  A2BC2_oF40_22_ej_ac_fi-001    **PNOD**
- Predicted Phase IV Cd$_2$Re$_2$O$_7$:
  A2B7C2_oF88_22_k_acefghij_k-001    **71CA**

**I222 (23):**
- BPS$_4$:
  ABC4_oI12_23_a_b_k-001    **ZCCT**
- NaFeS$_2$:
  ABC2_oI16_23_ac_e_k-001    **J362**
- H$_3$S (5 GPa):
  A3B_oI32_23_ef2k_k-001    **4LGD**
- Stannoidite (Cu$_8$(Fe,Zn)$_3$Sn$_2$S$_{12}$):
  A8B2C12D2E_oI50_23_acgk_e_3k_f_b-001    **7NS2**

**I2$_1$2$_1$2$_1$ (24):**
- Weberite (Na$_2$MgAlF$_7$):
  AB7CD2_oI44_24_a_c3d_b_ab-001    **Q6TN**

**Pmm2 (25):**
- High-pressure CdTe:
  AB_oP2_25_a_b-001    **CPSQ**

**Pmc2$_1$ (26):**
- H$_2$S 70 GPa:
  A2B_oP12_26_abc_ab-001    **PC8F**
- $\beta$-SeO$_2$:
  A2B_oP12_26_abc_ab-002    **2MYY**
- Metastable Au$_5$Zn$_3$:
  A5B3_oP16_26_a2c_a2b-001    **PD7K**
- TlP$_5$:
  A5B_oP24_26_3a3b2c_ab-001    **AHE7**
- $\gamma$-SeO$_2$:
  A2B_oP24_26_2a2b2c_2a2b-001    **PVJF**
- Low Temperature SmBaMn$_2$O$_6$:
  AB2C6D_oP40_26_2a_2a2b4c_2b-001    **LDXU**

**Pcc2 (27):**
- Ca$_4$Al$_6$O$_{16}$S:
  A6B4C16D_oP108_27_abcd4e_ae_16e_e-001    **TBPX**

**Pma2 (28):**
- Krennerite (AuTe$_2$):
  AB2_oP24_28_acd_2c3d-001    **A9SX**

**Pca2$_1$ (29):**
- ZrO$_2$:
  A2B_oP12_29_2a_a-001    **UZKW**
- Low Temperature Pyrite (FeS$_2$):
  A2B_oP12_29_a_2a-001    **O9UM**
- Cobaltite (CoAsS):
  ABC_oP12_29_a_a_a-001    **8UWB**
- Mercury (II) Azide [Hg(N$_3$)$_2$]:
  AB6_oP28_29_a_6a-001    **KSW9**
- Koechlinite (Room temperature Bi$_2$MoO$_6$):
  A2BC6_oP36_29_2a_a_6a-001    **UQQ3**
- Low Temperature Bornite (Cu$_5$FeS$_4$):
  A5BC4_oP160_29_20a_4a_16a-001    **H7LS**
- Orthorhombic 2-4-6 Trinitrotoluene (C$_7$H$_5$N$_3$O$_6$):
  A7B5C3D6_oP168_29_14a_10a_6a_12a-001    **EY4Y**
- Low Temperature (NH$_3$CH$_3$)Al(SO$_4$)$_2$ · 12H$_2$O:

AB30DE20F2_oP220_29_a_a_30a_a_20a_2a-001    2B7C

**Pnc2 (30):**
- ○ $Pnc2$ CuBrSe$_3$:
  AB3C3_oP20_30_2a_c_3c-001    2SVA
- ○ Bi$_5$Nb$_3$O$_{15}$:
  A5B3C15_oP46_30_a2c_bc_a7c-001    ABT7

**Pmn2$_1$ (31):**
- ● Guyanaite ($\beta$-CrOOH):
  ABC2_oP8_31_a_a_2a-001    TZ68
- ● WTe$_2$:
  A2B_oP12_31_4a_2a-001    ZLWB
- ○ Shcherbinaite ($D8_7$, V$_2$O$_5$) (*Obsolete*):
  A5B2_oP14_31_a2b_b-001    T4NG
- ● Enargite (AsCu$_3$S$_4$, $H2_5$):
  AB3C4_oP16_31_a_ab_2ab-001    L0GQ
- ● $\beta$-Li$_3$PO$_4$:
  A3B4C_oP16_31_ab_2ab_a-002    0L2F
- ● B$_4$SrO$_7$:
  A4B7C_oP24_31_2b_a3b_a-001    DLX0
- ● Clinobarylite (BaBe$_2$Si$_2$O$_7$):
  A2C7D2_oP24_31_a_b_a3b_b-001    WL10
- ● Seligmannite (PbCuAsS$_3$):
  ABCD3_oP24_31_2a_b_2a_2a2b-001    2RFK
- ○ Mg(ClO$_4$)$_2$·6H$_2$O ($H4_{11}$):
  A2B6CD8_oP34_31_2a_2a2b_a_4a2b-001    ZN0T
- ○ Orthorhombic Co$_4$Al$_{13}$ (Approximate Quasicrystal):
  A13B4_oP102_31_17a11b_8a2b-001    DHFF

**Pba2 (32):**
- ○ Re$_2$O$_5$(SO$_4$)$_2$:
  A13B2C2_oP34_32_a6c_c_c-001    16XX
- ○ Mo$_{17}$O$_{47}$:
  A17B47_oP128_32_a8c_a23c-001    484W

**Pna2$_1$ (33):**
- ○ Modderite (CoAs):
  AB_oP8_33_a_a-001    8Z7Q
- ○ LiGaO$_2$:
  ABC2_oP16_33_a_a_2a-003    9S7L
- ○ $\gamma$-LiIO$_3$:
  ABC3_oP20_33_a_a_3a-001    P3HQ
- ○ Cervantite ($\alpha$-Sb$_2$O$_4$):
  A2B_oP24_33_4a_2a-001    49UC
- ● $\beta$-CuAlCl$_4$:
  AB4C_oP24_33_a_4a_a-001    KHU7
- ● $\delta_1$-LiZnPO$_4$:
  AB4CD_oP28_33_a_4a_a_a-001    FVBE
- ○ AsK$_3$S$_4$:
  A3C4_oP32_33_a_3a_4a-001    H422
- ○ $\kappa$ alumina (Al$_2$O$_3$):
  A2B3_oP40_33_4a_6a-001    SGA3
- ○ LiZnPO$_4$·H$_2$O:
  A2BC5DE_oP40_33_2a_a_5a_a_a-001    SJPR
- ○ Possible $\delta$-Gd$_2$Si$_2$O$_7$:
  A2B7C2_oP44_33_2a_7a_2a-001    9ZSV
- ○ CsB$_4$O$_6$F:
  A4BCD6_oP48_33_4a_a_a_6a-001    Z0HT
- ○ CaB$_2$O$_4$ (III):
  A2BC4_oP84_33_6a_3a_12a-001    S26L
- ● $\alpha'_L$-Ca$_2$SiO$_4$:
  A2B4C_oP84_33_6a_12a_3a-001    2RUZ

**Pnn2 (34):**
- ○ FeSb$_2$:
  AB2_oP6_34_a_c-001    5KTQ
- ○ MnF$_{1-x}$(OH)$_x$:

A2BC2_oP10_34_c_a_c-001    5413
- ○ TiAl$_2$Br$_8$:
  A2B8C_oP22_34_c_4c_a-001    7UTD
- ● K$_2$AgSbS$_4$:
  AB2C4D_oP32_34_ab_abc_4c_c-001    Z6KJ

**Cmm2 (35):**
- ○ V$_2$MoO$_8$:
  AB8C2_oC22_35_a_ab3d_d-002    8B57

**Cmc2$_1$ (36):**
- ○ Low Temperature HCl:
  AB_oC8_36_a_a-001    SAL5
- ○ HgBr$_2$ ($C24$):
  A2B_oC12_36_2a_a-001    FWL2
- ● MoP$_2$:
  AB2_oC12_36_a_2a-002    JTSA
- ○ Si$_2$N$_2$O:
  A2BC2_oC20_36_b_a_b-001    8BR1
- ● K$_2$S$_3$:
  A2B3_oC20_36_2a_ab-001    36FH
- ● Orthorhombic B$_2$O$_3$:
  A2B3_oC20_36_b_ab-001    KLS9
- ● BaZnF$_4$:
  A4BC_oC24_36_a_4a_a-002    F77P
- ● Bi$_2$GeO$_5$:
  A2BC5_oC32_36_b_a_a2b-001    ZATP
- ● $\beta$-BiPd:
  AB_oC32_36_2ab_2ab-001    6W3L
- ● $\alpha$-NiI$_4$:
  A4B_oC40_36_4a2b_b-001    MTWW
- ● Ca$_3$Ti$_2$O$_7$:
  A3B7C2_oC48_36_ab_a3b_b-001    FD1H
- ○ Bertrandite (Be$_4$Si$_2$O$_7$(OH)$_2$, $S4_6$):
  A4B7C2D2_oC60_36_2b_a3b_2a_b-001    NZ59
- ○ Ni$_3$Si$_2$:
  A3B2_oC80_36_4a4b_2a3b-001    5H38
- ○ $\alpha$-Potassium Nitrate (KNO$_3$) II:
  ABC3_oC80_36_2ab_2ab_2a5b-001    3YA6
- ● Low Temperature BaBi$_4$Ti$_4$O$_{15}$:
  AB4C15D4_oC96_36_a_2b_a7b_2b-001    8N75

**Ccc2 (37):**
- ○ Li$_2$Si$_2$O$_5$:
  A2B5C2_oC36_37_d_c2d_d-001    JE1T
- ● Bi$_2$Sr$_2$Ca$_2$Cu$_3$O$_{10+x}$ (Bi-2223):
  A2B2C3D10E2_oC76_37_d_d_cd_5d_d-001    ZBUL

**Amm2 (38):**
- ○ C$_2$CeNi:
  A2BC_oC8_38_d_b_a-001    37DH
- ● Orthorhombic BaTiO$_3$:
  AB3C_oC10_38_a_ae_b-002    8STH
- ● Au$_2$V:
  A2B_oC12_38_de_ab-001    PCHF
- ● GdSn$_3$:
  AB3_oC16_38_ab_3a3b-001    L5S8
- ○ Ta$_3$Ti$_5$ (BCC SQS-16):
  A3B5_oC32_38_abcd_abcef-001    0NF7
- ○ Ta$_3$Ti$_{13}$ (BCC SQS-16):
  A3B13_oC32_38_ac_a2bcdef-001    A2U8
- ○ NaNb$_6$O$_{15}$F:
  ABC6D15_oC46_38_b_b_2a2d_2ab4d2e-001    MYRE

**Aem2 (39):**
- ○ Ta$_3$S$_2$:
  A2B3_oC40_39_2d_2c2d-001    4YG0
- ○ VPCl$_9$:

A9BC_oC44_39_3c3d_a_c-001     5ZH7

**Ama2 (40):**
- ○ K$_2$CdPb:
- AB2CdC_oC16_40_a_2b_b-001     K8N1
- ● CeTe$_3$:
- AB3_oC16_40_b_3b-001     WA7J
- ○ Orthorhombic CrO$_3$:
- AB3_oC16_40_b_a2b-002     8E60
- ○ Rb$_2$Mo$_2$O$_7$:
- A2B7C2_oC88_40_abc_2b6c_a3b-001     CBWE

**Aea2 (41):**
- ○ PtSn$_4$ ($D1_c$):
- AB4_oC20_41_a_2b-001     MALE
- ○ PdSn$_2$ ($C_c$):
- AB2_oC24_41_2a_2b-001     BVS7
- ● Orthorhombic Bi$_4$Ti$_3$O$_{12}$ $m = 3$ Aurivillius (*Obsolete*):
- A4B12C3_oC76_41_2b_6b_ab-001     5TXE
- ○ Santite (KB$_5$O$_8$·4H$_2$O, $K3_5$):
- A5B8CD12_oC104_41_a2b_4b_a_6b-001     C7EQ

**Fmm2 (42):**
- ○ BN (High-pressure, high-temperature):
- AB_oF8_42_a_a-001     0Q2T
- ○ W$_3$O$_{10}$ (WO$_3$ · $\frac{1}{3}$H$_2$O):
- A10B3_oF52_42_2abce_ab-001     TH7F
- ● Bi$_7$(Fe,Ti)$_6$O$_{21}$ $m = 6$ Aurivillius:
- A7B6C21_oF136_42_a3c_3c_ab3c3e-001     M0MD

**Fdd2 (43):**
- ○ Cs$_2$Se:
- A2B_oF24_43_b_a-001     DUC7
- ○ Ag$_2$O$_3$:
- A2B3_oF40_43_b_ab-001     X02L
- ○ Zr$_2$Al$_3$:
- A3B2_oF40_43_ab_b-001     7D48
- ○ Archerite (KH$_2$PO$_4$):
- A2BC4D_oF64_43_b_a_2b_a-001     DS2R
- ○ GeS$_2$ ($C44$):
- AB2_oF72_43_ab_3b-001     4KUZ
- ○ Blossite ($\alpha$-Cu$_2$V$_2$O$_7$):
- A2B7C2_oF88_43_b_a3b_b-001     D9YZ
- ○ Natrolite (Na$_2$Al$_2$Si$_3$O$_{10}$·2H$_2$O, $S6_{10}$):
- A2B4C2D12E3_oF184_43_b_2b_b_6b_ab-001     FT02
- ● Al$_{22}$Mo$_5$:
- A22B5_oF216_43_11b_a2b-001     15LF

**Imm2 (44):**
- ○ High-pressure GaAs:
- AB_oI4_44_a_b-001     JZUV
- ○ Ferroelectric NaNO$_2$ ($F5_5$):
- ABC2_oI8_44_a_a_c-001     JZVB
- ○ AgNO$_2$ ($F5_{12}$):
- ABC2_oI8_44_a_a_c-002     VTNW
- ○ Hemimorphite (Zn$_4$Si$_2$O$_7$(OH)$_2$·H$_2$O, $S2_2$):
- A2B5CD2_oI40_44_2c_abcde_d_e-001     W7BE
- ● B30 (MgZn?) (*Problematic*):
- AB_oI48_44_6c_abc2de-001     WDVJ

**Iba2 (45):**
- ○ MnGa$_2$Sb$_2$:
- A2BC2_oI20_45_c_a_c-001     LNLG
- ● Ca$_{11}$InSb$_9$:
- A11BC9_oI84_45_a5c_a_b4c-001     DZ16

**Ima2 (46):**
- ● Room Temperature AgC$_4$N$_3$:
- AB4C3_oI32_46_b_2bc_bc-001     77VH
- ● Low Temperature AgC$_4$N$_3$:

AB4C3_oI32_46_b_2bc_bc-002     AS99
- ○ TiFeSi:
- ABC_oI36_46_ac_ac_3b-001     7FJR
- ○ Nb$_2$Zr$_6$O$_{17}$:
- A2B17C6_oI100_46_ab_b8c_3c-001     DKTF

**Pmmm (47):**
- ● AuMn:
- AB_oP2_47_a_h-001     JM3Q
- ○ 1212C [YBa$_2$Cu$_3$O$_{7-x}$] High-$T_c$:
- A2B3C7D_oP13_47_k_cj_aijl_f-001     MZYC

**Pnnn (48):**
- ○ $\alpha$-RbPr[MoO$_4$]$_2$:
- A2B8CD_oP24_48_h_2m_a_b-001     7YK7

**Pccm (49):**
- ○ $\beta$-Ta$_2$O$_5$:
- A5B2_oP14_49_cehq_ab-001     XVUH
- ● $\delta$-V$_4$D$_3$:
- A3B4_oP14_49_ej_2q-001     6B3F
- ○ CsPr(MoO$_4$)$_2$:
- A2C8D_oP24_49_e_q_2qr_f-001     M8U7

**Pban (50):**
- ○ Orthorhombic La$_2$NiO$_4$:
- A2BC4_oP28_50_gh_ac_ghm-001     69JP
- ○ $\alpha$-Tl$_2$TeO$_3$:
- A3BC2_oP48_50_3m_m_2m-001     O6EV

**Pmma (51):**
- ○ $\beta'$-AuCd ($B19$):
- AB_oP4_51_e_f-001     WLS6
- ○ Parkerite (Ni$_3$Bi$_2$S$_4$):
- AB2C_oP8_51_e_be_f-001     LY1V
- ● $\alpha$-UB$_2$C:
- A2BC_oP8_51_i_a_f-001     DRSV
- ● TaRh:
- AB_oP12_51_ei_fj-001     1M54
- ● Kenhsuite ($\gamma$-Hg$_3$S$_2$Cl$_2$):
- AB2C_oP16_51_2e_bfi_j-001     LQ5X
- ● $\alpha$-BiPd$_3$:
- AB3_oP16_51_af_behk-001     9W5L
- ○ LiNb$_6$O$_{15}$F:
- ABC6D15_oP46_51_f_b_2e2i_cef4i2j-001     0JVV

**Pnna (52):**
- ○ Sr$_2$Bi$_3$:
- A3B2_oP20_52_de_cd-001     12L3
- ○ High-Pressure GaCl$_2$:
- A2B_oP24_52_2e_cd-001     PL6D
- ○ Carnallite [Mg(H$_2$O)$_6$KCl$_3$]:
- A3B12CDE6_oP276_52_d4e_18e_ce_de_2d8e-001     40W9

**Pmna (53):**
- ○ TaNiTe$_2$:
- ABC2_oP16_53_h_e_gh-001     AVNP
- ○ NH$_4$HF$_2$ ($F5_8$):
- A2BC_oP16_53_eh_ab_g-001     DCYY
- ○ Eriochalcite (CuCl$_2$·2H$_2$O, $C45$):
- A2BC4D2_oP18_53_h_a_i_e-001     UFSK
- ○ *Pmna* CuBrSe$_3$:
- ABC3_oP20_53_e_g_hi-001     SET9

**Pcca (54):**
- ● AgClO$_2$:
- ABC2_oP16_54_c_c_f-001     RAQG
- ○ BiGaO$_3$:
- ABC3_oP20_54_e_d_cf-001     WWU9

**Pbam (55):**
- ○ Rh$_5$Ge$_3$:

A3B5_oP16_55_ah_cgh-001     WXAV
- ○ R-carbon:
  A_oP16_55_2g2h-001     QPSD
- ● $ScB_2C_2$:
  A2B2C_oP20_55_2g_2g_h-001     93YX
- ○ $GeAs_2$:
  A2B_oP24_55_2g2h_gh-001     X52Z
- ● $YCrB_4$:
  A4BC_oP24_55_4g_h_h-001     67TW
- ○ $Nb_2Pd_3Se_8$:
  A2B3C8_oP26_55_h_ag_2g2h-001     X0K9
- ● $Ca_5Ga_2As_6$:
  A6B5C2_oP26_55_g2h_a2g_h-001     JZ3U
- ● Ambient pressure $Bi_2Fe_4O_9$:
  A2B4C9_oP30_55_h_fg_aghi-001     7EHM
- ○ $K_2HgCl_4 \cdot H_2O$ ($E3_4$):
  A4BCD2_oP32_55_ghi_e_f_gh-001     07TT
- ○ $HoMn_2O_5$:
  A2BC5_oP32_55_g_eh_fghi-001     KF2N
- ● $Ta_4SiTe_4$:
  A4BC4_oP36_55_e_2g2h_2g2h-001     UXGR
- ● Ludwigite ($Mg_2FeBO_5$):
  A2BC2D_oP36_55_g_g_adh_3g2h-001     A5YT
- ○ $Ru_{11}B_8$:
  A8B11_oP38_55_3gh_b2g3h-001     6FNW
- ● $Nb_8P_5$:
  A17B10_oP54_55_a3g5h_3g2h-001     BWRM
- ○ Orthorhombic $Sr_4Ru_3O_{10}$:
  A10B3C4_oP68_55_2e2fgh2i_adef_2e2f-001     PEKW

**Pccn (56):**
- ○ Valentinite ($Sb_2O_3$, $D5_{11}$):
  A3B2_oP20_56_ce_e-001     HKUU
- ● Calciborite ($CaB_2O_4$ II):
  A2BC4_oP56_56_2e_e_4e-001     ZD64

**Pbcm (57):**
- ● TlF-II:
  AB_oP8_57_d_d-001     E914
- ○ KNCS ($F5_9$):
  ABCD_oP16_57_d_c_d_d-001     K4A8
- ○ $D0_{10}$ ($WO_3$) (*Obsolete*):
  A3B_oP16_57_a2d_d-001     5B3E
- ● DyAl:
  AB_oP16_57_cd_2d-001     GHY9
- ● $SrUO_4$:
  A4BC_oP24_57_cde_d_a-001     JM5X
- ● $Ca_4Al_3Mg$:
  A3B4C_oP32_57_c2d_4d_a-001     6529
- ● $Ta_2S$:
  A2B_oP36_57_ce_2d2e-001     4L1U
- ● Lueshite ($NaNbO_3$):
  ABC3_oP40_57_cd_e_cd2e-001     1JM3
- ● $Hg_7Hg_5$:
  A7B5_oP48_57_c3e_5d-001     9PR5

**Pnnm (58):**
- ○ Hydrophilite ($CaCl_2$, $C35$):
  AB2_oP6_58_a_g-001     YTPF
- ○ $\eta$-$Fe_2C$:
  AB2_oP6_58_a_g-002     FF5Q
- ● Marcasite ($FeS_2$, $C18$):
  AB2_oP6_58_a_g-003     USY5
- ○ $\alpha$-$PdCl_2$ ($C50$):
  A2B_oP6_58_g_a-001     HU31
- ○ InS:

AB_oP8_58_g_g-001     A7M6
- ○ Kotoite ($Mg_3(BO_3)_2$):
  A2B3C6_oP22_58_g_af_gh-001     59YW
- ● $\gamma$-Alane ($AlH_3$):
  AB3_oP24_58_ag_c2gh-001     WUSJ
- ● $S_{12}$ Sulfur:
  A_oP24_58_eg2h-001     7TXG
- ○ $In_4Se_3$:
  A4B3_oP28_58_4g_3g-001     B01P
- ○ Andalusite ($Al_2SiO_5$, $S0_2$):
  A2B5C_oP32_58_eg_3gh_g-001     V93J
- ○ Adamite [$Zn_2(AsO_4)(OH)$, $H2_7$]:
  ABC5D2_oP36_58_g_g_3gh_eg-001     YE34
- ● $Ta_2P$:
  A2B_oP36_58_3g_6g-001     9JBK
- ○ Protoanthophyllite ($H_2Mg_7Si_8O_{24}$):
  A2B7C24D8_oP82_58_g_ae2f_2g5h_2h-001     APMF

**Pmmn (59):**
- ○ Vulcanite (CuTe):
  AB_oP4_59_a_b-001     9JE2
- ○ CNCl:
  ABC_oP6_59_a_a_a-001     Z1TY
- ● FeOCl ($E0_5$):
  ABC_oP6_59_a_b_a-001     DN69
- ○ $RuB_2$:
  A2B_oP6_59_e_a-001     BRW1
- ● $\beta$-$TiCu_3$ ($D0_a$):
  A3B_oP8_59_ae_b-001     CJHD
- ● Orthorhombic $LiFeO_2$ (o-$LiFeO_2$):
  ABC2_oP8_59_a_a_2b-004     WEQC
- ● Shcherbinaite ($V_2O_5$) (*Revised*):
  A5B2_oP14_59_a2e_e-001     KZMB
- ○ $NH_4NO_3$ IV ($G0_{11}$):
  A4B2C3_oP18_59_ef_ab_ae-001     NABB
- ● $\beta$-$NbPd_3$:
  AB3_oP24_59_ae_befg-001     5BSF
- ● $RbAlF_4$ III:
  AB4C_oP24_59_c_efg_ab-001     Q57Q

**Pbcn (60):**
- ○ $\zeta$-$Fe_2N$:
  A2B_oP12_60_d_c-001     5GP6
- ○ $\alpha$-$PbO_2$:
  A2B_oP12_60_d_c-003     9BKR
- ○ $Rh_2O_3$:
  A2B3_oP20_60_d_cd-001     WEMD
- ● $E3_2$ ($CaB_2O_4$ I):
  A2BC4_oP28_60_d_c_2d-001     44A7
- ○ $\beta$-$WO_3$:
  A3B_oP32_60_3d_d-001     S00Q
- ● Columbite ($FeNb_2O_4$, $E5_1$):
  A2C6_oP36_60_c_d_3d-001     ZB5U
- ● $Ru_2Ge_3$ Nowotny Chimney-Ladder:
  A3B2_oP40_60_3d_2cd-001     EU3H
- ○ $Cr_5O_{12}$:
  A5B12_oP68_60_c2d_6d-001     NF0B
- ● $Na_2FeSbO_5$:
  A2C5D_oP72_60_d_2cd_5d_2c-001     2H6S
- ○ $\beta$-Toluene:
  A7B8_oP120_60_7d_8d-001     7NVJ

**Pbca (61):**
- ● $\beta$-$HgO_2$:
  AB2_oP12_61_a_c-001     JQ93
- ● $AgF_2$:

AB2_oP12_61_a_c-002    F3J5
- PdSe$_2$:
 AB2_oP12_61_a_c-003    WV1J
- CdSb ($B_e$):
 AB_oP16_61_c_c-001    RL87
- Cyanogen [(CN)$_2$]:
 AB_oP16_61_c_c-002    NEGW
- Brookite (TiO$_2$, $C21$):
 A2B_oP24_61_2c_c-001    ZJ0C
- Tellurite ($\beta$-TeO$_2$, $C52$):
 A2B_oP24_61_2c_c-002    GK2A
- COCl:
 ABC_oP24_61_c_c_c-001    EHTG
- Pararammelsbergite ($\alpha$-NiAs$_2$):
 A2B_oP24_61_2c_c-003    ZBG0
- AuSn$_2$:
 AB_oP24_61_c_2c-001    FZA4
- Ca$_2$RuO$_4$:
 A2B4C_oP28_61_c_2a_a-001    NDWA
- ReP$_4$:
 A4B_oP40_61_4c_c-001    3PB1
- Benzene:
 AB_oP48_61_3c_3c-001    PWJE
- Tl$_3$AsS$_3$:
 AB3C3_oP56_61_c_3c_3c-001    233Q
- Hambergite [Be$_2$BO$_3$(OH) ($G7_2$)]:
 AB2CD4_oP64_61_c_2c_c_4c-001    PKYK
- Ca$_4$Ti$_3$O$_{10}$:
 A4B10C3_oP68_61_2c_5c_ac-001    R6BH
- O-Carbamoylhydroxylamine (CH$_4$N$_2$O$_2$):
 AB4C2D2_oP72_61_c_4c_2c_2c-001    K3UE
- (TiCl$_4$·POCl$_3$)$_2$:
 A7BCD_oP80_61_7c_c_c_c-001    Q18C
- Enstatite (MgSiO$_3$, $S4_3$):
 AB3C_oP80_61_2c_6c_c-001    N16Y
- CuMo$_3$I$_7$:
 AB7C3_oP88_61_c_7c_3c-001    WLBF
- SrZn(VO)(PO$_4$)$_2$:
 A9B2CDE_oP112_61_9c_2c_c_c_c-001    VHVW
- Room Temperature Bornite (Cu$_5$FeS$_4$):
 A5BC4_oP160_61_10c_2c_8c-001    L8SE

## Pnma ($62$):

- GeS ($B16$):
 AB_oP8_62_c_c-001    Y0CA
- MnP ($B31$):
 AB_oP8_62_c_c-002    36ZK
- FeB ($B27$):
 AB_oP8_62_c_c-003    GH4J
- $\alpha$-SnS ($B29$):
 AB_oP8_62_c_c-004    QZKW
- $\alpha$-Np ($A_c$):
 A_oP8_62_2c-001    L3MB
- Westerveldite (FeAs, $B14$):
 AB_oP8_62_c_c-005    P0GV
- $\eta$-NiSi ($B_d$):
 AB_oP8_62_c_c-010    4N6D
- High-Pressure oP8 Sodium:
 A_oP8_62_2c-002    3BFR
- Co$_2$Si ($C37$):
 A2B_oP12_62_2c_c-001    6EYJ
- HgCl$_2$ ($C25$):
 A2B_oP12_62_2c_c-002    DNGY
- Cotunnite (PbCl$_2$, $C23$):

A2B_oP12_62_2c_c-003    GSD1
- SrH$_2$ ($C29$):
 A2B_oP12_62_2c_c-004    4GNV
- $C53$ (SrBr$_2$) (*Obsolete*):
 A2B_oP12_62_2c_c-010    NJOX
- MnCuP:
 ABC_oP12_62_c_c_c-003    BN2H
- Lautite (CuAsS):
 ABC_oP12_62_c_c_c-017    DUB2
- Chalcostibite (CuSbS$_2$, $F5_6$):
 AB2C_oP16_62_c_2c_c-001    H6D5
- Cementite (Fe$_3$C, $D0_{11}$):
 AB3_oP16_62_c_cd-001    KYLC
- $\epsilon$-Al$_3$Ni ($D0_{20}$):
 A3B_oP16_62_cd_c-001    TJ6A
- Molybdite (MoO$_3$, $D0_8$):
 AB3_oP16_62_c_3c-001    2C2N
- NH$_4$I$_3$ (D0$_{16}$):
 AB3_oP16_62_c_3c-002    DE18
- Diaspore (AlOOH, $E0_2$):
 ABC2_oP16_62_c_c_2c-002    XPSY
- NH$_4$ClBrI ($F5_{14}$):
 ABCD_oP16_62_c_c_c_c-001    NOR6
- UMoC$_2$:
 A2BC_oP16_62_2c_c_c-003    69FA
- NiBi$_3$:
 A3B_oP16_62_3c_c-003    J16T
- Orthorhombic ClF$_3$:
 AB3_oP16_62_c_cd-005    AZN3
- CaMnSb$_2$:
 ABC2_oP16_62_c_c_2c-003    JHY5
- ThNi:
 AB_oP16_62_2c_2c-002    QLCJ
- Stibnite (Sb$_2$S$_3$, $D5_8$):
 A3B2_oP20_62_3c_2c-001    1WMN
- MgB$_4$:
 A4B_oP20_62_2cd_c-001    8U8B
- CaTiO$_3$ Pnma Perovskite:
 AB3C_oP20_62_c_cd_a-001    2KF1
- Tongbaite (Cr$_3$C$_2$, $D5_{10}$):
 A2B3_oP20_62_2c_3c-001    KS86
- (NH$_4$)CdCl$_3$ ($E2_4$):
 AB3C_oP20_62_c_3c_c-001    Z6H8
- $\alpha$-Potassium Nitrate (KNO$_3$) I:
 ABC3_oP20_62_c_c_cd-001    3DEN
- NH$_4$NO$_3$ III ($G0_{10}$):
 ABC3_oP20_62_c_c_cd-002    8RDG
- Aragonite ($G0_2$, CaCO$_3$):
 ABC3_oP20_62_c_c_cd-003    7S19
- K$_2$Te$_3$:
 A2B3_oP20_62_d_3c-001    GCVK
- Pt$_2$Ge$_3$:
 A3B2_oP20_62_3c_2c-002    GDA9
- Ottemannite (Sn$_2$S$_3$):
 A3B2_oP20_62_3c_2c-003    1ZQE
- Au$_4$Zr:
 A4B_oP20_62_4c_c-001    KLZ3
- CuTaS$_3$:
 AB3C_oP20_62_c_3c_c-002    S2FJ
- Monoclinic Vaterite (CaCO$_3$):
 ABC3_oP20_62_c_a_cd-001    RGRE
- Cubanite (CuFe$_2$S$_3$, $E9_e$):
 A2C3_oP24_62_c_d_cd-001    7QNH
- Barite (BaSO$_4$, $H0_2$):

A4BC_oP24_62_c_2cd_c-001    CXQX

○ Cs$_2$Sb:
A2B_oP24_62_4c_2c-001    9L36

○ Chalcocyanite (CuSO$_4$):
AB4C_oP24_62_a_2cd_c-001    QVKJ

● SrAu$_3$Al$_2$:
A2B3C_oP24_62_d_3c_c-001    DAG0

● K$_2$CuCl$_3$:
A3BC2_oP24_62_3c_c_2c-001    WK7H

● IrSe$_2$:
A2B_oP24_62_2c_4c-001    UH0Z

● SrCuYSe$_3$ (Eu$_2$CuS$_3$):
AB3CD_oP24_62_c_3c_c_c-001    SFSU

● SrZn$_5$:
AB5_oP24_62_c_3cd-001    BRA8

○ Forsterite (Mg$_2$SiO$_4$, $S1_2$):
A2B4C_oP28_62_ac_2cd_c-001    V88M

○ Arcanite (K$_2$SO$_4$, $H1_6$):
A2B4C_oP28_62_2c_2cd_c-001    HEYH

○ VOSO$_4$:
A5BC_oP28_62_3cd_c_c-001    3N5K

○ Berthierite (FeSb$_2$S$_4$, $E3_3$):
A4C2_oP28_62_c_4c_2c-001    CAYJ

○ Copper (II) Azide [Cu(N$_3$)$_2$]:
AB6_oP28_62_c_6c-001    VP85

○ Galenobismutite (PbBi$_2$S$_4$):
A2BC4_oP28_62_2c_c_4c-001    PUZH

○ Orthorhombic Ni$_4$B$_3$:
A3B4_oP28_62_3c_4c-001    49QQ

● Rh$_4$P$_3$:
A3B4_oP28_62_3c_4c-002    71TB

○ Monticellite (CaMgSiO$_4$):
ABC4D_oP28_62_c_a_2cd_c-001    TYTV

● Warwickite (FeCoBO$_4$):
ABCD4_oP28_62_c_c_c_4c-001    G1FS

○ Sillimanite (Al$_2$SiO$_5$, $S0_3$):
A2B5C_oP32_62_ac_3cd_c-001    XCRT

○ Original $\beta$-WO$_3$ (*Obsolete*):
A3B_oP32_62_ab4c_2c-001    KPH8

○ $E3_5$ (K$_2$SnCl$_4$·H$_2$O):
A4BC2D_oP32_62_2cd_a_2c_b-001    EVKH

○ K$_2$SnCl$_4$·H$_2$O:
A4BC2D_oP32_62_2cd_c_d_c-001    ZADN

● Li$_2$CdSiO$_4$:
A2C4D_oP32_62_d_2cd_c-001    J1AT

● Empressite (AgTe):
AB_oP32_62_2cd_2cd-001    JH5H

○ Topaz (Al$_2$SiO$_4$F$_2$, $S0_5$):
A2B2C4D_oP36_62_d_d_2cd_c-001    7RBD

○ Atacamite (Cu$_2$(OH)$_3$Cl):
A3C3D3_oP36_62_c_ac_cd_cd-001    4QUY

○ Rynersonite (Orthorhombic CaTa$_2$O$_6$):
AB6C2_oP36_62_c_2c2d_d-001    O7R8

● PbV$_2$O$_6$:
A6BC2_oP36_62_6c_c_2c-001    XXTA

○ BaCuSm$_2$O$_5$:
ABC5D2_oP36_62_c_c_c2d_2c-001    W9F8

○ C$_3$Cr$_7$ ($D10_1$):
A3B7_oP40_62_cd_3c2d-001    9Z5M

○ Norbergite [Mg(F,OH)$_2$· Mg$_2$SiO$_4$, $S0_7$]:
A2B3C4D_oP40_62_d_cd_2cd_c-001    J7L1

○ Ta$_2$Ni$_3$Te$_5$:
A3B2C5_oP40_62_3c_2c_5c-001    72A0

● K$_2$SnCl$_4$·H$_2$O:

A4B2C2DE_oP40_62_2cd_2c_d_c_c-001    AZ89

○ K$_2$S$_3$O$_6$ ($K5_1$):
A2B6C3_oP44_62_2c_2c2d_3c-001    PD3R

○ Possible $\delta$-Y$_2$Si$_2$O$_7$:
A7B2C2_oP44_62_3c2d_2c_d-001    Y82X

○ SbCl$_5$·POCl$_3$:
A8BCD_oP44_62_4c2d_c_c_c-001    XAT4

○ $\alpha$-HBO$_2$ (orthorhombic):
ABC2_oP48_62_3c_3c_6c-001    XYXT

○ Danburite (CaB$_2$Si$_2$O$_8$, $S6_3$):
A2BC8D2_oP52_62_d_c_2c3d_d-001    APE4

● $\alpha'_H$-Ca$_2$SiO$_4$:
A4B8C_oP52_62_2d_4d_c-001    1R45

○ Mo$_4$P$_3$:
A4B3_oP56_62_8c_6c-001    S9HM

● High pressure Bi$_2$Fe$_4$O$_9$:
A2B4C9_oP60_62_d_2cd_3c3d-001    BYQL

○ Approximate Cu$_3$(TeO$_4$)(SO$_4$)·H$_2$O:
A3B2C9DE_oP64_62_cd_2c_5c2d_c_c-001    WLK8

● CdSnP$_{14}$ (HgPbP$_{14}$):
AB14C_oP64_62_c_4c5d_c-001    LZG8

● Ag(tcm)(pyz) [AgC$_8$N$_5$H$_4$]:
A8BC4D5_oP72_62_c_4c2d_2d_3d-001    L9Y6

● Ba$_5$AlIr$_2$O$_{11}$:
A5C2D11_oP76_62_c_5c_2c_5c3d-001    CZW8

● Room Temperature SmBaMn$_2$O$_6$:
A2C6D_oP80_62_2c_2d_2c5d_d-001    8TUC

○ RhCl$_2$(NH$_3$)$_5$Cl ($J1_8$):
A3B15C5D_oP96_62_cd_3c6d_3cd_c-001    3AAY

○ K$_4$[Mo(CN)$_8$]· 2H$_2$O ($F2_1$):
A8B4C4DE8F2_oP108_62_4c2d_2d_2cd_c_4c2d_d-001    BH4W

○ P$_4$Se$_3$:
A4B3_oP112_62_8c4d_4c4d-001    DMDS

○ Epididymite (BeHNaO$_8$Si$_3$, $S4_7$):
ABCD8E3_oP112_62_d_2c_d_4c6d_3d-001    QQ21

○ Anthophyllite (Mg$_5$Fe$_2$Si$_8$O$_{22}$(OH)$_2$, $S4_4$):
A2B5C22D2E8_oP156_62_d_c2d_2c10d_2c_4d-001    P9BG

● Al$_3$Mn:
A9B4_oP156_62_5c11d_6c3d-001    WUDY

○ Autunite Ca[(UO$_2$)(PO$_4$)]$_2$(H$_2$O)$_{11}$:
A2B22C23D2E2_oP200_62_c_11d_3c10d_d_d-001    PYMC

**Cmcm (63):**

○ $\alpha$-U ($A20$):
A_oC4_63_c-001    AFG5

○ CrB ($B33$):
AB_oC8_63_c_c-001    QWV1

● $\beta$-SnS:
AB_oC8_63_c_c-004    8MTK

○ ZrSi$_2$ ($C49$):
A2B_oC12_63_2c_c-001    HPCM

○ UBC:
ABC_oC12_63_c_c_c-004    NQGP

● MoAlB:
ABC_oC12_63_c_c_c-005    XSM0

○ SrCuO$_2$:
AB2C_oC16_63_c_2c_c-001    WL5U

○ MgCuAl$_2$ ($E1_a$):
A2BC_oC16_63_f_c_c-002    ECJE

○ Mn$_3$As ($D0_d$):
AB3_oC16_63_c_3c-001    DPFG

○ Re$_3$B:
AB3_oC16_63_c_cf-001    EWOY

- NaHg:
  AB_oC16_63_g_2c-001    BVCU
- Post-perovskite ($MgSiO_3$):
  AB3C_oC20_63_c_cf_a-001    BCSA
- Lepidocrocite [$\gamma$-FeO(OH), $E04$]:
  AB2C2_oC20_63_c_f_2c-001    WBGV
- $ThFe_2SiC$:
  AB2CD_oC20_63_a_f_c_c-001    UF5S
- $V_3AsC$:
  ABC3_oC20_63_c_a_cf-003    2T8B
- $SrPdGa_3$:
  A3BC_oC20_63_ce_c_c-001    2P2W
- Rasvumite ($KFe_2S_3$):
  A2BC3_oC24_63_e_c_cg-001    88NB
- Anhydrite ($CaSO_4$, $H0_1$):
  AB4C_oC24_63_c_fg_c-001    EHW6
- $MgSO_4$:
  ABC4_oC24_63_c_a_fg-001    FNXT
- $ZrTe_5$:
  A5B_oC24_63_c2f_c-001    EE51
- $Si_{24}$ Clathrate:
  A_oC24_63_3f-001    TPXT
- $URe_2$:
  A2B_oC24_63_acg_f-001    RKF7
- $YNiAl_4$:
  A4BC_oC24_63_acf_c_c-002    ULYP
- $\eta$-$Fe_2Al_5$:
  A5B_oC24_63_afg_c-001    US2Q
- $BaZn_5$:
  AB5_oC24_63_c_ceg-001    BGG4
- $KCuZrS_3$:
  ABC3D_oC24_63_c_c_cf_a-001    3902
- $MnAl_6$ ($D2_h$):
  A6B_oC28_63_efg_c-001    386R
- $Na_2CrO_4$ ($H1_8$):
  AB2C4_oC28_63_c_ac_fg-001    A786
- $NbNiTe_5$:
  ABC5_oC28_63_c_a_c2f-001    BN2Z
- Pseudobrookite ($Fe_2TiO_5$, $E4_1$):
  A2B5C_oC32_63_f_c2f_c-001    NJWX
- $Pd_5Pu_3$:
  A5B3_oC32_63_cfg_ce-001    9AGN
- $Ta_2NiS_5$:
  AB5C2_oC32_63_c_c2f_f-004    FETU
- Refined $Tl_2AlF_5$:
  AB5C2_oC32_63_a_cef_g-001    KA1Q
- $Ag_7Ca_2$:
  A7B2_oC36_63_cgh_f-001    PMU3
- $CaFe_3O_5$:
  AB3C5_oC36_63_c_af_c2f-001    8XTL
- Pinalite ($Pb_3WO_5Cl_2$):
  A2B5C3D_oC44_63_ac_ch_cf_c-001    S4ZE
- $CaFe_4O_6$:
  AB4C6_oC44_63_c_2f_ac2f-001    8HU9
- $\beta$-$AlF_3$:
  AB3_oC48_63_ad_cfgh-001    W7TS
- $Cu_2Pb(SeO_3)_2Br_2$:
  A2B2C6DE2_oC52_63_g_e_fh_c_f-001    RZ26
- $CaFe_5O_7$:
  AB5C7_oC52_63_c_a2f_c3f-001    KAT1
- $Nb_4As_3$:
  A3B4_oC56_63_2c2f_ac3f-001    J9UY
- $Y_2Ga_9Co_3$:
  A3B9C2_oC56_63_ae_cfgh_g-001    1YQJ

- $\alpha$-$Ni_7S_6$:
  A9B5_oC56_63_c4f_c2f-001    T4VC
- $S0_4$ (Staurolite, $Fe(OH)_2Al_4SiO_{10}$) (*Obsolete*):
  A4BC12D2_oC76_63_eg_c_f3gh_g-001    CXFF
- $\beta$-$Bi_4V_2O_{11}$:
  A4B12C3_oC76_63_eg_fg2h_cf-001    AB76
- Orthorhombic $Nb_{12}O_{29}$:
  A12B29_oC164_63_6f_3c13f-001    V0KX
- $La_{43}Ni_{17}Mg_5$:
  A43B5C17_oC260_63_c8fg6h_cfg_ce3f2h-001    Z34R

**Cmce (64)**:

- $\alpha$-Ga ($A11$):
  A_oC8_64_f-001    JS9R
- Black Phosphorus ($A17$):
  A_oC8_64_f-002    7VYH
- Molecular Iodine ($A14$):
  A_oC8_64_f-003    234K
- KO:
  AB_oC16_64_e_f-002    HZUD
- $La_2Ni_3$:
  A2B3_oC20_64_f_ae-001    4LJZ
- $H_2S$ (170 GPa):
  A2B_oC24_64_2f_f-001    7FZR
- $SmSb_2$:
  A2B_oC24_64_ef_f-001    370K
- $CoGe_2$:
  AB2_oC24_64_d_ef-002    QB3A
- $Gd_2CuO_4$:
  AB2C4_oC28_64_a_d_ef-001    UBCB
- Base-Centered Orthorhombic $La_2CuO_4$:
  AB2C4_oC28_64_a_f_ef-003    Y6BX
- $R_2$ $Au_3Zn$:
  A3B_oC32_64_def_d-001    GVGD
- $PdSn_3$:
  AB3_oC32_64_d_fg-001    J38S
- Low Temperature $FeSi_2$:
  A2B_oC48_64_df_2g-001    MR28
- Experimental $Li_3AlP_2$:
  AB3C2_oC48_64_d_dg_ef-001    94E5
- Base-centered Orthorhombic $Sr_4Ru_3O_{10}$:
  A10B3C4_oC68_64_2dfg_ad_2d-001    XC67
- $Cs_4Mg_3F_{10}$:
  A4B10C3_oC68_64_2f_e2fg_af-001    B9LX
- $Zr_7Ni_{10}$:
  A10B7_oC68_64_f2g_adef-001    7BKZ
- $MgB_2C_2$:
  A2B2C_oC80_64_efg_efg_df-001    BJ3W
- $Na_2Mo_2O_7$:
  A2B2C7_oC88_64_ef_df_3f2g-001    0H5D
- Zektzerite ($NaLiZrSi_6O_{15}$):
  ABC15D6E_oC192_64_d_e_3f6g_3g_e-001    MZR4

**Cmmm (65)**:

- $CdPt_3$ ("New" $L1_3$):
  AB3_oC8_65_a_bf-001    JBK1
- $\alpha$-IrV:
  AB_oC8_65_g_j-001    S2CY
- $Mn_2AlB_2$:
  AB2C2_oC10_65_a_g_h-001    XPEY
- $Li_2PrO_3$:
  A2B3C_oC12_65_h_ah_b-001    2B54
- $Ga_2Zr$:
  A2B_oC12_65_acg_h-001    V3JJ
- $Ag_3Te_2Tl$:

A3B2C_oC12_65_ah_g_c-001  `35VG`

○ Ga$_3$Pt$_5$:
A3B5_oC16_65_ah_bej-001  `36DL`

○ Nb$_3$O$_7$F:
A3B8_oC22_65_bg_ac2gh-001  `YUGM`

● Cr$_4$AlB$_6$:
AB6C4_oC22_65_a_3g_2h-001  `6ZPS`

● ThMoB$_4$:
A4BC_oC24_65_gip_h_j-001  `B95F`

○ Mg(NH$_3$)$_2$Cl$_2$ ($E1_3$):
A2B8CD2_oC26_65_h_r_a_i-001  `5BTF`

● Tb$_3$Sn$_7$:
A11B3_oC28_65_c4gh_ah-001  `2ATW`

● "124 Superconductor" (YBa$_2$Cu$_4$O$_8$):
A2B4C8D_oC30_65_h_2g_3gh_c-001  `2EBJ`

● Li$_7$Ge$_2$:
A2B7_oC36_65_gj_achipq-001  `3WXK`

● High Temperature SmBaMn$_2$O$_6$:
A2B2C6D_oC40_65_g_n_ijklm_h-001  `VMAN`

● "247 Superconductor" (Y$_2$Ba$_4$Cu$_7$O$_{15}$):
A4B7C16D2_oC58_65_2h_b3g_ac5g2h_h-001  `TTW9`

**Cccm (66):**

● NbD:
AB_oC8_66_a_e-001  `O87H`

○ SrAl$_2$Se$_4$:
A2B4C_oC28_66_l_kl_a-001  `QSR1`

● H$_3$S (60 GPa):
A3B_oC64_66_gi2lm_2l-001  `DQEU`

○ β-ThI$_3$:
A3B_oC64_66_kl2m_acl-001  `7JBJ`

● γ-Li$_2$IrO$_3$:
A2C3_oC96_66_ik_cdj2k_gl2m-001  `9VYY`

**Cmme (67):**

○ α-FeSe:
AB_oC8_67_a_g-001  `JFD1`

○ α-PbO:
AB_oC8_67_a_g-002  `TSNN`

○ Al$_2$CuIr:
A2BC_oC16_67_ag_b_g-001  `O8UR`

○ HoCuP$_2$:
ABC2_oC16_67_a_g_bg-001  `BX6K`

○ NH$_4$H$_2$PO$_2$ ($F5_7$):
A2BC2D_oC24_67_m_a_n_g-001  `ZLA7`

● RbPaF$_6$ (V):
A6BC_oC32_67_no_c_g-001  `PLRV`

**Ccce (68):**

○ PdSn$_4$:
AB4_oC20_68_a_i-001  `LGPW`

**Fmmm (69):**

○ TlF ($B24$) (*Obsolete*):
AB_oF8_69_a_b-001  `XT4M`

○ β-SrRh$_2$As$_2$:
A2B2C_oF20_69_g_f_a-001  `DY16`

● Face-Centered Orthorhombic La$_2$CuO$_4$:
A2BC4_oF28_69_a_g_cg-001  `LB5P`

● Rb$_2$P$_3$:
A3B2_oF40_69_hm_fg-001  `QBNS`

● Orthorhombic Bi$_3$NbTiO$_9$ $m = 2$ Aurivillius:
A3B2C9_oF56_69_ag_g_bfgl-001  `FSA1`

● Orthorhombic Bi$_4$Ti$_3$O$_{12}$ $m = 3$ Aurivillius:
A4B12C3_oF76_69_2g_cf2gl_ag-001  `Y2ES`

**Fddd (70):**

○ γ-Pu:
A_oF8_70_a-001  `7VYN`

○ TiSi$_2$ ($C54$) Nowotony Chimney-Ladder:
A2B_oF24_70_e_a-001  `KN7L`

○ Mn$_2$B ($D1_f$):
AB2_oF48_70_e_ef-001  `RZ74`

○ Mg$_2$Cu ($C_b$):
AB2_oF48_70_e_ef-002  `S336`

○ Thenardite [Na$_2$SO$_4$ (V), $H1_7$]:
A2B4C_oF56_70_e_h_a-001  `8CBQ`

● RbVSe$_2$:
AB2C_oF64_70_e_h_ab-001  `1RMB`

● Sc$_2$S$_3$:
A3B2_oF80_70_fh_2e-001  `HQ18`

● β-Na$_2$PtO$_3$:
A2B3C_oF96_70_2e_fh_e-001  `7WUP`

● α-S ($A16$):
A_oF128_70_4h-001  `MYP4`

**Immm (71):**

● MoPt$_2$:
AB2_oI6_71_a_e-001  `EKX0`

○ ReSi$_2$:
AB2_oI6_71_a_e-002  `9FBX`

○ CsO:
AB_oI8_71_e_g-001  `2UZJ`

○ NbPS:
ABC_oI12_71_e_h_f-001  `B529`

● UTe$_2$:
A2B_oI12_71_eh_f-001  `7HXW`

○ Ta$_3$B$_4$ ($D7_b$):
A4B3_oI14_71_ef_af-001  `3XYE`

○ CsFeS$_2$ (100K):
ABC2_oI16_71_e_g_fi-001  `R5TH`

● Sc$_3$CoC$_4$:
A4BC3_oI16_71_m_a_bf-001  `KR8A`

● High-Temperature Cryolite (Na$_3$AlF$_6$):
AB6C3_oI20_71_a_el_bf-001  `6BL9`

● La$_3$Al$_{11}$:
A11B3_oI28_71_bf2m_ai-001  `TNYL`

● Nb$_6$Sn$_5$:
A6B5_oI44_71_egkl_fghl-001  `Q45R`

● Orthorhombic Fullerene (Cs$_3$C$_{60}$):
A15B_oI128_71_lmn6o_eg-001  `88NV`

**Ibam (72):**

○ SiS$_2$ ($C42$):
A2B_oI12_72_j_a-001  `PYBC`

○ Ga$_2$Mg$_5$ ($D8_9$):
A2B5_oI28_72_j_afj-001  `AB10`

● U$_2$Co$_3$Si$_5$:
A3B5C2_oI40_72_aj_bfj_j-001  `LD1K`

**Ibca (73):**

○ KAg[CO$_3$]:
ABCD3_oI48_73_d_c_c_cf-001  `HOD0`

● Predicted Li$_3$AlP$_2$:
AB3C2_oI96_73_f_3f_acde-001  `QKN5`

**Imma (74):**

○ KHg$_2$:
A2B_oI12_74_h_e-001  `TWQQ`

○ CeCu$_2$:
AB2_oI12_74_e_h-001  `237T`

○ GdSi$_2$:
AB2_oI12_74_e_2e-001  `EQ3C`

○ Al$_4$U ($D1_b$):
A4B_oI20_74_aeh_e-001  `FNF2`

○ LiCuVO$_4$:
ABC4D_oI28_74_a_c_hi_e-001     `50NO`

○ Zn(NH$_3$)$_2$Cl$_2$ ($E1_2$):
A2B6C2D_oI44_74_i_hj_h_e-001     `XWSU`

● Hg$_3$S$_2$I$_2$:
A3B2C2_oI56_74_fhi_2ei_j-001     `EXRV`

● MgAlB$_{14}$:
AB14C2_oI68_74_a_3i2j_h-001     `U35P`

● Moskvinite (Na$_2$KYSi$_6$O$_{15}$):
AB2C15D6E_oI100_74_e_g_e2hi2j_hj_a-001     `F0JR`

**P4 (75)**:

○ Hexagonal Hollandite (BaRu$_4$Cr$_2$O$_{12}$):
A2B2C12D4_tP76_75_2a2b_2d_12d_4d-001     `MXLF`

**P4$_1$ (76)**:

○ LaRhC$_2$:
A2BC_tP16_76_2a_a_a-001     `5CQR`

○ Cs$_3$P$_7$:
A3B7_tP40_76_3a_7a-001     `DCAK`

● $\beta$-Ca$_2$P$_2$O$_7$:
A2B7C2_tP88_76_4a_14a_4a-001     `BAJJ`

**P4$_2$ (77)**:

○ H$_2$S III:
A2B_tP48_77_8d_4d-001     `GRWY`

○ Pinnoite (MgB$_2$O[OH]$_6$):
A2B6CD7_tP64_77_2d_6d_d_ab6d-001     `Z918`

○ Gwihabaite (NH$_4$NO$_3$ V):
A4B2C3_tP72_77_8d_ab2c2d_6d-001     `JS41`

**P4$_3$ (78)**:

○ Sr$_2$As$_2$O$_7$:
A2B7C2_tP88_78_4a_14a_4a-001     `SCRS`

**I4 (79)**:

○ TlZn$_2$Sb$_2$:
A2BC2_tI20_79_c_2a_c-001     `T892`

**I4$_1$ (80)**:

○ $\beta$-NbO$_2$:
AB2_tI48_80_2b_4b-001     `2XGR`

**P$\bar{4}$ (81)**:

○ GeSe$_2$ (High Pressure):
AB2_tP12_81_adg_2h-001     `1L9R`

**I$\bar{4}$ (82)**:

○ BPO$_4$ ($H0_7$):
AB4C_tI12_82_c_g_a-001     `C601`

● InPS$_4$:
ABC4_tI12_82_a_c_g-001     `SUXC`

○ CdAl$_2$S$_4$ ($E3$):
A2BC4_tI14_82_bc_a_g-001     `S1AD`

○ Kesterite [Cu$_2$(Zn,Fe)SnS$_4$]:
A2B4CD_tI16_82_ac_g_b_d-001     `92JU`

○ Ni$_3$P ($D0_e$):
A3B_tI32_82_3g_g-001     `WHNY`

**P4/m (83)**:

○ Ti$_2$Ga$_3$:
A3B2_tP10_83_adj_k-001     `FCWX`

● Au$_{31}$Mn$_9$:
A31B9_tP40_83_ae3j4k_cjk-001     `FCR5`

**P4$_2$/m (84)**:

○ Vysotskite (PdS, $B34$):
AB_tP16_84_cej_k-001     `KXWJ`

● Cu$_4$Pd:
A4B_tP20_84_afjk_j-001     `L3AR`

**P4/n (85)**:

○ MoPO$_5$:
AB5C_tP14_85_c_cg_a-001     `N6T3`

○ $\alpha$-SrBr$_2$:
A2B_tP30_85_ab2g_cg-001     `G8GD`

○ Bromocarnallite ($E2_6$, KMg(H$_2$O)$_6$(Cl,Br)$_3$):
A3B6CD_tP44_85_acg_3g_bc_d-001     `PV6C`

● Provisional $\delta$-Alane (AlH$_3$):
AB3_tP64_85_2ceg_2cf5g-001     `U4T0`

● BaRu$_6$O$_{12}$:
AB12C6_tP76_85_2c_6g_3g-001     `QUVT`

**P4$_2$/n (86)**:

○ Ti$_3$P:
AB3_tP32_86_g_3g-001     `ENA6`

○ PNCl$_2$ ($E1_4$):
A2BC_tP32_86_2g_g_g-001     `QLZ6`

○ NaSb(OH)$_6$ ($J1_{11}$):
AB6C_tP32_86_c_3g_d-001     `NR43`

○ $\beta$-LiIO$_3$:
ABC3_tP40_86_g_3g-001     `9SQG`

● "Double-Double" Perovskite CaMnCrSbO$_6$:
ABCD6E_tP40_86_e_c_ab_3g_d-001     `UT10`

● Nd$_4$Re$_2$O$_{11}$:
A4B11C2_tP68_86_2g_ab5g_g-001     `8MQB`

**I4/m (87)**:

○ Ni$_4$Mo ($D1_a$):
AB4_tI10_87_a_h-001     `VMNQ`

● Ga$_2$Te$_5$:
A2B5_tI14_87_d_ah-001     `4RP0`

○ Ti$_5$Te$_4$:
A4B5_tI18_87_h_ah-001     `E7DL`

○ Sr$_2$NiWO$_6$:
AB6C2D_tI20_87_a_eh_d_b-001     `QP36`

● High Temperature Metastable VO$_2$:
A2B_tI24_87_2h_h-001     `MHEM`

● Ni$_{12}$P$_5$:
A12B5_tI34_87_hi_ah-001     `13WH`

● Ba$_5$Yb$_8$Ni$_4$O$_{21}$:
A5B4C21D8_tI76_87_ah_h_bh2i_2h-001     `M4UL`

○ Marialite Scapolite [Na$_4$Cl(AlSi$_3$)$_3$O$_{24}$, $S64$]:
AB4C24D12_tI82_87_a_h_2h2i_hi-001     `8NAD`

**I4$_1$/a (88)**:

○ $\alpha$-ThCl$_4$:
A4B_tI20_88_f_a-001     `5VT7`

○ Scheelite (CaWO$_4$, $H0_4$):
AB4C_tI24_88_a_f_b-003     `Y9T0`

○ Copper (I) Azide (CuN$_3$):
AB3_tI32_88_c_df-001     `LOWX`

○ $\alpha$-NbO$_2$:
AB2_tI96_88_2f_4f-001     `BLKR`

○ Na$_4$Ge$_9$O$_{20}$:
A9B4C20_tI132_88_a2f_f_5f-001     `JMG5`

**P422 (89)**:

○ (CH)$_{17}$FeO$_4$Pt (Original Page):
A17BC4D_tP184_89_17p_p_4p_il-001     `EK6P`

● (CH)$_{17}$FeO$_4$Pt (Revised):
A17BC17D4E_tP320_89_17p_p_17p_4p_il-001     `HZJM`

**P42$_1$2 (90)**:

○ $G7_5$ (PbCO$_3$·PbCl$_2$, Phosgenite) (Obsolete):
A2B3C3D2_tP16_90_c_f_ce_e-001     `ZUNT`

○ Na$_4$Ti$_2$Si$_8$O$_{22}$[H$_2$O]$_4$:
A4B2C13D4E_tP48_90_g_d_cef2g_g_c-001     `9JHP`

● BaCu$_4$[VO]|[PO$_4$]$_4$:
A4B4C17D4E_tP54_90_a_g_c4g_g_c-001     `61ZR`

**P4$_1$22 (91)**:

○ ThBC:

`ABC_tP24_91_d_d_d-001`  `S6UF`

**P4₂2₁2 (92):**
- ○ α-Cristobalite (SiO₂, low, *C*30):
  `A2B_tP12_92_b_a-001`  `VHRN`
- ○ Paratellurite (αTeO₂):
  `A2B_tP12_92_b_a-002`  `6BKE`
- ● γ-LiAlO₂:
  `ABC2_tP16_92_a_a_b-001`  `TYEK`
- ● Tetragonal (red) ZnP₂:
  `A2B_tP24_92_2b_b-001`  `WWPP`
- ● Zr₅Si₄:
  `A4B5_tP36_92_2b_a2b-001`  `C3GS`
- ● Intermediate Temperature Tetragonal TlS:
  `AB_tP64_92_2a3b_4b-001`  `X71C`
- ● Maucherite (Ni₁₁As₈):
  `A8B11_tP76_92_2a3b_a5b-001`  `CBJ0`
- ● Y₃Ni₂:
  `A2B3_tP80_92_4b_2a5b-001`  `J821`
- ○ Retgersite (α-NiSO₄·6H₂O, *H*4₅):
  `A12BC10D_tP96_92_6b_a_5b_a-001`  `JAKV`

**P4₂22 (93):**
- ○ AsPh₄CeS₈P₄Me₈:
  `AB32CD4E8_tP184_93_i_16p_af_2p_4p-001`  `PSUB`

**P4₃2₁2 (94):**
- ○ Li₂MoF₆:
  `A6B2C_tP18_94_eg_c_a-001`  `1ZGT`
- ○ Na₅Fe₃F₁₄:
  `A14B3C5_tP44_94_c3g_ad_bg-001`  `CH2B`

**P4₃22 (95):**
- ○ ThBC:
  `ABC_tP24_95_d_d_d-001`  `8RZF`

**P4₃2₁2 (96):**
- ○ "ST12" of Si:
  `A_tP12_96_ab-001`  `YUU3`
- ○ Keatite (SiO₂):
  `A2B_tP36_96_3b_ab-001`  `J0BF`
- ● α-AlB₁₂:
  `A5B22_tP216_96_5b_2a21b-001`  `ZF6C`

**I422 (97):**
- ○ NaGdCu₂F₈:
  `A2B8CD_tI24_97_d_k_a_b-001`  `FH59`
- ○ Ta₂Se₈I:
  `AB8C2_tI44_97_e_2k_cd-001`  `4HFE`

**I4₁22 (98):**
- ○ CdAs₂:
  `A2B_tI12_98_f_a-001`  `G4QQ`
- ○ Phase III Cd₂Re₂O₇:
  `A2B7C2_tI44_98_f_acde_f-001`  `XLYF`

**P4mm (99):**
- ○ Tetragonal PZT [Pb(Zr$_x$Ti$_{1-x}$)O₃]:
  `A3BC_tP5_99_ac_b_a-002`  `Z2SB`
- ● Room Temperature Tetragonal BaTiO₃:
  `A3C_tP5_99_b_ac_a-001`  `3BZX`

**P4bm (100):**
- ● *F*5₄ (NH₄ClO₂) (*Obsolete*):
  `ABC2_tP8_100_b_a_c-001`  `YZB2`
- ○ NH₄NO₃ II (*G*0₉):
  `ABC3_tP10_100_b_a_bc-001`  `80PK`
- ○ Fresnoite (Ba₂TiSi₂O₈):
  `A2B8C2D_tP26_100_c_abcd_c_a-001`  `QC0E`
- ○ Ce₃Si₆N₁₁:
  `A3B11C6_tP40_100_ac_bc2d_cd-001`  `LZ7C`

**P4₂cm (101):**

○ Y-MgNiSn:
`A7B7C2_tP32_101_ade_bde_d-001`  `LR8K`

**P4₂nm (102):**
- ○ Gd₃Al₂:
  `A2B3_tP20_102_2c_b2c-001`  `AY9W`

**P4cc (103):**
- ○ High Temperature NbTe₄:
  `AB4_tP10_103_a_d-001`  `MRXL`
- ○ VSe₂O₆:
  `A6B2C_tP72_103_abc5d_2d_abc-001`  `9ZX1`

**P4nc (104):**
- ○ Tl₄HgI₆:
  `AB6C4_tP22_104_a_2ac_c-001`  `Z85J`
- ○ Ba₅In₄Bi₅:
  `A5B5C4_tP28_104_ac_ac_c-001`  `GFEJ`

**P4₂mc (105):**
- ○ BaGe₂As₂:
  `A2BC2_tP20_105_f_bc_2d-001`  `W96U`

**P4₂bc (106):**
- ○ NaZn[OH]₃:
  `A3BC3D_tP64_106_3c_c_3c_c-001`  `SP8Q`

**I4mm (107):**
- ○ GeP (High-pressure, superconducting):
  `AB_tI4_107_a_a-001`  `B4LQ`
- ○ BaNiSn₃:
  `ABC3_tI10_107_a_a_ab-001`  `N3SR`
- ○ Co₅Ge₇:
  `A5B7_tI24_107_ac_abd-001`  `Y7TL`
- ● UP₂:
  `A2B_tI24_107_2abc_2ab-001`  `3JHK`

**I4cm (108):**
- ○ Sr₅Si₃ (*Obsolete*):
  `A3B5_tI32_108_ac_a2c-001`  `BY2X`

**I4₁md (109):**
- ○ NbAs:
  `AB_tI8_109_a_a-001`  `M2PT`
- ○ LaPtSi:
  `ABC_tI12_109_a_a_a-001`  `RTEY`

**I4₁cd (110):**
- ○ Be[BH₄]₂:
  `A2BC8_tI176_110_2b_b_8b-001`  `29CG`

**P4̄2m (111):**
- ○ *E*3₁ (β-Ag₂HgI₄):
  `A2BC4_tP7_111_e_b_n-001`  `34W5`
- ○ VN (Low-temperature):
  `AB_tP8_111_n_n-001`  `GUN1`
- ○ MnF₂:
  `A2B_tP12_111_2n_bce-001`  `YKCX`
- ● Mooihoekite (Cu₉Fe₉S₁₆):
  `A9B9C16_tP34_111_ajn_bcdek_2no-001`  `2KD6`

**P4̄2c (112):**
- ○ α-CuAlCl₄:
  `AB4C_tP12_112_b_n_e-001`  `FV74`

**P4̄2₁m (113):**
- ○ BaS₃ (*D*0₁₇):
  `AB3_tP8_113_a_ce-001`  `0A77`
- ○ Ammonium Chlorite (NH₄ClO₂):
  `AB4CD2_tP16_113_c_f_a_e-001`  `RYUC`
- ○ Akermanite (Ca₂MgSi₂O₇, *S*5₃):
  `A2BC7D2_tP24_113_e_a_cef_e-001`  `QV01`

**P4̄2₁c (114):**
- ○ Pd₄Se:

A4B_tP10_114_e_a-001    C2PK

○ SeO₃:
A3B_tP32_114_3e_e-001    U620

○ Ag₂SO₄·4NH₃ ($H4_{17}$):
A2B12C4D4E_tP46_114_d_3e_e_e_a-001    PABV

○ C₁₉Sc₁₅:
A19B15_tP68_114_ac4e_bc3e-001    PVAV

**P4̄m2 (115):**
● $F6_1$ Chalcopyrite (CuFeS₂) (*Obsolete*):
ABC2_tP4_115_a_c_g-001    88N3

○ Rh₃P₂:
A2B3_tP5_115_g_ag-001    EXDQ

● Zr₂CuSb₃:
AB3C2_tP6_115_a_bg_g-001    NQAB

○ Orange (averaged) HgI₂:
AB2_tP12_115_j_egi-001    RJ86

● Deltalumite (δ-alumina, Al₂O₃):
A3B4_tP84_115_acef3g3j3k_6j6k-001    YL8L

**P4̄c2 (116):**
○ Ru₂Sn₃:
A2B3_tP20_116_adi_ej-001    RJFD

● Mn₄Ge₇ Nowotny Chimney-Ladder:
A4B7_tP44_116_ach2i_e3j-001    2NXF

● Rh₁₀Ga₁₇:
A17B10_tP108_116_e8j_ad2g2h5i-001    D8JE

**P4̄b2 (117):**
○ β-Bi₂O₃ ($D5_{12}$):
A2B3_tP20_117_i_adgh-001    B48H

**P4̄n2 (118):**
○ RuIn₃:
A3B_tP16_118_ei_f-001    0ZUR

○ Ir₃Ga₅:
A5B3_tP32_118_f2i_aceh-001    DCPJ

**I4̄m2 (119):**
○ GaSb (II):
AB_tI4_119_c_a-001    K2V9

○ Tetragonal TlFeS₂:
ABC2_tI8_119_c_e_a-001    J41Y

● Mn₂RhSn Tetragonal Heusler:
A2BC_tI8_119_ac_b_d-001    4BRY

○ RbGa₃:
A3B_tI24_119_a2i_bf-001    6TQK

○ Phase II Cd₂Re₂O₇:
A2B7C2_tI44_119_i_acefgh_i-001    AZDB

**I4̄c2 (120):**
○ KAu₄Sn₂:
A4BC2_tI28_120_i_a_h-001    KAKD

○ BeSO₄·4H₂O ($H4_3$):
AB8C8D_tI72_120_b_2i_2i_c-001    KR2S

**I4̄2m (121):**
○ C17 (Fe₂B) (*Obsolete*):
AB2_tI12_121_ab_i-001    1BSH

○ Stannite (Cu₂FeS₄Sn, $H2_6$):
A2BC4D_tI16_121_d_a_i_b-001    JRUG

● Luzonite (Cu₃AsS₄):
AB3C4_tI16_121_a_bd_i-001    3FZ9

● Tl₂CdGeTe₄:
ABC4D2_tI16_121_a_b_i_c-001    RB89

○ K₃CrO₈:
A3C8_tI24_121_a_bd_2i-001    FR9M

○ α-V₃S:
A3B_tI32_121_f_g2i-001    QHX2

○ SrCu₂(BO₃)₂:

A2B2C6D_tI44_121_i_i_ij_c-001    USSD

**I4̄2d (122):**
○ Chalcopyrite (CuFeS₂, $E1_1$):
ABC2_tI16_122_a_b_d-001    XWJF

● Mn₁.₄PtSn:
A3B2C2_tI28_122_ad_c_d-001    1YSS

○ Mercury Cyanide [Hg(CN)₂, $F1_1$]:
A2BC2_tI40_122_e_d_e-001    O5M4

○ KH₂PO₄ ($H2_2$):
A4BC4D_tI40_122_e_b_e_a-001    TLFW

● NaS₂:
AB2_tI48_122_cd_2e-001    Q5KE

○ NH₄H₂PO₄:
A8BC4D_tI56_122_2e_b_e_a-001    PGWH

● Ba₁₉Li₄₄:
A19B44_tI252_122_ac4e_2d10e-001    HB2V

**P4/mmm (123):**
○ CuAu(I) ($L1_0$):
AB_tP2_123_a_d-001    9JUZ

○ δ-CuTi ($L2_a$):
AB_tP2_123_a_d-002    C9G2

● Linzhiite (High Temperature FeSi₂):
AB2_tP3_123_a_h-001    N538

○ FeNi₃:
ABC_tP3_123_c_a_b-001    C14E

○ CuTi₃ ($L6_0$):
AB3_tP4_123_a_ce-001    B4JD

○ CaCuO₂:
ABC2_tP4_123_a_d_e-001    ACQJ

○ NH₄HgCl₃ ($E2_5$):
A3BC_tP5_123_ah_c_b-001    5MRT

● Mn₂Co₂C:
AB2C2_tP5_123_b_e_ac-001    8ZWP

○ TlAlF₄ ($H0_8$):
AB4C_tP6_123_b_eg_c-001    416G

● CeCr₂Si₂C:
ABC2D2_tP6_123_b_a_e_h-001    G8WV

○ HoCoGa₅:
AB5C_tP7_123_b_ci_a-001    SBCX

● K₂PtCl₄ ($H1_5$):
A4B2C_tP7_123_j_e_a-001    3SRH

● CePdGa₆:
AB6C_tP8_123_a_hi_b-001    1ALT

○ CaRbFe₄As₄ (Superconducting):
A4B4C4D_tP10_123_gh_a_i_d-001    X7PG

○ $E6_1$ [Sr(OH)₂(H₂O)₈] (*Obsolete*):
AB2C8_tP11_123_r_e_b-001    QEVQ

○ $E6_2$ [SrO₂(H₂O)₈] (Possibly Obsolete):
AB2C8_tP11_123_r_h_a-001    E0G5

● YBaCuFeO₅:
A2BC2D5E_tP11_123_a_h_h_ci_b-001    K0VP

● Ho₂CoGa₈:
AB8C2_tP11_123_a_ehi_g-003    G7WP

● AuMn₃:
AB3_tP12_123_ae_cghi-001    U1FT

● HgBa₂CaCu₂O₆₊δ:
A2BC2DE7_tP13_123_g_b_h_c_ahi-001    BESE

● Ce₃PdIn₁₁:
A3B11C_tP15_123_ag_ch2i_b-001    7DQQ

● Ce₅Pd₂In₁₉:
A5B19C2_tP26_123_a2g_ce2h3i_g-001    RR9W

**P4/mcc (124):**
○ Room Temperature NbTe₄:

AB4_tP10_124_a_m-001     XFHX

○ Nb$_4$CoSi:
AB4C_tP12_124_a_m_c-001     LAN3

○ CaO$_2$(H$_2$O)$_8$:
AB8C2_tP22_124_a_n_h-001     QCQG

**P4/nbm (125):**

○ PtPb$_4$ ($D1_d$):
A4B_tP10_125_m_a-001     OBXW

○ KCeSe$_4$:
ABC4_tP12_125_a_b_m-001     575A

● PuGa$_6$:
A6B_tP14_125_gm_c-001     PCUV

● SrFe$_2$S$_4$:
AB2C4D_tP16_125_a_cd_m_b-001     DV64

● Sm$_2$NiGa$_{12}$:
A12BC2_tP30_125_2g2m_c_h-001     UYGN

**P4/nnc (126):**

○ BiAl$_2$S$_4$:
A2BC4_tP28_126_cd_e_k-001     4WPC

○ Ag[Co(NH$_3$)$_2$(NO$_2$)$_4$] ($J1_9$):
ABC4D2E8_tP32_126_a_b_h_e_k-001     EW36

○ Vesuvianite (Ca$_{10}$Al$_4$(Mg,Fe)$_2$Si$_9$O$_{34}$(OH)$_4$, $S2_3$):
A4B10C2D34E4F9_tP252_126_k_ce2k_f_h8k_
k_d2k-001     QSJ3

**P4/mbm (127):**

○ Si$_2$U$_3$ ($D5_a$):
A2B3_tP10_127_g_ah-001     6QGS

● Mo$_2$FeB$_2$:
A2BC2_tP10_127_g_a_h-002     5SWM

● RbAlF$_4$ II:
AB4C_tP12_127_a_eg_c-001     QJN8

● Mn$_2$Hg$_5$:
A5B2_tP14_127_cj_g-001     BAHM

○ Pd(NH$_3$)$_4$Cl$_2$·H$_2$O ($H4_9$):
A2BC2D_tP16_127_g_c_j_b-001     Y171

○ ThB$_4$ ($D1_e$):
A4B_tP20_127_ehj_g-001     GQBH

○ Phosgenite [Pb$_2$Cl$_2$(CO$_3$)]:
A2C3D2_tP32_127_g_eh_gk_k-001     URTV

● $\kappa$-AlF$_3$:
AB3_tP40_127_di_cg2ij-001     46QF

● Tetragonal Potassium Bronze (K$_3$W$_5$O$_{15}$):
A3B15C5_tP46_127_bh_cg2ij-001     LCR4

**P4/mnc (128):**

○ Phase II K$_2$SnCl$_6$:
A6B2C_tP18_128_eh_d_a-001     AD9M

○ FeCu$_2$Al$_7$ ($E9_a$):
A7B2C_tP40_128_egi_h_e-001     YX16

● U$_2$Mn$_3$Si$_5$:
A3B5C2_tP40_128_dh_egh_h-001     P4UA

○ Chiolite (Na$_5$Al$_3$F$_{14}$, $K7_5$):
A3B14C5_tP44_128_ac_ehi_bg-001     GFQ6

○ Apophyllite (KCa$_4$Si$_8$O$_{20}$F·8H$_2$O, $S5_2$):
A4BC16D28F8_tP116_128_h_a_2i_b_g3i_i-001     DPWF

**P4/mmm (129):**

○ Litharge (tetragonal PbO, $B10$):
AB_tP4_129_c_a-001     CWTN

○ $\gamma$-CuTi ($B11$):
AB_tP4_129_c_c-001     EKF2

○ $\beta$-Np ($A_d$):
A_tP4_129_ac-001     UNPK

○ Cu$_2$Sb ($C38$):
A2B_tP6_129_ac_c-001     YG4M

○ Matlockite ($E0_1$, PbFCl):
ABC_tP6_129_c_a_c-001     OVRY

● LaOF:
ABC_tP6_129_a_c_b-001     JX34

○ AsCuSiZr:
ABCD_tP8_129_c_b_a_c-001     1T51

○ LaOAgS:
ABCD_tP8_129_b_c_a_c-001     VCTW

● UCoC$_2$:
A2BC_tP8_129_2c_a_c-002     K32Y

● HfCuSi$_2$:
ABC2_tP8_129_a_c_bc-002     3Z42

○ CaBe$_2$Ge$_2$:
A2BC2_tP10_129_ac_c_bc-001     HOMM

● Ti$_2$Cu$_3$:
A3B2_tP10_129_3c_2c-001     T7XE

● SrPt$_3$P:
AB3C_tP10_129_c_ce_a-001     YXBN

○ NH$_4$Br ($B25$):
AB4C_tP12_129_c_i_a-001     G1GD

● CsFeF$_4$ II:
AB4C_tP24_129_bc_ijd_d-001     K69U

● Meta-autunite (I) [Ca(UO$_2$)$_2$(PO$_4$)$_2$·6H$_2$O, $H51_0$]:
AB4C6DE_tP26_129_c_j_2ci_a_c-001     VQBJ

● $\theta$-AlF$_3$:
AB3_tP64_129_2cdi_2cfhijk-001     HQ3Z

**P4/ncc (130):**

○ $\alpha$-WO$_3$:
A3B_tP16_130_cf_c-001     HPJW

○ CuBi$_2$O$_4$:
A2BC4_tP28_130_f_c_g-001     N4MV

○ Ba$_5$Si$_3$:
A5B3_tP32_130_cg_cf-001     OVYM

● Room Temperature Metastable VO$_2$:
A2B_tP48_130_2g_g-001     P5YW

○ Sr(OH)$_2$(H$_2$O)$_8$:
A18B10C_tP116_130_2c4g_2c2g_a-001     JDN7

**P4$_2$/mmc (131):**

○ Cooperite (PtS, $B17$):
AB_tP4_131_c_e-001     8GEP

● LaB$_2$C$_2$:
A2B2C_tP10_131_j_l_f-001     XLRT

**P4$_2$/mcm (132):**

○ AgUF$_6$:
AB6C_tP16_132_b_io_c-001     U3ZM

○ Rb$_2$TiCu$_2$S$_4$:
A2B2C4D_tP18_132_e_i_o_b-001     BD46

**P4$_2$/nbc (133):**

○ $\beta$-V$_3$S:
AB3_tP32_133_h_i2j-001     2PY4

● Zr$_3$PD$_3$:
A4BC3_tP64_133_2k_h_i2j-001     XBGP

**P4$_2$/nnm (134):**

○ T-50 B ($A_g$):
A_tP50_134_a2m2n-001     R7N6

**P4$_2$/mbc (135):**

○ Downeyite ($\alpha$-SeO$_2$, $C47$):
A2B_tP24_135_gh_h-001     YNQK

○ ZnSb$_2$O$_4$:
A4B2C_tP28_135_gh_h_d-001     2OMM

● Bi$_4$Fe$_5$O$_{13}$F:
A4B4C5D14_tP108_135_i_i_dfh_egh2i-001     55HD

**P4$_2$/mnm (136):**

- ○ $\gamma$-N:
  `A_tP4_136_f-001`  `VJ5B`
- ○ Rutile (TiO$_2$, $C4$):
  `A2B_tP6_136_f_a-001`  `D4ZH`
- ○ $\beta$-BeO:
  `AB_tP8_136_f_g-001`  `FJT4`
- ● $\alpha$-Li$_3$BN$_2$:
  `ABC2_tP12_136_a_bd_f-001`  `2PDY`
- ○ ZrFe$_2$Si$_2$:
  `A4B2C_tP14_136_i_f_a-001`  `U6YK`
- ● IrIn$_3$:
  `A3B_tP16_136_cj_f-001`  `YFFG`
- ● Ordoñezite (ZnSb$_2$O$_6$):
  `A6B2C_tP18_136_fj_e_a-001`  `8DHD`
- ○ $\alpha$'-V$_5$O:
  `AB8_tP18_136_a_2fi-001`  `WVZU`
- ○ Zr$_3$Al$_2$:
  `A2B3_tP20_136_j_dfg-001`  `QSNP`
- ● Na$_3$Hg$_2$:
  `A2B3_tP20_136_j_cfg-001`  `T3N4`
- ○ K$_2$CuCl$_4 \cdot$ 2H$_2$O ($H4_1$):
  `A4BC4D2E2_tP26_136_fg_a_j_d_e-001`  `40BA`
- ○ $\sigma$-CrFe ($D8_b$):
  `A4B4C4DE2_tP30_136_i_j_i_a_f-001`  `6EHC`
- ○ $\beta$-U ($A_b$):
  `A_tP30_136_af2ij-001`  `5T44`
- ● Nd$_2$Fe$_{14}$B:
  `AB14C2_tP68_136_f_ce2j2k_fg-001`  `FUR0`

**P4$_2$/nmc (137):**
- ○ ZrO$_2$ (High-temperature):
  `A2B_tP6_137_d_a-001`  `T9FR`
- ○ Coccinite (Red HgI$_2$, $C13$):
  `A2B_tP6_137_a_d-001`  `2F93`
- ● Low Temperature KBD$_4$:
  `A4BC4_tP12_137_a_g_b-001`  `Q06Z`
- ○ CeCo$_4$B$_4$:
  `A4BC4_tP18_137_g_a_g-001`  `EUHG`
- ○ Li$_6$CoO$_4$:
  `AB6C4_tP22_137_a_df_g-001`  `ZQRP`
- ● Orange (II) HgI$_2$:
  `AB2_tP24_137_g_cdf-001`  `L566`
- ○ Zn$_3$P$_2$ ($D5_9$):
  `A2B3_tP40_137_cdf_3g-001`  `68SX`

**P4$_2$/ncm (138):**
- ○ C (T12 Group IV):
  `A_tP12_138_bi-001`  `HZJ9`
- ○ Cl ($A18$) (*Obsolete*):
  `A_tP16_138_j-001`  `ZBJR`
- ● Tetragonal La$_2$NiO$_4$:
  `A2BC4_tP28_138_i_c_aei-001`  `J1M9`

**I4/mmm (139):**
- ○ In ($A6$):
  `A_tI2_139_a-001`  `F5BG`
- ○ Pa ($A_a$):
  `A_tI2_139_a-002`  `YF8N`
- ○ Hypothetical BCT5 Si:
  `A_tI4_139_e-001`  `ZSDQ`
- ○ "Martensite Type" FeC$_x$ ($x < 0.06$) ($L'2_0$):
  `AB_tI4_139_a_b-002`  `ZN4Y`
- ● $\beta$-LiFeO$_2$:
  `AB_tI4_139_a_b-003`  `G0K3`
- ○ ThH$_2$ ($L'2_b$):
  `A2B_tI6_139_d_a-001`  `R0C0`

- ○ MoSi$_2$ ($C11_b$):
  `AB2_tI6_139_a_e-001`  `NG7F`
- ○ CaC$_2$-I ($C11_a$):
  `A2B_tI6_139_e_a-001`  `RDQF`
- ○ Al$_3$Ti ($D0_{22}$):
  `A3B_tI8_139_ad_b-001`  `7W3W`
- ○ Hypothetical Tetrahedrally Bonded Carbon with 4-Member Rings:
  `A_tI8_139_h-001`  `TE0L`
- ○ Calomel (Hg$_2$Cl$_2$, $D3_1$):
  `AB_tI8_139_e_e-001`  `T1KN`
- ○ $D1_3$ (BaAl$_4$):
  `A4B_tI10_139_de_a-001`  `676C`
- ○ TlCo$_2$S$_2$:
  `A2B2C_tI10_139_d_e_a-002`  `D85X`
- ○ ThCr$_2$Si$_2$:
  `A2B2C_tI10_139_d_e_a-003`  `LS62`
- ○ Au$_2$Nb$_3$:
  `A2B3_tI10_139_e_ae-001`  `5STC`
- ○ Li$_2$CN$_2$:
  `A2B2C_tI10_139_a_d_e-002`  `5K3T`
- ○ $H5_9$ [Autunite Ca(UO$_2$)$_2$(PO$_4$)$_2 \cdot 10\frac{1}{2}$H$_2$O] (*Obsolete*):
  `A2B2C_tI10_139_a_d_e-003`  `SC3Z`
- ● Na$_2$HgO$_2$:
  `A2C2_tI10_139_a_e_e-001`  `6LGV`
- ● LuNi$_2$B$_2$C:
  `A2BCD2_tI12_139_e_a_b_d-001`  `4LX6`
- ○ 0201 [(La,Ba)$_2$CuO$_4$] High-$T_c$:
  `A2BC4_tI14_139_a_e_ce-001`  `V35A`
- ○ K$_2$NiF$_4$:
  `A4B2C_tI14_139_ce_e_a-001`  `SENW`
- ● Sr$_2$CuO$_2$Cl$_2$:
  `A2BC2D2_tI14_139_e_a_c_e-001`  `0HCF`
- ● Ti$_3$Cu$_4$:
  `A4B3_tI14_139_2e_ae-001`  `N44F`
- ● ('T') Nd$_2$CuO$_4$:
  `AB2C4_tI14_139_a_e_cd-003`  `1V8W`
- ○ Al$_3$Zr ($D0_{23}$):
  `A3B_tI16_139_cde-001`  `CSTH`
- ● BaZnSb$_2$:
  `AB2C_tI16_139_e_ce_d-003`  `00JM`
- ○ V$_4$Zn$_5$:
  `A4B5_tI18_139_i_ah-001`  `84H6`
- ● Pt$_5$Ti:
  `A8B_tI18_139_hi_a-001`  `XTZU`
- ○ K$_2$OsO$_2$Cl$_4$ ($J1_5$):
  `A4B2C2D_tI18_139_h_d_e_a-001`  `1KN8`
- ● Fe$_8$N ($D2_9$):
  `A8B_tI18_139_deh_a-001`  `6VR8`
- ● Sr$_2$Mn$_3$As$_2$O$_2$:
  `A2B3C2D2_tI18_139_e_ad_c_e-001`  `JBSJ`
- ● AuCsCl$_3$ ($K7_6$):
  `AB3C_tI20_139_ab_eh_d-001`  `QC4Q`
- ● CePt$_2$In$_7$:
  `AB7C2_tI20_139_a_bdg_e-001`  `P6JF`
- ● Ba$_2$Ti$_2$Fe$_2$As$_4$O:
  `A4B2C2DE2_tI22_139_2e_e_d_a_c-001`  `8C3Q`
- ● Sr$_3$Ti$_2$O$_7$:
  `A7B3C2_tI24_139_aeg_be_e-001`  `PDHZ`
- ● Mn$_{12}$Th ($D2_b$):
  `A12B_tI26_139_fij_a-001`  `AJGV`
- ● KCa$_2$Fe$_4$As$_4$F$_2$ (12442-type superconductor):
  `A4B2C2D4E_tI26_139_2e_e_d_g_a-001`  `YU5Q`

- $Bi_2Sr_2CaCu_2O_8$ (BSCCO):
  A2BC2D8E2_tI30_139_e_a_e_2eg_e-001    BZRA
- $Nd_4Ni_3O_8$:
  A4B3C8_tI30_139_2e_ae_cdg-002    W8GW
- $Sr_4Ti_3O_{10}$:
  A10B4C3_tI34_139_c2eg_2e_ae-001    TJ17
- High Temperature $BaBi_4Ti_4O_{15}$ $m = 4$ Aurivillius:
  A4C15D4_tI48_139_a_2e_bd2e2g_2e-001    Y186
- $\gamma$-$Bi_4V_2O_{11}$:
  A5B9C2_tI64_139_em_d2n_h-001    ON6E
- $Ho_{11}Ge_{10}$:
  A10B11_tI84_139_dehim_eh2n-001    HN10
- $K_3TlCl_6 \cdot 2H_2O$ ($J3_1$):
  A6B2C3D_tI168_139_egikl2m_ejn_bh2n_acf-001    8AAX

**I4/mcm (140):**

- Khatyrkite ($Al_2Cu$, $C16$):
  A2B_tI12_140_h_a-001    PARL
- $\alpha$-$SiU_3$ ($D0_c$):
  AB3_tI16_140_b_ah-001    6CWC
- $\beta$-$SeTl$ ($B37$):
  AB_tI16_140_ab_h-001    BFE0
- $KHF_2$ ($F5_2$):
  A2BC_tI16_140_h_d_a-001    B7FR
- $PdGa_5$:
  A5B_tI24_140_cl_a-001    528Z
- $V_4SiSb_2$:
  A2BC4_tI28_140_h_a_k-001    VENL
- $U_6Mn$ ($D2_c$):
  AB6_tI28_140_a_hk-001    1Q6T
- $W_5Si_3$ ($D8_m$):
  A3B5_tI32_140_ah_bk-001    4K4U
- $Cr_5B_3$ ($D8_l$):
  A3B5_tI32_140_ah_cl-001    AK18
- $(NH_4)Pb_2Br_5$ ($K3_4$) (*Erroneous*):
  A5BC2_tI32_140_bl_a_h-001    HL9W
- $(NH_4)Pb_2Br_5$ ($K3_4$) (*Revised*):
  A5BC2_tI32_140_cl_a_h-001    REV7
- $Cs_3CoCl_5$ ($K3_1$):
  A5BC3_tI36_140_cl_b_ah-001    GHEH
- $BaLa_2ZnO_5$:
  A2BC5D_tI36_140_a_h_cl_b-001    4VUW
- $Pu_{31}Rh_{20}$:
  A31B20_tI204_140_b2gh3m_ac2fh3l-001    TBHL

**I4₁/amd (141):**

- $\beta$-$Sn$ ($A5$):
  A_tI4_141_a-001    2BUF
- $NbP$ ("40"):
  AB_tI8_141_a_b-001    R06D
- Anatase ($TiO_2$, $C5$):
  A2B_tI12_141_e_a-001    1TUW
- $\alpha$-$ThSi_2$ ($C_c$):
  A2B_tI12_141_e_a-002    NF8F
- $MoB$ ($B_g$):
  AB_tI16_141_e_e-001    B9A8
- $\gamma$-$LiFeO_2$:
  ABC2_tI16_141_a_b_e-003    XJE5
- $\beta$-$ThCl_4$:
  A4B_tI20_141_h_a-001    NG21
- $Ga_2Hf$:
  A2B_tI24_141_2e_e-001    DBUS
- Zircon ($ZrSiO_4$, $S1_1$):
  A4BC_tI24_141_h_a_b-001    0FQT
- $\beta$-$LiSn$:

---

- AB_tI24_141_ae_be-001    1GTY
- Haussmannite ($Mn_3O_4$):
  A3B4_tI28_141_ad_h-001    KP7A
- $CuCr_2O_4$:
  A2BC4_tI28_141_c_b_h-001    763J
- Paramelaconite ($Cu_4O_3$):
  A4B3_tI28_141_cd_ae-002    P57T
- $BaCd_{11}$:
  AB11_tI48_141_b_aci-001    EV03
- Orange (I) $HgI_2$:
  AB2_tI48_141_h_2eg-001    Q9V3
- $\beta$-$In_2S_3$:
  A2B3_tI80_141_ceh_3h-001    PJGC
- $V_{13}O_{16}$:
  A16B13_tI116_141_2hi_a2fh-001    NCQA

**I4₁/acd (142):**

- S-III:
  A_tI16_142_f-001    GC8R
- $\alpha$-$PdSn_2$:
  A2B_tI48_142_d_ef-001    FUUU
- $Sr_2IrO_4$:
  AB4C2_tI56_142_a_df_d-001    EHYC
- $R_1$ $Au_3Zn$:
  A3B_tI64_142_def_d-001    Q6FW
- $NaPb$:
  AB_tI64_142_ef_g-001    X6RS
- $PPrS_4$:
  ABC4_tI96_142_e_ab_2g-001    B2Y6
- $BaCo_2V_2O_8$:
  A2B8C2D_tI104_142_a_f_2g_e-001    BMKB
- $Cd_3As_2$:
  A2B3_tI160_142_deg_3g-001    1P80
- Tetragonal $Be_3P_2$:
  A3B2_tI160_142_3g_abcef-001    4A1E
- Analcime ($NaAlSi_2O_6 \cdot H_2O$, $S6_1$):
  A2B2C3D12E4_tI184_142_f_f_be_3g_g-001    PFBA
- $Er_5Rh_6Sn_{18}$:
  A5B6C18_tI232_142_bg_dg_e2f3g-001    21TS

**P3 (143):**

- $ScRh_6P_4$:
  A4B6C_hP11_143_ad_2d_b-001    HTHS
- Trigonal $MoS_2$:
  AB2_hP12_143_ad_bc2d-001    1G0H
- Simpsonite ($Ta_3Al_4O_{13}[OH]$):
  A4B14C3_hP21_143_ad_bc4d_d-001    TY99
- $P3$ $La_3BWO_9$:
  AB3C9D_hP28_143_2a_2d_6d_bc-001    OXN3
- Trigonal (h') $Al_5Mo$:
  A5B_hP60_143_7a7b6c10d_3a3b4c-001    TWK7

**P3₁ (144):**

- $ZnTe$ (high-pressure):
  AB_hP6_144_a_a-001    LX4R
- $IrGe_4$:
  A4B_hP15_144_4a_a-001    0VM9
- $RbNO_3$ (IV):
  AB3C_hP45_144_3a_9a_3a-001    UV9A

**P3₂ (145):**

- Sheldrickite ($NaCa_3[CO_3]_2F_3[H_2O]$):
  A2B3C3DE7_hP48_145_2a_3a_3a_a_7a-001    3V4Q

**R3 (146):**

- $\gamma$-$Ag_3SI$ (Low-Temperature):
  A3BC_hR5_146_b_a_a-001    YGSU
- $FePSe_3$:

ABC3_hR10_146_2a_2a_2b-001  EARG
- • $Li_2ZrTeO_6$:
  A2B6CD_hR10_146_2a_2b_a_a-001  T7PA
- • $Pd_7P_3$:
  A3B7_hR20_146_2b_2a4b-001  35AG
- Carlinite ($Tl_2S$):
  AB2_hR27_146_3a2b_6b-001  X2PG
- • Rhombohedral $LiZnPO_4$:
  A4CD_hR42_146_2b_8b_2b_2b-001  BGS4

## P3 (147):
- ○ $\gamma$-AgZn ($B_b$):
  A2B_hP9_147_g_ad-001  RQCB
- • $PtBi_2$:
  A2B_hP9_147_g_ad-002  0RUP
- ○ $Na_2SO_3$ ($G3_2$):
  A2B3C_hP12_147_abd_g_d-001  H9Z3
- • $Na_2BaNi(PO_4)_2$:
  A2BD2C8E2_hP14_147_a_d_b_dg_d-001  B64B
- • $URu_3B_2$:
  A2B3C_hP48_147_2d2g_4g_abef-001  B1DH

## R3 (148):
- • $S_6$ ($\rho$-S) Sulfur:
  A_hR6_148_f-001  XW0B
- ○ $BiI_3$ ($D0_5$):
  AB3_hR8_148_c_f-001  TZC0
- • $\beta$-$CuZrF_6$:
  A6C_hR8_148_a_f_b-001  7RW9
- ○ Ilmenite ($FeTiO_3$, $E2_2$):
  AB3C_hR10_148_c_f_c-001  4YUA
- ○ Dolomite [$MgCa(CO_3)_2$, $G1_1$]:
  A2BCD6_hR10_148_c_a_b_f-001  WCJU
- • Rhombohedral $Cr_2S_3$:
  A2B3_hR10_148_abc_f-001  CVKE
- • $FePSe_3$:
  ABC3_hR10_148_c_c_f-001  CFP9
- • $BaNi_2As_2O_8s$:
  A2B2C2D8_hR13_148_c_a_c_cf-001  VJPX
- ○ $Ni(H_2O)_6SnCl_6$ ($I6_1$):
  A6B6CD_hR14_148_f_f_a_b-001  5954
- ○ $K_2Sn(OH)_6$ ($H6_2$):
  A6B2C6D_hR15_148_f_c_f_a-001  FKGA
- • $Li_7TaO_6$:
  A8B6C_hR15_148_cf_f_a-001  L1GH
- • $BaMo_6S_8$:
  A6C8_hR15_148_a_f_cf-001  CY1Y
- ○ Solid Cubane ($C_8H_8$):
  AB_hR16_148_cf_cf-001  8Z24
- • $Pd_{15}P_2$:
  A2B15_hR17_148_c_ac2f-001  Z055
- • $\beta$-$PdCl_2$:
  A2B_hR18_148_2f_f-001  8GMP
- • $Mg_3TeO_6$:
  A3B6C_hR20_148_f_2f_ab-001  K70X
- • $SrMn_7O_{12}$:
  A7B12C_hR20_148_ade_2f_b-001  5YFD
- • Trechmannite ($AgAsS_2$):
  ABC2_hR24_148_f_f_2f-001  3VWQ
- ○ PdAl:
  AB_hR26_148_a2f_b2f-001  5EQ4
- • Mikasaite (Rhombohedral $Fe_2(SO_4)_3$):
  A2B12C3_hR34_148_2c_4f_f-001  GHHO
- Phenakite ($Be_2SiO_4$, $S1_3$):
  A2B4C_hR42_148_2f_4f_f-001  BQNX

- • Dioptase [$Cu_6(Si_6O_{18})$·$6H_2O$]:
  AB2C4D_hR48_148_f_2f_4f_f-001  4C7V
- • $Mg_{23}Al_{30}$:
  A30B23_hR53_148_5f_a2c3f-001  DB0A
- • $Y_3Cu_9(OH)_{19}Cl_8$:
  A8B9C24D19E3_hR63_148_cf_df_4f_a3f_bc-001  GGPS

## P312 (149):
- • $PbMnTeO_6$:
  AB6CD_hP9_149_a_l_d_e-001  63FP
- ○ $Ti_3O$ (Room-Temperature):
  AB3_hP24_149_acgi_3l-001  9DMX

## P321 (150):
- ○ Original $Fe_2P$ ($C2_2$):
  A2B_hP9_150_ef_ad-001  6CGG
- • $\beta$-$Na_2ThF_6$:
  A6B2C_hP9_150_ef_d_a-001  R3Z7
- ○ Steklite [$KAl(SO_4)_2$, $H3_2$]:
  ABC8D2_hP12_150_a_b_dg_d-001  SDL2
- ○ $Cs_3As_2Cl_9$ ($K7_3$):
  A2B9C3_hP14_150_d_eg_ad-001  4FK1
- • $SrCl_2$·$(H_2O)_6$:
  A2B12C6D_hP21_150_d_2g_ef_a-001  FX47
- • Dugganite ($Pb_3Zn_3TeAs_2O_{14}$):
  A2B14C3DE3_hP23_150_d_d2g_e_a_f-001  8A06
- Paralstonite ($BaCa(CO_3)_2$):
  A2C2D6_hP30_150_e_c2d_f_3g-001  FL2F
- $KSO_3$ ($K1_1$):
  A3BC_hP30_150_ef_3g_c2d-001  YVPA

## P3₁12 (151):
- ○ $CrCl_3$ ($D0_4$):
  A3B_hP24_151_3c_2a-001  5EPG
- • $V_6C_5$:
  A5B6_hP33_151_3a2b_3c-001  V1KT

## P3₁21 (152):
- ○ $\gamma$-Se ($A8$):
  A_hP3_152_a-001  SM1V
- ○ $\alpha$-Quartz (low Quartz):
  A2B_hP9_152_c_a-001  P53L
- • Trigonal $B_2O_3$:
  A2B3_hP15_152_c_ac-001  MKFS
- • Alarsite ($AlAsO_4$):
  ABC4_hP18_152_a_b_2c-001  P55Q
- • $Ni_3Ga_3Ge_6$:
  A3B6C13_hP66_152_ac_3c_a2b5c-001  A5TG

## P3₂12 (153):
- ○ $CrCl_3$:
  A3B_hP24_153_3c_2a-001  Z0S9

## P3₂21 (154):
- ○ Cinnabar ($HgS$, $B9$):
  AB_hP6_154_a_b-001  A3TK
- ○ S-II:
  A_hP9_154_ac-001  GAZQ
- • $EuIr_2P_2$:
  AB2C2_hP15_154_a_ab_c-001  3CT8
- • $RbAg_2SbS_4$:
  A2BC4D_hP24_154_c_a_2c_b-001  8R3H

## R32 (155):
- ○ Hazelwoodite ($Ni_3S_2$, $D5_e$):
  A3B2_hR5_155_e_c-001  WQGS
- ○ $AlF_3$ ($D0_{14}$):
  AB3_hR8_155_c_de-001  QG2U
- ○ $KBe_2BO_3F_2$:
  A2C2DE3_hR9_155_a_c_c_b_d-001  Q3TB

- Huntite [CaMg$_3$(CO$_3$)$_4$]:
  A4BC3D12_hR20_155_ad_b_e_2df-001    **1B5F**
- ErNi$_3$Al$_9$:
  A9BC3_hR26_155_3cdef_c_f-001    **9G5F**
- Low Temperature GdBO$_3$:
  ABC3_hR30_155_de_f_de2f-001    **5YKD**

**P3m1 (156):**
- $\beta$-CuI (Sakuma):
  AB_hP4_156_ab_ab-001    **1M22**
- CdI$_2$ (Polytype 6H$_1$):
  AB2_hP9_156_2ab_a2b3c-001    **0FK1**
- $\beta$-CuI (Kurdyumova):
  AB_hP12_156_3a2bc_3a2bc-001    **ZMXD**
- KNaSO$_4$:
  ABC4D_hP14_156_ac_bc_ab2d_ab-001    **Z1Y1**

**P31m (157):**
- Ag$_5$Pb$_2$O$_6$:
  A5B6C2_hP13_157_2ac_2c_b-001    **1YM0**

**P3c1 (158):**
- $\beta$-RuCl$_3$:
  A3B_hP8_158_d_a-001    **KC70**

**P31c (159):**
- HP-Bi$_2$O$_3$:
  A2B3_hP20_159_bc_2c-001    **XT2J**
- YbBaCo$_4$O$_7$:
  A4C7D_hP26_159_b_ac_a2c_b-001    **ZUBF**
- Nierite ($\alpha$-Si$_3$N$_4$):
  A4B3_hP28_159_ab2c_2c-001    **9UA2**

**R3m (160):**
- GeTe:
  AB_hR2_160_a_a-002    **8379**
- Carbonyl Sulphide (COS, F0$_2$):
  ABC_hR3_160_a_a_a-001    **V77F**
- Rhombohedral MoS$_2$:
  AB2_hR3_160_a_2a-001    **EW06**
- H$_3$S (130 GPa):
  A3B_hR4_160_b_a-001    **P2LA**
- Grimaldiite ($\alpha$-CrOOH):
  ABC2_hR4_160_a_a_2a-002    **LP35**
- CrCuS$_2$:
  ABC2_hR4_160_a_a_2a-003    **BC9T**
- $\gamma$-GaSe:
  AB_hR4_160_2a_2a-001    **M3V5**
- KBrO$_3$ (G0$_7$):
  ABC3_hR5_160_a_a_b-001    **0ZSQ**
- $\gamma$-Potassium Nitrate (KNO$_3$):
  ABC3_hR5_160_a_a_b-001    **V4KY**
- Rhombohedral BaTiO$_3$:
  ABC3_hR5_160_a_b_a-001    **X85U**
- Moissanite 9R (CSi):
  AB_hR6_160_3a_3a-001    **T8Y7**
- Millerite (NiS, B13):
  AB_hR6_160_b_b-001    **MV4U**
- Cronstedtite Fe(Fe,Si)[(OH)$_2$O]O$_3$OH, S5$_7$:
  AB3C2D_hR7_160_a_b_2a_a-001    **H9NP**
- Sn$_4$As$_3$:
  A3B4_hR7_160_3a_4a-001    **RN9C**
- Fe$_{5-\delta}$GeTe$_2$:
  A5BC2_hR8_160_5a_a_2a-001    **6V4S**
- Moissanite-15R (SiC, B7):
  AB_hR10_160_5a_5a-001    **KX5G**
- Fe$_3$PO$_7$:
  A3B7C_hR11_160_b_a2b_a-001    **JN4A**

- Low Temperature GaMo$_4$S$_8$:
  AB4C8_hR13_160_a_ab_2a2b-001    **5V5H**
- $\gamma$-Na$_2$Ti$_3$Cl$_8$:
  A8B2C3_hR13_160_2a2b_2a_b-001    **L333**
- Cr$_5$Al$_8$ (D8$_{10}$):
  A8B5_hR26_160_a3bc_a3b-001    **X9QA**
- SbI$_3$S$_{24}$:
  A3B24C_hR28_160_b_2b3c_a-001    **8TMQ**

**R3c (161):**
- Ferroelectric LiNbO$_3$:
  ABC3_hR10_161_a_a_b-001    **FZF0**
- Proustite (Ag$_3$AsS$_3$):
  A3BC3_hR14_161_b_a_b-001    **TXGA**
- $\alpha$-BaB$_2$O$_4$ (Low Temperature):
  A2BC4_hR42_161_2b_b_4b-001    **UFMT**
- Stepanovite (NaMgFe(C$_2$O$_4$)$_3$·9H$_2$O):
  A6BC18DEF21_hR96_161_2b_a_6b_a_a_7b-001    **KKZD**

**P31m (162):**
- $\beta$-V$_2$N (L'3$_2$):
  AB2_hP9_162_ad_k-001    **RBZR**
- I1$_3$ [SrCl$_2$·(H$_2$O)$_6$] (*Obsolete*):
  A2B6C_hP9_162_c_k_b-001    **WZ1P**
- Rosiaite (PbSb$_2$O$_6$):
  A6BC2_hP9_162_k_a_d-001    **8OKE**
- Trigonal Au$_7$P$_{10}$I:
  A7BC10_hP18_162_ceg_a_hk-001    **RQ68**

**P31c (163):**
- NaSbF$_4$(OH)$_2$ (J1$_{12}$):
  A6BC_hP16_163_i_b_c-001    **KLNT**
- Colquiriite (LiCaAlF$_6$):
  ABC6D_hP18_163_c_b_i_d-001    **8D1V**
- $\beta$-Na$_2$GeTeO$_6$:
  A2C6D_hP20_163_c_f_i_a-001    **UZBW**
- Mn$_3$Si$_2$Te$_6$:
  A3B2C6_hP22_163_cf_e_i-001    **ELB0**
- Trigonal Cr$_5$S$_6$:
  A5B6_hP22_163_abcf_i-001    **Z6CB**
- KAg(CN)$_2$ (F5$_{10}$):
  A2CD2_hP36_163_h_i_bf_i-001    **JLGL**
- Si$_2$Te$_3$:
  A7B3_hP40_163_e2i_i-001    **MVBJ**

**P3m1 (164):**
- $\omega$ Phase (CdI$_2$, C6):
  AB2_hP3_164_a_d-001    **32EZ**
- Zlatogorite (CuNiSb$_2$):
  ABC2_hP4_164_a_b_d-001    **PUFU**
- D0$_{13}$ (AlCl$_3$) (*Obsolete*):
  AB3_hP4_164_a_bd-001    **K2MJ**
- Al$_3$Ni$_2$ (D5$_{13}$):
  A3B2_hP5_164_ad_d-001    **EUFT**
- La$_2$O$_3$ (D5$_2$):
  A2B3_hP5_164_d_ad-001    **D3QE**
- Ce$_2$O$_2$S:
  A2B2C_hP5_164_d_d_a-002    **FG79**
- Brucite [Mg(OH)$_2$]:
  A2BC2_hP5_164_d_a_d-001    **N2MR**
- $\beta$-CuI (Keen-Hull):
  A2B_hP6_164_2d_d-001    **TVVQ**
- Nukundamite [(Cu$_{1-x}$Fe$_x$)$_4$S$_4$]:
  ABC2_hP8_164_d_a_cd-001    **CG4B**
- $\delta_H^{II}$-NW$_2$:
  A2B_hP9_164_ad_c2d-001    **Z07R**
- K$_2$Pt(SCN)$_6$ (H6$_3$):

A2BC6_hP9_164_d_a_i-001    F3SY

○ Bararite [Trigonal $(NH_4)_2SiF_6$, $J1_6$]:
A6B2C_hP9_164_i_d_a-001    4BR9

○ $K_2GeF_6$ ($J1_{13}$):
A6BC2_hP9_164_i_a_d-001    2HX1

● $Li_7Pb_2$:
A7B2_hP9_164_ac2d_d-001    FCSA

● $K_2Pb_7$:
A7B2_hP9_164_ai_d-001    QCMW

○ Jacutingaite ($Pt_2HgSe_3$):
AB2C3_hP12_164_d_a_ei-001    DCW7

○ Nevskite (BiSe):
AB_hP12_164_c2d_c2d-001    UEMM

● Kapellasite ($Cu_3Zn(OH)_6Cl_2$):
A2B3C6D_hP12_164_d_e_i_a-001    K216

● $MnBi_4Te_7$:
A4BC7_hP12_164_2d_a_bc2d-001    SPWJ

● $AgBiSe_2$:
ABC2_hP12_164_ad_bd_c2d-001    1TV2

● Trigonal $\alpha$-$Ca_2SiO_4$:
A2B4C_hP14_164_abd_di_d-001    DBUP

● Aphthitalite/Glaserite [$K_3Na(SO_4)_2$]:
A3BC8D2_hP14_164_ad_b_di_d-001    9RQB

● $Ba_3SrTa_2O_9$:
A3B9C2D_hP15_164_ad_ei_b_d-001    4LGS

● $Li_{13}Sn_5$:
A13B5_hP18_164_a2c4d_b2d-001    G8A6

○ Predicted $Li_2MgH_{16}$ 300 GPa:
A16B2C_hP19_164_2d2i_d_a-001    LLDQ

● $YCu_3(OH)_6Cl_3$:
ABC2D2E_hP21_164_bd_e_i_ac-001    1PRT

**P3c1 (165):**

○ $H_3Ho$:
A3B_hP24_165_adg_f-001    V86P

○ $Cu_3P$ ($D0_{21}$):
A3B_hP24_165_bdg_f-001    Z2X0

● $Nb_2Mn_4O_9$:
A4B2C9_hP30_165_2d_c_fg-001    GPQH

● $Li_9Al_3(P_2O_7)_3(PO_4)_2$:
A3B9C29D8_hP98_165_f_bdg_df4g_dg-001    JRX6

**R3m (166):**

○ $\beta$-Po ($A_i$):
A_hR1_166_a-001    RUSN

○ $\alpha$-Hg (A10):
A_hR1_166_a-002    8ZQP

○ Rhombohedral CuPt ($L1_1$):
AB_hR2_166_a_b-001    3A9R

○ $\alpha$-As ($A7$):
A_hR2_166_c-001    F2V5

○ Rhombohedral Graphite:
A_hR2_166_c-002    NXUA

○ $\beta$-$O_2$:
A_hR2_166_c-003    KTH0

○ KSH ($B22$):
AB_hR2_166_a_b-003    HA1H

○ $\alpha$-Sm ($C19$):
A_hR3_166_ac-001    MPB6

● $CdCl_2$:
A2B_hR3_166_a_c-004    TKCJ

○ Caswellsilverite ($CrNaS_2$, $F5_1$):
ABC2_hR4_166_a_b_c-001    J6ND

○ Rhombohedral Delafossite ($CuFeO_2$):
ABC2_hR4_166_a_b_c-004    Y8C3

○ $\alpha$-$NaFeO_2$:
ABC2_hR4_166_a_b_c-009    5FG5

● $Bi_2Te_3$ ($C33$):
A2B3_hR5_166_c_ac-001    NGV5

○ SmSI:
ABC_hR6_166_c_c_c-001    YTD0

○ $CaSi_2$ ($C12$):
AB2_hR6_166_c_2c-002    85MW

○ $CaUO_4$:
AB4C_hR6_166_a_2c_b-001    MMVG

● YOF:
ABC_hR6_166_c_c_c-002    DGDK

● $Mo_2B_5$ ($D8_5$):
A5B2_hR7_166_a2c_c-001    WZKM

● $CaC_6$:
A6B_hR7_166_f_b-001    N45M

● $Al_4C_3$ ($D7_1$):
A4B3_hR7_166_2c_ac-001    YAHB

● $MnBi_2Te_4$:
A2BC4_hR7_166_c_a_2c-002    MU2V

● Shandite ($Ni_3Pb_2S_2$):
A3B2C2_hR7_166_d_ab_c-001    SD1B

● $CaCu_4P_2$:
AB4C2_hR7_166_a_2c_c-001    SFS1

● $YbFe_2O_4$:
A2B4C_hR7_166_c_2c_a-001    EVRK

● $In_3Se_4$:
A3B4_hR7_166_ac_2c-001    DJRQ

● $Bi_4Te_3$:
A4B3_hR7_166_2c_ac-002    MVZB

● $GeSb_2Te_4$:
A2C4_hR7_166_a_c_2c-002    5J2E

○ $\beta$-Potassium Nitrate ($KNO_3$):
ABC6_hR8_166_a_b_h-001    XD6C

● $SnP_3$:
A3B_hR8_166_h_c-001    JF25

● $Fe_3Sn_2$:
A3B2_hR10_166_h_2c-001    KQQW

● $Al_6C_3N_2$:
A6B3C2_hR11_166_3c_ac_c-001    T8T4

● $Li_8Pb_2$:
A8B3_hR11_166_4c_ac-001    758U

○ $\alpha$-B (R-12):
A_hR12_166_2h-001    1Y2H

● $ThB_2C$:
A2BC_hR12_166_g_d_ac-001    AFH9

● $NbBe_3$:
A3B_hR12_166_ach_bc-001    3QUZ

● $BaPb_3$:
AB3_hR12_166_ac_eh-001    NHXQ

● $Fe_7W_6$ ($D8_5$) $\mu$-phase:
A7B6_hR13_166_ah_3c-001    MXB3

● $Ba_3Cr_2O_8$:
A3B2C8_hR13_166_ac_c_ch-002    7UDV

● $Na_2Mn_3Cl_8$:
A8B3C2_hR13_166_ch_e_c-001    799E

● $B_{13}C_2$ "$B_4C$" ($D1_g$):
A13B2_hR15_166_a2h_c-001    PZDF

● $Al_8C_3N_4$:
A8B3C4_hR15_166_4c_ac_2c-001    HGYT

● $TaTi_7$ (BCC SQS-16):
AB7_hR16_166_c_c2h-001    56NK

● $MnBi_6Te_{10}$:
A6BC10_hR17_166_3c_a_5c-001    LD13

- Er$_2$Co$_7$:
  A7B2_hR18_166_a2cdh_2c-001     VGKE
- Th$_2$Zn$_{17}$:
  A2B17_hR19_166_c_cdfh-001     PZ2T
- Nb$_2$Be$_{17}$:
  A17B2_hR19_166_cegh_c-001     3JAE
- Ba$_4$NbRu$_3$O$_{12}$:
  A4BC12D3_hR20_166_2c_a_2h_bc-001     9CFG
- High Temperature K$_2$LiAlF$_6$:
  A6C2D_hR20_166_ab_2h_2c_c-001     KTF5
- Mn$_2$La$_3$Sb$_3$O$_{14}$:
  A3B2C14D3_hR22_166_d_ab_c2h_e-001     6NJM
- Rhombohedral CuTi$_2$S$_4$:
  A4C2_hR28_166_2c_2c2h_abh-001     8BYU
- Chabazite (Ca$_{1.4}$Sr$_{0.3}$Al$_{3.8}$Si$_{8.3}$O$_{24}$·13H$_2$O, $S3_4$ (I)):
  A5B21C24D12_hR62_166_a2c_ehi_fg2h_i-001     2K92
- $\beta$-B (R-105):
  A_hR105_166_ac9h4i-001     Y6TF

**R3c (167):**
- FeF$_3$ ($D0_{12}$):
  A3B_hR8_167_e_b-001     GPY3
- Corundum ($\alpha$-alumina, Al$_2$O$_3$, $D5_1$):
  A2B3_hR10_167_c_e-001     CBQY
- Paraelectric LiNbO$_3$:
  ABC3_hR10_167_a_b_e-001     TXPE
- Calcite (CaCO$_3$, $G0_1$):
  ABC3_hR10_167_a_b_e-002     UVUF
- PrNiO$_3$:
  AB3C_hR10_167_b_e_a-001     XMCY
- Rhombohedral Al$_5$Mo:
  A5B_hR12_167_ce_b-001     584Z
- CrCl$_3$(H$_2$O)$_6$ ($J2_2$):
  A3BC6_hR20_167_e_b_f-001     YFN8
- AgRuO$_3$:
  AB3C_hR20_167_c_f_c-001     QQE1
- Rinneite (K$_3$NaFeCl$_6$):
  A6BC3D_hR22_167_f_b_e_a-001     PYAM
- ScRh$_3$Si$_7$:
  A3BC7_hR22_167_e_b_af-001     NS5K
- KBO$_2$ ($F5_{13}$):
  ABC2_hR24_167_e_e_2e-001     L33F
- Cs$_3$Tl$_2$Cl$_9$ ($K7_2$):
  A9B3C2_hR28_167_ef_e_c-001     7M03
- $\beta$-BaB$_2$O$_4$ (High Temperature):
  A2BC4_hR42_167_f_ac_2f-001     RYPV
- MoP$_3$SiO$_{11}$ (Rhombohedral Model):
  AB11C3D_hR64_167_c_be3f_f_c-001     A76C
- Zr$_7$Re$_{25}$:
  A25B21_hR92_167_b2e3f_e3f-001     1BVC

**P6 (168):**
- K$_2$Ta$_4$O$_9$F$_4$:
  A2B13C4_hP57_168_d_c6d_2d-001     6140
- Al[PO$_4$]:
  ABC4_hP72_168_2d_8d_2d-001     5X6W

**P6$_1$ (169):**
- Al$_2$S$_3$:
  A2B3_hP30_169_2a_3a-001     0WYL

**P6$_5$ (170):**
- Al$_2$S$_3$:
  A2B3_hP30_170_2a_3a-001     G6GN

**P6$_2$ (171):**
- Sr[S$_2$O$_6$][H$_2$O]$_4$:
  A10B2C_hP39_171_5c_c_a-001     WCPM

**P6$_4$ (172):**
- Sr[S$_2$O$_6$][H$_2$O]$_4$:
  A10B2C_hP39_172_5c_c_a-001     XXQ8

**P6$_3$ (173):**
- PI$_3$:
  A3B_hP8_173_c_b-001     RB2M
- $\alpha$-LiIO$_3$:
  ABC3_hP10_173_b_a_c-001     DF6Y
- $\beta$-Si$_3$N$_4$:
  A4B3_hP14_173_bc_c-001     Y4EE
- LiKSO$_4$ ($H1_4$):
  ABC4D_hP14_173_a_b_bc_b-001     F86C
- La$_3$CuSiS$_7$:
  AB3C7D_hP24_173_a_c_b2c_b-001     H1TU
- P6$_3$ La$_3$BWO$_9$:
  AB3C9D_hP28_173_a_c_3c_b-001     DVKH
- Crancrinite (Na$_6$Ca$_2$Al$_6$Si$_6$O$_{24}$(CO$_3$)$_2$, $S3_3$(I)):
  A3BCD3E15F3_hP52_173_c_b_b_c_5c_c-001     7U5Z

**P6 (174):**
- GdSI:
  ABC_hP12_174_aj_dk_ej-001     4BHM
- Fe$_{12}$Zr$_2$P$_7$:
  A12B7C2_hP21_174_2j2k_ajk_cf-001     EE97
- Tl$_3$Ga$_9$S$_{13}$O$_2$:
  A9B2C13D3_hP27_174_jl_i_ajkl_bcd-001     Q3WS
- Stützite (Ag$_{5-x}$Te$_3$):
  A12B7_hP57_174_2j2k4l_ghi3j2k-001     T2R2

**P6/m (175):**
- Nb$_7$Ru$_6$B$_8$:
  A8B7C6_hP21_175_ck_aj_k-001     F86A
- Mg[NH]:
  ABC_hP36_175_jk_jk_jk-001     12DH
- Na$_4$Ge$_{13}$:
  A19B10_hP58_175_e2j2kl_djl-001     XPLN
- Ag$_{51}$Gd$_{14}$:
  A27B7_hP68_175_chjk3l_ejk-001     V9AH

**P6$_3$/m (176):**
- UCl$_3$:
  A3B_hP8_176_h_c-001     CZNF
- Er$_3$Ru$_2$:
  A3B2_hP10_176_h_bc-001     1QZH
- LaRu$_3$Si$_2$:
  AB3C2_hP12_176_b_h_f-001     JSY8
- TlFe$_3$Te$_3$:
  A3B3C_hP14_176_h_h_c-001     BQR4
- $\beta$-Si$_3$N$_4$:
  A4B3_hP14_176_ch_h-001     YEJK
- Nb$_3$Te$_4$:
  A3B4_hP14_176_h_ch-002     FUM3
- Th$_7$S$_{12}$ (D8$_k$):
  A3B2_hP20_176_2h_ah-001     X9DF
- $\zeta$-Cu$_{10}$Sn$_3$:
  A10B3_hP26_176_bcfi_h-001     YGNT
- Cu$_{10}$Sb$_3$:
  A10B3_hP26_176_c3h_h-001     4QXQ
- K$_3$W$_2$Cl$_9$ ($K7_1$):
  A9B3C2_hP28_176_hi_af_f-001     XV4Y
- Rh$_{20}$Si$_{13}$:
  A10B7_hP34_176_c3h_b2h-001     L6DT
- Fe$_2$(CO)$_9$ ($F4_1$):
  A9B2C9_hP40_176_hi_f_hi-001     858V
- Fluorapatite [Ca$_5$F(PO$_4$)$_3$, $H5_7$]:
  A5BC12D3_hP42_176_fh_a_2hi_h-001     KHCK

- Lead Apatite [Pb$_{10}$(PO$_4$)$_6$O]:
  A14B3C5_hP44_176_e2hi_h_fh-001          4302
- YBa$_3$B$_9$O$_{18}$:
  A9B3C18D_hP62_176_hi_af_2h2i_b-001      PC4V
- Na$_5$Co$_{15.5}$Te$_6$O$_{36}$ (NCTO):
  A7B5C18D3_hP66_176_ci_bef_2h2i_h-001    KL15
- Nb$_7$Rh$_6$B$_8$:
  A8B7C6_hP126_176_2h3i_acd6h_3i-001      01SA

**P622 (177):**
○ Hypothetical Hexagonal SiO$_2$:
  A2B_hP36_177_j2lm_n-001                 XNAH
● [Fe(OMe)$_2$(proline)]$_{12}$[ClO$_4$]$_{12}$:
  A7BCD15EF12_hP444_177_7n_n_jl_15n_
  n_12n-001                               TXX6

**P6$_1$22 (178):**
○ Sc-V (High-Pressure):
  A_hP6_178_a-001                         Z993
○ AuF$_3$:
  AB3_hP24_178_b_ac-001                   UBNE
● CsCuCl$_3$:
  A3BC_hP30_178_bc_b_a-001                8791
● Zr$_5$Ir$_3$:
  A3B5_hP48_178_ac_a2bc-001               15N7

**P6$_5$22 (179):**
● AuF$_3$:
  AB3_hP24_179_b_ac-001                   45S0
● Sc(H$_2$O)$_2$[BP$_2$O$_8$]·H$_2$O:
  AB4C12D2E_hP120_179_b_2c_6c_c_b-001     1CPH

**P6$_2$22 (180):**
○ β-SiO$_2$ (C8):
  A2B_hP9_180_i_d-001                     MK1H
○ CrSi$_2$ (C40):
  AB2_hP9_180_c_i-001                     3K6M
○ Mg$_2$Ni (C$_a$):
  A2B_hP18_180_fj_ac-001                  TDY8
○ Hg$_2$O$_2$NaI:
  A2BCD2_hP18_180_f_c_b_i-001             TQ0B
● Rhadophane (CePO$_4$):
  AB4C_hP18_180_c_k_d-001                 QL18
● CsC$_8$:
  A8B_hP27_180_2ik_d-001                  T338

**P6$_4$22 (181):**
○ β-SiO$_2$ (C8):
  A2B_hP9_181_i_d-001                     VH37
● Co[Au(CN)$_2$]$_2$:
  A2B4CD4_hP66_181_k_2k_f_2k-001          R3D2
● β-Eucryptite (LiAlSiO$_4$):
  ABC4D_hP84_181_gi_bcf_4k_hj-001         N1MQ

**P6$_3$22 (182):**
○ Bainite (Fe$_3$C):
  AB3_hP8_182_c_g-001                     JRHU
○ E2$_3$ (LiIO$_3$) (Obsolete):
  ABC3_hP10_182_c_b_g-001                 6B26
● WAl$_5$:
  A5B_hP12_182_bcg_d-001                  BQ6N
○ BaAl$_2$O$_4$ (H2$_8$):
  A2BC6_hP18_182_f_b_gh-001               6VW9
● CoNb$_3$S$_6$:
  AB3C6_hP20_182_c_af_i-001               A0VJ
● Na$_2$Co$_2$TeO$_6$:
  A2B13C6D_hP44_182_bc_a2i_i_d-001        1XST

**P6mm (183):**
○ AuCN:

- ABC_hP3_183_a_a_a-001                   Q8EL
○ CrFe$_3$NiSn$_5$:
  AB_hP6_183_c_ab-001                     C57U
● Ta$_{21}$Te$_{13}$:
  A21B13_hP136_183_abc3d6e2f_2ab3d5e-001  7BEE

**P6cc (184):**
○ Al[PO$_4$] (Framework type AFI):
  AB4C_hP72_184_d_4d_d-001                72JS
○ NaTi$_2$(PS$_4$)$_3$:
  A13B6C24D4_hP188_184_2a4d_2d_8d_bd-001  7PRY

**P6$_3$cm (185):**
○ β-RuCl$_3$:
  A3B_hP8_185_c_a-001                     5RVU
○ Cu$_3$P:
  A3B_hP24_185_ab2c_c-001                 L8U5
○ Na$_3$As:
  AB3_hP24_185_c_ab2c-001                 3ERL
○ KNiCl$_3$:
  A3BC_hP30_185_cd_c_ab-001               PKFR
● LuMnO$_3$:
  ABC3_hP30_185_ab_c_ab2c-001             YGS5
● Stibiopalladinite (Pd$_5$Sb$_2$):
  A5B2_hP42_185_ab4c_abc-001              O4LT
● Sr$_8$Os$_{6.3}$O$_{24}$:
  A36B11C12_hP118_185_4c4d_a2b2c_ab3c-001 AL94

**P6$_3$mc (186):**
○ Original BN (B12) (Obsolete):
  AB_hP4_186_b_a-001                      QUDC
○ Wurtzite (ZnS, B4):
  AB_hP4_186_b_b-001                      6CGK
○ Buckled Graphite:
  A_hP4_186_ab-001                        F6SX
○ C27 (CdI$_2$) (Questionable):
  AB2_hP6_186_b_ab-001                    G4EJ
● LiGaGe Crystal:
  ABC_hP6_186_b_b_a-003                   ZZ7U
○ Moissanite-4H SiC (B5):
  AB_hP8_186_ab_ab-001                    FSH2
○ Cd(OH)Cl (E0$_3$):
  ABCD_hP8_186_b_b_a_a-001                R9CA
● O2-LiCoO$_2$:
  ABC2_hP8_186_b_b_ab-001                 5BYD
○ Moissanite-6H SiC (B6):
  AB_hP12_186_a2b_a2b-001                 NYRK
● O4-LiCoO$_2$:
  ABC2_hP16_186_ab_ab_a3b-001             3DYH
● δ-GaSe:
  AB_hP16_186_2a2b_2a2b-001               W6UW
● Al$_5$C$_3$N (E9$_4$):
  A5B3C_hP18_186_2a3b_2ab_b-001           KHLK
● Fe$_3$Th$_7$ (D10$_2$):
  A3B7_hP20_186_c_b2c-001                 VGVG
● HPC-Bi$_2$O$_3$:
  A2B3_hP20_186_bc_2c-001                 14M3
● Ti$_3$Al$_2$N$_2$:
  A2B4C5_hP22_186_ab_4b_a4b-001           7JMK
● Zn$_3$Mo$_3$O$_8$:
  A3B8C2_hP26_186_c_ab2c_2b-001           DG1Q
● Swedenborgite (NaBe$_4$SbO$_7$, E9$_2$):
  A4BC7D_hP26_186_ac_b_a2c_b-001          QF4S
● Al$_7$C$_3$N$_3$:
  A7B3C3_hP26_186_3a4b_2ab_a2b-001        JALT
○ LiClO$_4$·3H$_2$O (H4$_{18}$):

AB6CD7_hP30_186_b_d_a_b2c-001     TXNW

∘ $Nd(BrO_3)_3 \cdot 9H_2O$ ($G2_2$):
A3B9CD9_hP44_186_c_3c_b_cd-001     776H

• $Ca_5Pb_3$:
A5B3_hP48_186_3cd_3c-001     XPM6

• $Ce_{24}Co_{11}$:
A24B11_hP70_186_2ab7c_ab3c-001     D6S6

## P6m2 (187):
∘ Tungsten Carbide (WC, $B_h$):
AB_hP2_187_a_d-001     X4K0

∘ BaPtSb:
ABC_hP3_187_b_c_e-001     NKMF

∘ $Re_3N$:
AB3_hP4_187_a_bh-001     6BD5

• ZrTaNO:
ABCD_hP4_187_c_b_a_f-001     84DL

• $Ti_2InB_2$:
A2BC2_hP5_187_ac_b_i-001     AJFU

• $\beta$-CuI [Bührer-Hälg]:
A2B_hP6_187_gi_ad-001     P31P

• $\epsilon$-GaSe:
AB_hP8_187_gh_gh-001     M3JC

• $LiCo_6P_4$:
A6BC4_hP11_187_jk_a_ck-001     NFL9

∘ Cr-233 Quasi-One-Dimensional Superconductor ($K_2Cr_3As_3$):
A3B3C2_hP16_187_jk_jk_ak-001     XP6Z

∘ $Cs_7O$:
A7B_hP24_187_ah2j2kn_j-001     BZC9

## P6c2 (188):
∘ $LiScI_3$:
A3BC_hP10_188_k_a_c-001     BL96

∘ $Sr_2Be_2B_2O_7$:
A2B2C7D2_hP26_188_i_h_cl_ab-001     FFAL

∘ $BaSi_4O_9$ ($S3_2$):
AB9C4_hP28_188_a_kl_ck-001     2DJ4

## P62m (189):
∘ $Th_3Pd_5$:
A5B3_hP8_189_cf_g-001     K8CX

∘ Barringerite (Revised $Fe_2P$, $C22$) Crystal:
A2B_hP9_189_fg_ad-001     6UK2

∘ ZrNiAl:
ABC_hP9_189_f_bc_g-003     0T4U

• $\beta_1$-$K_2UF_6$:
A6B2C_hP9_189_fg_c_b-001     XTDJ

• NaO:
AB_hP12_189_fg_eh-001     PQWK

∘ $\pi$-$FeMg_3Al_8Si_6$ ($E9_b$):
A8BC3D6_hP18_189_agh_b_f_i-001     WS3B

∘ $\pi$-$FeMg_3Al_9Si_5$:
A9BC3D5_hP18_189_fi_a_g_bh-001     0WD7

• Hexagonal $Au_7P_{10}I$:
A7BC10_hP18_189_ceg_a_hi-001     16HD

• $CsCrF_4$:
ABC4_hP18_189_f_g_fgj-001     6BNK

• $Ca_5Ir_3O_{12}$:
A5B3C12_hP20_189_dg_f_2gj-001     8TFL

## P62c (190):
∘ $\alpha$-$Sm_3Ge_5$ (High-temperature):
A5B3_hP16_190_bch_g-001     PJF4

∘ $Li_2Sb$:
A2B_hP18_190_gh_bf-001     DA98

∘ $CsSO_3$ ($K1_2$):

∘ Troilite (FeS):
AB_hP24_190_i_afh-001     6ZXP

∘ Bastnäsite [$CeF(CO_3)$, $G7_1$]:
ABCD3_hP36_190_h_g_af_hi-001     XFRG

## P6/mmm (191):
∘ Simple Hexagonal ($HgSn_{6-10}$ $A_f$):
A_hP1_191_a-001     3V7H

∘ Hexagonal $\omega$ ($C32$):
AB2_hP3_191_a_d-001     75K4

∘ $Li_3N$:
A3B_hP4_191_bc_a-001     RYZL

∘ $Cu_2Te$ ($C_h$):
A2B_hP3_191_h_e-001     X9YK

∘ $AlB_4Mg$:
AB4C_hP6_191_a_h_b-001     PDN7

∘ $CaCu_5$ ($D2_d$):
AB5_hP6_191_a_cg-001     F7Y6

∘ CoSn ($B35$):
AB_hP6_191_f_ad-001     Z7YH

• $Zr_4Al_3$:
A3B4_hP7_191_f_de-001     AMP4

• $KV_3Sb_5$:
AB5C3_hP9_191_b_ah_f-001     F3TR

• Hexagonal $WO_3$:
A3B_hP12_191_gl_f-001     SQAX

• Lucabindiite ($KAs_4O_6Cl$):
A4BCD6_hP12_191_h_a_b_i-001     4NVE

• $BaFe_2Al_9$:
AB2C9_hP12_191_fm_a_c-001     CSJ8

∘ $D2_a$ (approximate $TiBe_{12}$):
A12B_hP13_191_cdei_a-001     FEUM

• $HfFe_6Ge_6$:
A6B6C_hP13_191_i_cde_a-002     4E59

## P6/mcc (192):
∘ Beryl ($Be_3Al_2Si_6O_{18}$, $S3_1$):
A2B3C18D6_hP58_192_c_f_lm_l-001     52T5

∘ $AlPO_4$:
AB2_hP72_192_m_j2kl-001     M1VB

• Osumilite ($KMg_2Al_3Si_{12}O_{30}$):
A3BC2D30E12_hP96_192_f_a_c_l2m_m-001     FPN8

## P6₃/mcm (193):
∘ Mavlyanovite ($Mn_5Si_3$, $D8_8$):
A5B3_hP16_193_dg_g-001     AQOJ

∘ $Ti_5Ga_4$:
A4B5_hP18_193_bg_dg-001     443S

∘ $D0_6$ (Tysonite, $LaF_3$) (Obsolete):
A3B_hP24_193_ack_g-001     JJ07

• Ordered $TmBO_3$:
AB3C_hP30_193_g_gk_bd-001     AE3Q

## P6₃/mmc (194):
∘ Hexagonal Close Packed (Mg, $A3$, hcp):
A_hP2_194_c-001     LZW5

∘ Nickeline (NiAs, $B8_1$):
AB_hP4_194_c_a-001     Y3TN

∘ BN ($B_k$):
AB_hP4_194_c_d-001     SHYT

∘ $\alpha$-La ($A3'$):
A_hP4_194_ac-001     9631

∘ Hexagonal Graphite ($A9$):
A_hP4_194_bc-001     W4SC

∘ Lonsdaleite (Hexagonal Diamond):
A_hP4_194_f-001     4AMD

- ○ LiZn$_2$ ($C_k$):
  - AB_hP4_194_a_c-003   CY59
- ○ $L'3_0$ (approximate Fe$_2$N):
  - AB_hP4_194_c_a-003   Q425
- ○ CaIn$_2$:
  - AB2_hP6_194_b_f-001   21YK
- ○ InNi$_2$ ($B8_2$):
  - AB2_hP6_194_c_ad-001   X5ZH
- ○ Molybdenite (MoS$_2$, $C7$):
  - AB2_hP6_194_c_f-001   A8HJ
- ○ LiBC:
  - ABC_hP6_194_c_d_a-001   F0CN
- ○ Hypothetical Tetrahedrally Bonded Carbon with 3-Member Rings:
  - A_hP6_194_h-001   1BYJ
- ● ReB$_2$:
  - A2B_hP6_194_f_c-003   TH7D
- ● Hexagonal High-Temperature NbS$_2$:
  - AB2_hP6_194_b_f-002   Z5TL
- ○ Ni$_3$Sn ($D0_{19}$):
  - A3B_hP8_194_h_c-001   LJF8
- ○ Na$_3$As ($D0_{18}$):
  - AB3_hP8_194_c_bf-001   5M89
- ○ AlCCr$_2$:
  - ABC2_hP8_194_c_a_f-007   NEW2
- ○ AsTi ($B_i$):
  - AB_hP8_194_ac_f-004   E9RV
- ○ ReB$_3$:
  - A3B_hP8_194_af_c-001   72PY
- ○ Hexagonal Delafossite (CuFeO$_2$):
  - ABC2_hP8_194_a_c_f-005   Y6AL
- ● SnTaS$_2$:
  - A2BC_hP8_194_e_a_c-001   DG69
- ● NiMoP$_2$:
  - ABC2_hP8_194_b_a_f-001   D1PM
- ● LiO:
  - AB_hP8_194_ac_f-003   JV7G
- ● $\beta$-GaSe:
  - AB_hP8_194_f_f-003   BDG1
- ● Pt$_2$Sn$_3$ ($D5_b$):
  - A2B3_hP10_194_f_bf-001   Y79W
- ○ Na$_{0.74}$CoO$_2$:
  - AB2C2_hP10_194_a_bc_f-001   6X9E
- ○ EuIn$_2$P$_2$:
  - AB2C2_hP10_194_a_f_f-001   PL9Q
- ● $\beta$-Be$_3$N$_2$:
  - A3B2_hP10_194_bf_ac-001   CJUB
- ● BaNiO$_3$:
  - ABC3_hP10_194_c_a_h-001   UTCF
- ● High Temperature GdBO$_3$:
  - ABC3_hP10_194_c_a_h-002   R6VB
- ● $\beta$-Tridymite (SiO$_2$, $C10$):
  - A2B_hP12_194_cg_f-001   3KTM
- ○ MgZn$_2$ Hexagonal Laves ($C14$):
  - AB2_hP12_194_f_ah-001   LL0C
- ○ CMo:
  - AB_hP12_194_af_bf-001   Q5G8
- ○ Covellite (CuS, $B18$):
  - AB_hP12_194_cf_de-001   QP0Q
- ● Fe$_3$GeTe$_2$:
  - A3BC2_hP12_194_ce_d_f-001   ANDB
- ● W$_2$B$_5$ ($D8_h$):
  - A5B2_hP14_194_abcf_f-001   S78A
- ○ AlN$_3$Ti$_4$:

- ○ AB3C4_hP16_194_c_af_ef-001   53FU
- ○ Ni$_3$Ti ($D0_{24}$):
  - A3B_hP16_194_gh_ac-001   7D2A
- ○ OP4-LiNaCo$_2$O$_4$:
  - A2BC2D4_hP18_194_f_a_cd_ef-001   C4AH
- ● Disordered YBO$_3$:
  - A3B5C_hP18_194_h_fh_a-001   3718
- ● CeH$_9$:
  - AB9_hP20_194_c_bfk-001   B62K
- ○ Proposed 300 GPa HfH$_{10}$:
  - A10B_hP22_194_bhj_c-001   MK58
- ○ Hexagonal Ti$_6$Sn$_5$:
  - A5B6_hP22_194_ach_gh-001   O9JW
- ○ MgNi$_2$ Hexagonal Laves ($C36$):
  - AB2_hP24_194_ef_fgh-001   HV5V
- ○ Lu$_2$CoGa$_3$:
  - AB3C2_hP24_194_f_k_bh-001   OGA4
- ● VCo$_3$:
  - A3B_hP24_194_hk_bf-001   XEL5
- ● Hexagonal PuAl$_3$:
  - A3B_hP24_194_hk_bf-002   E109
- ● Low Temperature NbSe$_2$:
  - AB2_hP24_194_bh_fk-001   YM9E
- ● CeNi$_3$:
  - A3B_hP24_194_cf_abdk-001   1SSV
- ● Al$_9$Mn$_3$Si ($E9_c$):
  - A9B3C_hP26_194_hk_h_a-001   KM59
- ● Hexagonal Vaterite (CaCO$_3$):
  - A3BC9_hP26_194_h_a_hk-001   LCWU
- ● Na$_9$Pb$_4$ ($\delta'$-NaPb):
  - A9B4_hP26_194_ce3f_ef-001   J5ST
- ● Co$_2$Al$_5$ ($D8_{11}$):
  - A5B2_hP28_194_ahk_ch-001   VX03
- ● Cs$_3$Cr$_2$Cl$_9$:
  - A9B2C3_hP28_194_hk_f_bf-001   77Z6
- ● Hexagonal Ba$_3$CoIr$_2$O$_9$:
  - A3BC2D9_hP30_194_bf_a_f_hk-001   BHER
- ● Hexagonal BaTiO$_3$:
  - AB3C_hP30_194_bf_hk_af-001   VHHA
- ● BaLi$_4$:
  - AB4_hP30_194_h_afhk-001   JAJR
- ● Co$_3$W$_9$C$_4$:
  - A2B3C3_hP32_194_cg_2h_k-001   JSHF
- ● Hexagonal $\alpha$-Ca$_2$SiO$_4$:
  - A3B12C_hP32_194_af_2k_c-001   CEQR
- ● $S3_4$(II) (Catapleiite, Na$_2$Zr(SiO$_3$)$_3$·H$_2$O) (*Obsolete*):
  - A3B2C9D3E_hP36_194_g_f_hk_h_a-001   6TN8
- ● Ce$_2$Ni$_7$:
  - A2B7_hP36_194_2f_aefhk-001   AOTL
- ● Na$_{13}$Pb$_5$ ($\gamma$-NaPb):
  - A13B5_hP36_194_a2e4f_b2f-001   G1KV
- ● Th$_2$Ni$_{17}$:
  - A17B2_hP38_194_fgjk_bc-001   DQFR
- ● Barlowite [Cu$_4$FBr(OH)$_6$]:
  - A6BC6D6E_hP40_194_c_gh_b_k_k-001   GN34
- ● Al$_{23}$V$_4$:
  - A23B4_hP54_194_fh3k_ah-001   W4P8
- ● Ba$_6$Nd$_2$Ti$_4$O$_{17}$:
  - A6B2C17D4_hP58_194_ab2f_e_fh2k_2f-001   X8HZ
- ● $\beta$-Alumina ($D5_6$, Al$_2$O$_3$):
  - A2B3_hP60_194_3fk_cdef2k-001   T15Z
- ○ Magnetoplumbite (PbFe$_{12}$O$_{19}$):
  - A12B19C_hP64_194_ab2fk_efh2k_c-001   5L5V

- SrMg$_4$:
  A4B_hP90_194_e2fgh4k_hk-001     DU5B

**P23 (195)**:
- Cd$_8$As$_7$Cl:
  A7B12C_cP20_195_ag_3e_b-001     J7FX
- PrRu$_4$P$_{12}$:
  A12BC4_cP34_195_2j_ab_2e-001     8SJC

**F23 (196)**:
- Cu$_2$Fe[CN]$_6$:
  A12B2C_cF60_196_h_ac_b-001     DJ7F
- Ca$_{14}$Zn$_6$Al$_{10}$O$_{35}$:
  A12B14C35D4_cF260_196_abeg_2ef_cef2h_e-001     JELG
- Li$_{22}$Si$_5$:
  A22B5_cF432_196_abcd6efg4h_2efg-001     GA7M

**I23 (197)**:
- Ga$_4$Ni:
  A4B_cI40_197_cde_c-001     PBV9
- Hf$_{10}$Ta$_3$S$_3$:
  A11B3C2_cI64_197_cdf_e_c-001     8883
- γ-Bi$_2$O$_3$:
  A13B2O_cI66_197_af_2cf-001     8111

**P2$_1$3 (198)**:
- α-CO (B21):
  AB_cP8_198_a_a-001     LCVR
- FeSi (B20):
  AB_cP8_198_a_a-002     93P7
- α-N (P2$_1$3):
  A_cP8_198_2a-001     VHKA
- Ullmanite (NiSbS, F0$_1$):
  ABC_cP12_198_a_a_a-001     VHP8
- Ammonia (NH$_3$, D0$_1$):
  A3B_cP16_198_b_a-001     QTNQ
- Sodium Chlorate (NaClO$_3$, G0$_3$):
  ABC3_cP20_198_a_a_b-001     QEXW
- α-Carnegieite (NaAlSiO$_4$, S6$_5$):
  ABC4D_cP28_198_a_a_ab_a-001     M6W2
- Na$_2$Si$_2$Si$_5$ (S6$_6$):
  A2C4D_cP32_198_a_2a_ab_a-001     MFVM
- Cubic Cu$_2$OSeO$_3$:
  A2B4C_cP56_198_ab_2a2b_2a-001     PZ2V
- Low Temperature CsW$_2$O$_6$:
  A6C2_cP72_198_2a_4b_ab-001     WBDQ
- Langbeinite [K$_2$Mg$_2$(SO$_4$)$_3$]:
  A2B2C12D3_cP76_198_2a_2a_4b_b-001     98LD

**I2$_1$3 (199)**:
- CoU (B$_a$):
  AB_cI16_199_a_a-001     TFMD
- Cordeeoite (α-Hg$_3$S$_2$Cl$_2$):
  A2B3C2_cI28_199_a_b_a-002     VYHG
- C26$_a$ (NO$_2$) (Obsolete):
  AB2_cI36_199_b_c-001     VGMG

**Pm3 (200)**:
- CsHgN$_3$O$_6$:
  ABC3D6_cP11_200_a_b_c_f-001     4JP4
- C-AlRuNi (Al$_{20}$Ni$_3$Ru$_5$):
  A23B2C6_cP31_200_cij_ab_f-001     L1EN
- Mg$_2$Zn$_{11}$ (D8$_c$):
  A2B11_cP39_200_f_begik-001     8FHP

**Pn3 (201)**:
- KSbO$_3$ (High-Temperature):
  A3BC_cP60_201_be_fh_g-001     WXOM
- Bi$_3$Ru$_3$O$_{11}$:
  A3B11C3_cP68_201_be_efh_g-001     0WEU

- CuMo$_3$Cl$_7$:
  A7BC3_cP88_201_e2h_e_h-001     HF2G
- KW$_3$Br$_7$:
  A28B5C12_cP90_201_e2h_bd_h-001     RMGB
- Ce$_6$Cd$_{37}$:
  A41B6_cP188_201_b2efg5h_h-001     D917

**Fm3 (202)**:
- α-CuZrF$_6$:
  AB12C_cF56_202_a_h_b-001     G65V
- K$_3$Co(NO$_2$)$_6$ (J2$_4$):
  AB3C6D12_cF88_202_a_bc_e_h-001     G8R0
- KB$_6$H$_6$:
  A6B6C_cF104_202_h_h_c-001     PZB0
- FCC C$_{60}$ Buckminsterfullerene:
  A_cF240_202_h2i-001     TV0E

**Fd3 (203)**:
- Rb$_3$AsSe$_{16}$:
  AB3C16_cF160_203_a_bc_eg-001     HAGZ
- Pyrochlore (Na$_3$Co(CO$_3$)$_2$Cl):
  A2BCD3E6_cF208_203_e_c_d_f_g-001     LH7J
- Tychite (Na$_6$Mg$_2$SO$_4$(CO$_3$)$_4$):
  A4B2C6D16E_cF232_203_e_c_f_eg_b-001     LJXD

**Im3 (204)**:
- Al$_{12}$W:
  A12B_cI26_204_g_a-001     JDL2
- Skutterudite (CoAs$_3$, D0$_2$):
  A3B_cI32_204_g_c-001     9U6F
- LaFe$_4$P$_{12}$:
  A4BC12_cI34_204_c_a_g-001     GXVQ
- Modern C26 (NO$_2$):
  AB2_cI36_204_d_g-001     XJC7
- NaMn$_7$O$_{12}$:
  A7BC12_cI40_204_bc_a_g-001     M5NA
- CaCu$_3$Mn$_4$O$_{12}$:
  A3B4D12_cI40_204_a_b_c_g-001     RGPS
- Ru$_3$Be$_{17}$:
  A17B3_cI160_204_def2gh_g-001     92TS
- Bergman [Mg$_{32}$(Al,Zn)$_{49}$]:
  A32C48_cI162_204_a_2efg_2gh-001     9AQF
- YCd$_6$:
  A20B3_cI184_204_def3gh_g-001     HYXN

**Pa3 (205)**:
- α-N (Pa3):
  A_cP8_205_c-001     84Y3
- Pyrite (FeS$_2$, C2):
  AB2_cP12_205_a_c-001     M44S
- SC16 (CuCl):
  AB_cP16_205_c_c-001     F4G8
- NaSbF$_6$:
  A6BC_cP32_205_d_a_b-001     N7XD
- Pb(NO$_3$)$_2$ (G2$_1$):
  A2B6C_cP36_205_c_d_a-001     LPJD
- SnI$_4$ (D1$_1$):
  A4B_cP40_205_cd_c-001     MJPK
- ZrP$_2$O$_7$ (K6$_1$) High-Temperature:
  A7B2C_cP40_205_ad_c_b-001     NTUQ
- Zn(BrO$_3$)$_2$ · 6H$_2$O (J1$_{10}$):
  A2B6C6D_cP60_205_c_d_d_a-001     Q0QA
- H64 [Ni(NO$_3$)$_2$(NH$_3$)$_6$] (Obsolete):
  A2B6C6D_cP60_205_c_d_a_d-001     GED2
- CaB$_2$O$_4$ (IV):
  A2BC4_cP84_205_d_ac_2d-001     TZRB
- NaCr(SO$_4$)$_2$ · 12H$_2$O Alum:

AB12CD8E2_cP96_205_a_2d_b_cd_c-001     AUGT

○ $\gamma$–Alum [AlNa(SO$_4$)$_2$ · 12H$_2$O, $H4_{15}$]:
AB24CD20E2_cP192_205_a_4d_b_c3d_c-001     VBLW

○ $\alpha$–Alum [KAl(SO$_4$)$_2$ · 12H$_2$O, $H4_{13}$]:
AB24CD28E2_cP224_205_a_4d_b_2c4d_c-001     MHB8

○ Simple Cubic C$_{60}$ Buckminsterfullerene:
A_cP240_205_10d-001     3VYE

○ $\beta$–Alum [Al(NH$_3$CH$_3$)$_2$(SO$_4$)$_2$ · 12H$_2$O, $H4_{14}$]:
AB2C36D2E20F2_cP252_205_a_c_6d_c_c3d_c-001     SFC7

○ Ca$_3$Al$_2$O$_6$:
A2B3C6_cP264_205_2d_ab2c2d_6d-001     Q2JD

**Ia3 (206):**

○ BC8 (Si):
A_cI16_206_c-001     83MZ

○ Bixbyite (Mn$_2$O$_3$, $D5_3$):
AB3C6_cI80_206_a_d_e-001     G3EF

○ AlLi$_3$N$_2$ ($E9_d$):
AB3C2_cI96_206_c_e-ad-001     WEUP

● ZrTe$_3$O$_8$:
A8B3C_cI96_206_ce_d_a-001     3HHF

**P432 (207):**

○ Palladseite (Pd$_{17}$Se$_{15}$):
A17B15_cP64_207_acfk_eij-001     8LHO

**P4$_2$32 (208):**

○ H$_3$P:
A3B_cP16_208_i_c-001     CCPW

○ Cs$_2$ZnFe[CN]$_6$:
A6B2C2D6E_cP64_208_m_ad_b_m_c-001     OHGH

● Na$_{11}$U$_5$O$_{16}$:
A11B16C5_cP64_208_bfh_adm_ce-001     NFNJ

**F432 (209):**

○ KPF$_6$:
A24BC_cF104_209_j_a_b-001     XBGM

● PCN-20 [C$_{14}$CuO(H$_2$O)$_4$]:
A14BC8D5_cF1344_209_7j_g_4j_g2j-001     5WNB

**F4$_1$32 (210):**

○ MgB$_{12}$H$_{12}$[H$_2$O]$_{12}$:
A12B36CD12_cF488_210_h_3h_a_fg-001     X29X

○ Te[OH]$_6$:
A12B6C_cF608_210_4h_2h_e-001     7K4Q

**I432 (211):**

○ Hypothetical Cubic SiO$_2$:
A2B_cI72_211_hi_i-001     NT1Y

**P4$_3$32 (212):**

○ SrSi$_2$:
A2B_cP12_212_c_a-001     XX75

● Li$_2$Pd$_3$B:
AB2C3_cP24_212_a_c_d-001     K791

○ $\gamma$–Fe$_2$O$_3$ ($D5_7$):
A2B3_cP60_212_acd_bce-001     J7V3

**P4$_1$32 (213):**

○ $\beta$–Mn ($A13$):
A_cP20_213_cd-001     4VEQ

○ Mg$_3$Ru$_2$:
A3B2_cP20_213_d_c-001     6QE7

● Co$_8$Zn$_9$Mn$_3$:
A2B3_cP20_213_c_d-001     KXT1

● Al$_2$Mo$_3$C:
A2BC3_cP24_213_c_a_d-001     SRHO

● SrCuTe$_2$O$_6$:
AB6CD2_cP120_213_d_3e_ac_e-001     1AJT

**I4$_1$32 (214):**

○ Petzite (Ag$_3$AuTe$_2$):

A3B2C_cI48_214_f_a_e-001     3JDV

○ Ca$_3$PI$_3$:
A3BC_cI56_214_g_h_a-001     4XVN

● La$_3$Rh$_4$Sn$_{13}$:
A3B4C13_cI320_214_gh_abgh_e4i-001     1DPD

**P43m (215):**

● Fe$_4$C:
AB4_cP5_215_a_e-001     WNQX

○ Sulvanite (Cu$_3$S$_4$V, $H2_4$):
A3B4C_cP8_215_c_e_b-001     WCVT

○ Cubic Lazarevićite (AsCu$_3$S$_4$):
AB3C4_cP8_215_a_c_e-001     31D9

● Intermediate Temperature TmNi$_2$:
A2B_cP24_215_ei-ace-001     X751

○ $\gamma$–brass (Cu$_9$Al$_4$, $D8_3$):
A4B9_cP52_215_ei_3efgi-001     TRNX

● Palladseite (Pd$_{17}$Se$_{15}$):
A17B15_cP64_215_acg2i_f2i-001     55EV

**F43m (216):**

○ Zincblende (ZnAs, $B3$):
AB_cF8_216_a_c-001     TL8Z

○ Half–Heusler (AgAsMg, $C1_b$):
ABC_cF12_216_a_c_b-001     3BJR

○ Quaternary Heusler (LiMgAuSn):
ABCD_cF16_216_a_b_c_d-001     4CEX

○ Hg$_2$TiCu Inverse Heusler:
AB2C_cF16_216_a_bc_d-001     4M8A

○ AuBe$_5$ ($C15_b$):
AB5_cF24_216_a_ce-001     3454

○ High-Temperature Cubic KClO$_4$ ($H0_5$):
ABC4_cF24_216_a_b_e-001     Q1UZ

○ Theoretical cF40 AlN:
AB_cF40_216_ae_be-001     3QP3

○ Room Temperature GaMo$_4$S$_8$:
AB4C8_cF52_216_a_e_2e-001     BEEG

● LiGaCr$_4$O$_8$:
A4BCD8_cF56_216_e_a_c_2e-001     1XJV

● Al$_{13}$Cr$_4$Si$_4$:
A13B4C4_cF84_216_afg_e_e-001     5ADG

○ Cubic Cr$_4$PtGa$_{17}$:
A4B17C_cF88_216_aefg_c-001     9Q9K

● Gd$_4$RhIn:
A4BC_cF96_216_efg_e_e-001     39CF

● Ba$_3$In$_2$Zn$_5$O$_{11}$:
A3B2C11D5_cF168_216_f_e_ab2eh_eg-001     Z69R

● Low Temperature TmNi$_2$:
A2B_cF192_216_2e2h_ab2eg-001     5WAR

○ Zunyite [Al$_{13}$(OH,F)$_{18}$Si$_5$O$_{20}$Cl ($S0_8$)]:
A13BC18D20E5_cF228_216_ah_c_gh_2eh_be-001     R5TJ

○ Murataite [(Y,Na)$_6$(Zn,Fe)$_5$Ti$_{12}$O$_{29}$(O,F)$_{10}$F$_4$]:
A16B40C12D6E5_cF316_216_eh_e2g2h_h_f_ae-001     ZX49

● Pt$_3$Zn$_{10}$:
A2B5_cF392_216_4efg_4ef4h-001     V17U

● $\delta$–Cu$_{41}$Sn$_{11}$:
A41B11_cF416_216_7e2fg3h_egh-001     1MZW

● Li$_{17}$Pb$_4$:
A17B4_cF420_216_a6efg4h_2efg-001     4E2L

● Sm$_{11}$Cd$_{45}$:
A45B11_cF448_216_ac4efg5h_bd2eh-001     LZ4J

**I43m (217):**

○ SiF$_4$ ($D1_2$):
A4B_cI10_217_c_a-001     WBME

○ Theoretical cI16 AlN:
AB_cI16_217_c_c-001    `6RJH`

● Tl₃VS₄:
A4B3C_cI16_217_c_b_a-001    `YOMN`

○ γ-Brass (Cu₅Zn₈, $D8_2$):
A5B8_cI52_217_ce_cg-001    `AYKC`

● α-Ba₈Ga₁₆Sn₃₀ Clathrate:
A4B4C6D13_cI154_217_c_c_d_ag-001    `YXNP`

○ α-Mn (A12):
A_cI58_217_ac2g-001    `SV19`

● Mg₁₇Al₁₂:
A12B17_cI58_217_g_acg-001    `56N2`

○ Tennantite (Cu₁₂As₄S₁₃):
A4B24C13_cI82_217_c_deg_ag-001    `5PAV`

**P4̄3n (218):**

○ Ag₃(PO₄) ($H2_1$):
A3B4C_cP16_218_c_e_a-001    `X143`

○ Sodalite [Na₄(AlSiO₄)₃Cl, $S6_2$]:
A3BC4D12E3_cP46_218_c_a_e_i_d-001    `NYVB`

● KGe:
AB_cP64_218_ei_ei-001    `RUTB`

○ Hauyne
[(Na₀.₅Ca₀.₃K₀.₂)₈(Al₆Si₆O₂₄)(SO₄)₁.₅, $S6_9$]:
A3B4C4D4E16F4G3_cP76_218_c_e_e_e_
ei_e_d-001    `BCKL`

● Li₇MnN₄:
A7B4C96_cP96_218_bcefi_ad_ei-001    `HJWV`

○ γ-HBO₂ (cubic):
ABC2_cP96_218_i_i_2i-001    `W9M1`

**F4̄3c (219):**

● Si₃Cl₈:
A8B3_cF176_219_eh_abe-001    `ZRK3`

○ Boracite (Mg₃B₇ClO₁₃):
A7BC3D13_cF192_219_ce_a_d_bh-001    `OWNV`

● Sn[Co(CO)₄]₄:
A16B4C16D_cF296_219_eh_e_eh_a-001    `6J8K`

**I4̄3d (220):**

○ High-pressure cI16 Li:
A_cI16_220_c-001    `BTF5`

○ Theoretical cI24 AlN:
AB_cI24_220_a_b-001    `MH9E`

○ Th₃P₄ ($D7_3$):
A4B3_cI28_220_c_a-001    `JM9C`

○ Pu₂C₃ ($D5_c$):
A3B2_cI40_220_d_c-001    `M1ZX`

● Y₃Au₃Sb₄:
A3B4C4_cI40_220_a_c_b-001    `GET1`

○ Cu₁₅Si₄ ($D8_6$):
A15B4_cI76_220_ae_c-001    `MVX2`

○ Eulytine (Bi₄(SiO₄)₃, $S1_5$):
A4B12C3_cI76_220_c_e_a-001    `VQL6`

○ Mayenite (12 CaO·7Al₂O₃, $K74$, C12A7):
A7B12C19_cI152_220_ac_2d_bce-001    `L8NP`

○ Al(PO₃)₃ ($G5_2$):
AB9C3_cI208_220_c_3e_e-001    `2BX1`

**Pm3̄m (221):**

○ α-Po ($A_h$, simple cubic):
A_cP1_221_a-001    `NG7Y`

○ CsCl (B2):
AB_cP2_221_a_b-002    `QM6B`

○ NH₄NO₃ I ($G0_8$):
AB_cP2_221_a_b-001    `5YKM`

○ α-ReO₃ ($D0_9$):

A3B_cP4_221_c_b-001    `3DRV`

○ Bogdanovite (Cu₃Au, $L1_2$):
AB3_cP4_221_a_c-001    `NK1T`

○ Cubic Perovskite (CaTiO₃, $E2_1$):
AB3C_cP5_221_a_c_b-001    `QUKL`

*Erroneous $L'1_0$:*
A4B_cP5_221_bc_a-001    `Q52R`

● γ'-Fe₄N ($L'1_0$):
A4B_cP5_221_ac_b-002    `TEA9`

○ NbO:
AB_cP6_221_c_d-001    `H75X`

● CaB₆ ($D2_1$):
A6B_cP7_221_e_b-001    `GKJM`

● S₃U₄:
A3B4_cP7_221_d_ac-001    `62DR`

○ Predicted High-Pressure YCaH₁₂:
AB12C_cP14_221_a_h_b-001    `SQ16`

○ Model of Ferrite (cP16):
AB11CD3_cP16_221_a_dg_b_c-001    `729N`

○ Model of Austenite (cP32):
AB27CD3_cP32_221_a_dij_b_c-001    `3F7M`

○ Ca₃Al₂O₆ ($E9_1$):
A2B3C6_cP33_221_cd_ag_fh-001    `TZEN`

● Ce₈Pd₂₄Sb:
A8B24C_cP33_221_g_efh_a-001    `CLMU`

○ BaH₁₁ ($D2_e$):
AB11_cP36_221_c_agij-001    `2EDD`

● Palladseite (Pd₁₇Se₁₅):
A17B15_cP64_221_acfm_eij-001    `RWLL`

**Pn3̄n (222):**

○ Ce₈Mo₃O₁₆:
A5B3C16_cP96_222_ce_d_fi-001    `3NUA`

● Co(H₂O)₂(C₄O₄):
A4BC4D6_cP360_222_2i_h_2i_3i-001    `7XG8`

**Pm3̄n (223):**

○ Cr₃Si (A15):
A3B_cP8_223_c_a-001    `7MHE`

● SrB₃C₃ Clathrate:
A3B3C_cP14_223_c_d_a-001    `FACW`

● NaPt₃O₄:
AB4C3_cP16_223_a_e_c-001    `R1MG`

● β-UH₃:
A3B_cP32_223_k_ac-001    `4UMV`

● Yb₃Rh₄Sn₁₃:
A4B13C3_cP40_223_e_ak_c-001    `GYST`

○ Si₄₆ Clathrate:
A_cP46_223_cik-001    `VFEB`

● β-Ba₈Ga₁₆Sn₃₀ Clathrate:
A13B3C8D12_cP72_223_ak_c_i_k-001    `KF6U`

● "A15" Fullerene (Ba₃C₆₀):
AB20_cP126_223_c_k2l-001    `HVOS`

● β-Hg₃S₂Cl₂:
A2B3C2_cP224_223_abcdefk_j3k_il-001    `UWQ4`

**Pn3̄m (224):**

○ Cuprite (Cu₂O, C3):
A2B_cP6_224_b_a-001    `WUJH`

● Mg₃P₂ ($D5_5$):
A3B2_cP10_224_d_b-001    `LQ7A`

○ PW₁₂O₄₀·3H₃O:
A3B40CD12_cP112_224_d_e3k_a_k-001    `R1HF`

○ 12-phosphotungstic acid [H₃PW₁₂O₄₀·5H₂O]
($H4_{16}$):
A5B40CD12_cP116_224_bd_e3k_a_k-001    `E8XB`

○ Dodecatungstophosphoric Acid Hexahydrate ($H_3PW_{12}O_{40} \cdot 6H_2O$):
A27B52CD12_cP184_224_dl_eh3k_a_k-001    **XBJQ**

## Fm$\bar{3}$m (225):
○ Face-Centered Cubic (Cu, $A1$, fcc):
A_cF4_225_a-001    **6EYD**
○ Rock Salt/Halite (NaCl, $B1$):
AB_cF8_225_a_b-001    **Q53X**
○ Fluorite ($CaF_2$, $C1$):
AB2_cF12_225_a_c-001    **8VXQ**
○ Heusler ($Cu_2AlMn$, $L2_1$):
A2BC_cF16_225_a_c_b-001    **O2WQ**
○ $BiF_3$ ($D0_3$):
AB3_cF16_225_a_bc-001    **KFY2**
○ $Ca_7Ge$:
A7B_cF32_225_ad_b-001    **CAN8**
○ $L1_a$ (disputed $CuPt_3$):
A7B_cF32_225_a_bd-001    **3APN**
● $\alpha'$-$CuZrF_6$:
A6BC_cF32_225_a_e_b-001    **2016**
○ $K_2PtCl_6$ ($J1_1$):
A6B2C_cF36_225_e_c_a-001    **L6LF**
● $\delta$-$Bi_2O_3$:
AB8_cF36_225_a_f-001    **X2Q7**
○ Double Perovskite ($Ba_2MnWO_6$):
A2BC6D_cF40_225_c_a_e_b-001    **83J5**
● Room Temperature $KBD_4$:
AB8C_cF40_225_a_f_b-001    **WSA1**
○ $LaH_{10}$ High-$T_c$ Superconductor:
A10B_cF44_225_cf_a-001    **8F2Q**
○ $UB_{12}$ ($D2_f$):
A12B_cF52_225_h_b-001    **JTKK**
● $Mg_6MnO_8$:
A6BC8_cF60_225_d_a_ce-001    **4YGB**
● $Co_9S_8$ ($D8_9$):
A9B8_cF68_225_af_ce-001    **X1EM**
○ Sulphohalite [$Na_6ClF(SO_4)_2$, $H5_8$]:
ABC6D8E2_cF72_225_a_b_e_f_c-001    **8Y3K**
● High Temperature $Cu_2Se$:
A18B_cF76_225_c2f_a-001    **7K0D**
○ $Cu_3[Fe(CN)_6]_2 \cdot xH_2O$ ($J2_5$):
A6B9CD2E6_cF96_225_e_af_b_c_e-001    **N4GS**
○ Model of Austenite (cF108):
AB18C8_cF108_225_a_eh_f-001    **VR5Q**
○ $Cr_{23}C_6$ ($D8_4$):
A6B23_cF116_225_e_acfh-001    **EY4X**
○ $Th_6Mn_{23}$ ($D8_a$):
A23B6_cF116_225_ad2f_e-001    **W8MF**
● $Mg_6Si_7Cu_{16}$:
A16B6C7_cF116_225_2f_e_ad-001    **NHP2**
○ Model of Ferrite (cF128):
A9B16C7_cF128_225_acd_2f_be-001    **FENM**
○ $(NH_4)_3AlF_6$ ($J2_1$):
AB30C16D3_cF200_225_a_ej_2f_bc-001    **U8XN**
● Intermediate Temperature Bornite ($Cu_{5/6}Fe_{1/6}$)$_3S_2$:
A7B_cF256_225_f2k_ce-001    **EMLR**
● fcc Fullerene ($K_3C_{60}$):
A40B_cF492_225_j2l_ac-001    **54ZJ**

## Fm$\bar{3}$c (226):
○ $NaZn_{13}$ ($D2_3$):
AB13_cF112_226_a_bi-001    **DCLX**

## Fd$\bar{3}$m (227):
○ Diamond ($A4$):
A_cF8_227_a-001    **X5M8**
○ NaTl ($B32$):
AB_cF16_227_a_b-001    **CHL9**
○ Ideal $\beta$-Cristobalite ($SiO_2$, $C9$):
A2B_cF24_227_c_a-001    **QDSV**
○ $Cu_2Mg$ Cubic Laves ($C15$):
A2B_cF24_227_c_b-001    **8YL7**
○ Cubic CuPt [$L1_3(I)$, $D4$]:
AB_cF32_227_c_d-001    **OHDN**
● T-Carbon:
A_cF32_227_e-001    **WPN9**
○ $Ti_2C$:
AB2_cF48_227_c_e-001    **PR5K**
● Spinel ($Al_2MgO_4$, $H1_1$):
A2BC4_cF56_227_c_b_e-001    **MZYE**
● Spinel ($Co_3O_4$, $D7_2$):
A3B4_cF56_227_ad_e-001    **SV6X**
● $\beta$-Alane ($AlD_3$):
AB3_cF64_227_c_f-001    **9LSQ**
○ Senarmontite ($D6_1$, $Sb_2O_3$):
A3B2_cF80_227_f_e-002    **UHQ6**
○ Pyrochlore Iridate ($Eu_2Ir_2O_7$, $E8_1$):
A2B2C7_cF88_227_c_d_af-001    **OR9R**
○ $NiTi_2$:
AB2_cF96_227_e_cf-001    **AVWT**
○ $D6_2$ ($Sb_2O_4$, *Obsolete*):
A2B_cF96_227_abf_cd-001    **UGWY**
○ $\eta$-carbide ($Fe_3W_3C$, $E9_3$):
AB3C3_cF112_227_c_de_f-001    **WVM5**
○ $\gamma$-$Ga_2O_3$:
A11B4_cF120_227_acdf_e-001    **VTDR**
○ $Si_{34}$ Clathrate:
A_cF136_227_aeg-001    **U115**
● $Dy_5Pd_2$:
A7B2_cF144_227_2ef_e-001    **WYVV**
● Predicted $Li_2MgH_{16}$ High-$T_c$ Superconductor (250 GPa):
A16B2C_cF152_227_eg_c_b-001    **WZBC**
● $Al_{10}V$:
A10B_cF176_227_cfg_d-001    **WWX6**
● $Mg_3Cr_2Al_{18}$:
A18B2C3_cF184_227_fg_d_ac-001    **3DEV**
○ $Zn_{22}Zr$:
A22B_cF184_227_cdfg_a-001    **MX6R**
○ $G7_3$ [$Na_3MgCl(CO_3)_2$] (*Obsolete*):
A2BCD3E6_cF208_227_e_c_d_f_g-001    **8XP6**
○ $H5_6$ [Tychite, $Na_6Mg_2SO_4(CO_3)_4$)] (*Obsolete*):
A4B2C6D16E_cF232_227_e_c_f_eg_b-001    **Y05P**
○ $H_3PW_{12}O_{40} \cdot 29H_2O$ ($H4_{21}$):
A29B40CD12_cF656_227_ae2fg_e3g_b_g-001    **3VYY**

## Fd$\bar{3}$c (228):
○ $Te[OH]_6$ (*Obsolete*):
A6B_cF224_228_h_c-001    **MSL9**
○ $CuCrCl_5[NH_3]_6$:
A5BCD6_cF416_228_eg_c_b_h-001    **19XK**
○ Voltaite ($K_2Fe_8Al[SO_4]_{12} \cdot 18H_2O$):
AB8C24D2E84F12_cF2096_228_a_cg_2h_b_7h_h-001    **E4V9**

## Im$\bar{3}$m (229):
● Body-Centered Cubic (W, $A2$, bcc):
A_cI2_229_a-001    **7W8F**
○ High-pressure (200GPa) $H_3S$:
A3B_cI8_229_b_a-001    **VV0Z**

- $\beta$-Hg$_4$Pt:
  A4B_cI10_229_c_a-001                78LF
- Pt$_3$O$_4$:
  A4B3_cI14_229_c_b-001               BYND
- High Pressure Cubic CaH$_6$:
  AB6_cI14_229_a_d-001                N2LT
- Model of Ferrite (cI16):
  AB4C3_cI16_229_a_c_b-001            YRPB
- Model of Austenite (cI32):
  AB12C3_cI32_229_a_h_b-001           L6PB
- Ir$_3$Ge$_7$ ($D8_f$):
  A7B3_cI40_229_df_e-001              9SK5
- $\alpha$-AgI ($B23$):
  A21B_cI44_229_bdh_a-001             QV5K
- Ce$_3$Ni$_6$Si$_2$:
  A3B6C2_cI44_229_e_h_c-001           BC5E
- Dy$_6$Fe$_{16}$O:
  A6B16C_cI46_229_e_ch_a-001          Y3T5
- $\gamma$-brass (Fe$_3$Zn$_{10}$, $D8_1$):
  A3B10_cI52_229_e_fh-001             TJ2T
- Sb$_2$Tl$_7$ ($L2_2$):
  A2B7_cI54_229_e_afh-001             KM4R

**Ia3d (230):**
- Ga$_4$Ni$_3$:
  A4B3_cI112_230_af_g-001             K0Z6
- RhBi$_4$:
  A4B_cI120_230_h_c-001               V61M
- Garnet [$S1_4$, Co$_3$Al$_2$(SiO$_4$)$_3$]:
  A2B3C12D3_cI160_230_a_c_h_d-001     YMRU
- Ca$_3$Al$_2$(OH)$_{12}$ ($J2_3$):
  A2B3C12D12_cI232_230_a_c_h_h-001    LL0W

## CRediT authorship contribution statement

**Hagen Eckert:** Writing – original draft, Writing – review and editing, Web page development, Software development. **Simon Divilov:** Web page development, Software development. **Michael J. Mehl:** Writing – original draft, Writing – review and editing, Web page development, Software development. **David Hicks:** Symmetry analysis, Software development. **Adam C. Zettel:** Web page development, Software development. **Marco Esters:** Web page development, Software development. **Xiomara Campilongo:** Writing – review and editing. **Stefano Curtarolo:** Writing – review and editing, Supervision, Funding acquisition, Resources.

## Declaration of competing interest

The authors declare that they have no known competing financial interests or personal relationships that could have appeared to influence the work reported in this paper.

## Data availability

All of the crystallographic information and sources for the structures found in the Encyclopedia are available on the web page for each prototype. In addition, the structural information may be obtained from AFLOW using the command `aflow --proto=STRING` where STRING is the AFLOW label for that prototype.

## Code availability

The underlying software used in this study and its source code can be accessed via this link https://aflow.org/install-aflow/.

## Acknowledgments

This work has been supported by Office of Naval Research through project N00014-20-1-2525 and a Multidisciplinary University Research Initiative (MURI) program under project number N00014-21-1-2515, as well as the National Science Foundation (NSF NRT-HDR DGE-2022040) as well as by high-performance computer time and resources from the DoD High Performance Computing Modernization Program (Frontier).